\documentclass{article}
\usepackage[T1]{fontenc}
\usepackage[latin1]{inputenc}
\usepackage{graphics}
\usepackage{graphicx}
\usepackage{longtable}
\usepackage{color}
\usepackage{pslatex}
\usepackage{hyperref}
\usepackage{sverb}
\usepackage{subfigure}
\usepackage{epstopdf}

\usepackage{multirow}
\usepackage{adjustbox}
\usepackage{indentfirst}
\usepackage{mathrsfs}

\usepackage{a4p}
\usepackage{rotating}

\usepackage{graphicx}
\usepackage{atlasphysics}
\usepackage{subfigure}
\usepackage{longtable}
\usepackage{multirow}
\usepackage{textcomp}
\usepackage{authblk,colortbl}
\usepackage{units}
\usepackage{amsmath,amssymb,amsfonts} % Typical maths resource packages

\usepackage{authblk}

\usepackage{rotating}

\begin{document}

\begin{titlepage}
 
\begin{flushright} 
%{ \bf IFJPAN-VIII-2016-21 %\\  
%} 
\end{flushright}
 
\vskip 30 mm
\begin{center}
  {\bf\huge  Deep Neural Network application: }\\
  \vskip 4mm
  {\bf\huge  Higgs boson CP state mixing angle }\\
  \vskip 4mm
  {\bf\huge  in $H \to \tau \tau$ decay and at LHC}\\
\end{center}
\vskip 13 mm

\begin{center}
   {\bf K. Lasocha$^{a,b}$, E. Richter-Was$^{a}$, M. Sadowski$^{c}$ and Z. Was$^{d}$}\\
   \vskip 3 mm
   {\em $^a$ Institute of Physics, Jagellonian University, ul. Lojasiewicza 11, 30-348 Krak\'ow, Poland} \\
   {\em $^b$ CERN, Beam Department, 1211 Geneva 23, Switzerland } \\
   {\em $^c$ Faculty of Mathematics and Information Technologies, Jagellonian University, ul. Lojasiewicza 6, 30-348 Krak\'ow, Poland}\\
   {\em $^d$ Institute of Nuclear Physics, IFJ-PAN, 31-342, ul. Radzikowskiego 152, Krak\'ow, Poland}\\
\end{center}
\vspace{1.1 cm}
\begin{center}
{\bf   ABSTRACT  }
\end{center}
The consecutive steps of cascade decay initiated by $H\to \tau\tau$
can be useful for the measurement of Higgs couplings and in particular of 
the Higgs boson parity.  In the previous
papers we have found, that multi-dimensional signatures 
of the $\tau^\pm \to \pi^\pm \pi^0 \nu$  and $\tau^\pm \to 3\pi^\pm\nu$ decays can be used to distinguish between
scalar and pseudoscalar Higgs state. The Machine Learning techniques (ML) of binary classification,
offered break-through opportunities to manage such complex multidimensional signatures.

The classification between two possible CP states: scalar and pseudoscalar,
is now extended to the measurement of the hypothetical mixing angle of Higgs boson parity states.
The functional dependence of $H \to \tau \tau$ matrix element on the mixing angle is predicted by theory.
The potential to determine preferred mixing angle of the Higgs
boson events sample including $\tau$-decays is studied using  Deep Neural Network. The problem is addressed
as classification or regression  with the aim to determine the per-event:
a) probability distribution (spin weight) of the mixing angle;
b) parameters of the functional form of the spin weight;
c) the most preferred mixing angle.
 Performance of proposed methods is evaluated and compared.

\vskip 1 cm

%Draft as of: {\bf \today}. 

-
\vfill
{\small
\begin{flushleft}
{  IFJPAN-IV-2019-19\\
   December 2019 (Improved Jul. 2020)
}
\end{flushleft}
}
 
%%%%%%%%%%%%%%%%%%%%%%%%%%%%%%%%%%%%%%%%%%%%%%%%%%%%%%
\vspace*{1mm}
\footnoterule
\noindent
{\footnotesize % \noindent  $^{\dag}$

This project was supported in part from funds of Polish National Science
Centre under decisions DEC-2017/27/B/ST2/01391.

Majority of the numerical calculations were performed at the PLGrid Infrastructure of 
the Academic Computer Centre CYFRONET AGH in Krakow, Poland.
}
\end{titlepage}
\section{Introduction}

Machine Learning (ML) techniques find increasing number of applications in  High Energy Physics (HEP) phenomenology.
Being used at Tevatron and LHC experiments, they have became an analysis standards. 
For the recent reviews see e.g.~\cite{Guest:2018yhq,Carleo:2019ptp,Albertsson:2018maf}.
The most common approach is via classification routines, however the impact of regression methods is not negligible as well.
Let us  point to two such examples  in LHC   experimental analysis. The measurement of polarization fractions in $WW$ pair
production using Deep Neural Network ({\it DNN})~\cite{dlbook} explores both; the  classification~\cite{Lee:2018xtt} and
regression~\cite{Searcy:2015apa} approaches. The regression technique is also used in~\cite{Forte:2002fg} for parton
distribution functions.

In this paper we present how ML techniques can be helpful to exploit substructure of the hadronically
decaying $\tau$ leptons in  the measurement of the Higgs boson CP-state mixing angle in $H \to \tau \tau$  decay.
This problem has a long history~\cite{Kramer:1993jn,Bower:2002zx} and was studied  both for 
electron-positron~\cite{Rouge:2005iy,Desch:2003mw} and for hadron-hadron~\cite{Berge:2008dr,Berge:2015nua} colliders.
Despite these efforts, the Higgs boson  CP state  was so far not measured at LHC,
from $H \to \tau \tau$  decay. The ML has not been even proposed for the analysis design, contrary to the classical
experimental analysis strategies, see e.g.~\cite{ATL-PUB-2019-008}
for High Luminosity LHC. One of the reasons is that ML adds complexity to the data analysis.
ML solutions  need to be investigated in context of their suitability for  work on systematic
ambiguities.

On the other hand, theoretical basis for the  measurement is simple, the cross-section dependence on the parity mixing
angle has  the form of the first order single angle  trigonometric polynomial.
It can be implemented in the Monte Carlo simulations
as per event spin weight $wt$, % which parametrises this sensitivity,
see~\cite{Przedzinski:2014pla} for more details.
In~\cite{Jozefowicz:2016kvz, Lasocha:2018jcb} we have performed analysis for the three channels
of the $\tau$ lepton-pair decays, respectively  $\rho^\pm\nu_\tau\rho^\mp\nu_\tau$, $a_1^\pm\nu_\tau\rho^\mp\nu_\tau$ and $a_1^\pm\nu_\tau a_1^\mp\nu_\tau$ but
we limited ourselves to the scalar-pseudoscalar classification case. In the scope of our interest was the
kinematics of outgoing decay products of the $\tau$ leptons and geometry of decay vertices.

With these concerns in mind, in the following we extend  our previous work on the physics of the
Higgs CP parity scalar/pseudoscalar classification, to a measurement of  scalar-pseudoscalar 
mixing angle $\phi^{CP}$ of the $H\tau\tau$ coupling. We do not intend to investigate
possible extensions the Standard Model and avoid discussion on the motivations.
We constrain ourselves to the measurement of the coupling and the 
simplest channel of $H \to \tau^+ \tau^- \to \rho^+ \nu_\tau \rho^-\nu_\tau \to \pi^+ \pi^0 \nu_\tau \pi^- \pi^0 \nu_\tau $ decay.
and focus on comparative studies for potential of different ML techniques. 

Possible solutions are analyzed with {\it Deep Neural Network} ({\it DNN}) algorithms~\cite{dlbook} implemented in {\it Tensorflow}
environment~\cite{TensorFlow} which we have previously found working well for
the  binary classification~\cite{Jozefowicz:2016kvz, Lasocha:2018jcb} between  scalar
or  pseudoscalar Higgs boson variants (which correspond to $\phi^{CP}=0$ and $\phi^{CP}=\pi/2$).
%Now, we attempt to extend scope to the case of measurement of the parity mixing angle.
Our goals for the {\it DNN} algorithms are to determine per event:
\begin{itemize}
\item
Spin weight as a function of the mixing angle.
\item
  Decay configuration dependent coefficients, for the known functional form of the spin weight distribution.
\item
The most preferred mixing angle, i.e. where the spin weight is at a maximum.
\end{itemize}

These goals are complementary or even to large extent redundant, e.g. with functional form of the spin weight we can
easily find the mixing angle at which it has a maximum. But the precision of predicting that value would not be necessarily
the same for different methods. 
All three cases are studied as classification  and as regression problems. By this we mean, that the underlying
{\it DNN} cost functions is either designed for classification or regression tasks.
We show quantitative comparison of the performance of {\it DNN} learning on the distributions which are
relevant for physics analyses and then  draw some conclusions.

Paper is organized as follows. In Section \ref{Sec:CPtheory} we describe a basic phenomenology of the problem.
Properties of the matrix elements and distributions at the level of final, measurable quantities as well as unmeasurable
quantities are presented.
In Section~\ref{Sec:Features} we review lists of features (variables) used as an input to {\it DNN} and present samples
prepared for analyses.
As a straightforward extension of~\cite{Jozefowicz:2016kvz, Lasocha:2018jcb}, still using binary classification,
we analyze possibility to distinguish between scalar and arbitrary mix of scalar/pseudoscalar states.
This study is covered in Section \ref{Sec:binaryclas}. The multiclass classification approach is covered
in Section~\ref{Sec:ML-multiclass}.
The regression approach  is discussed in Section~\ref{Sec:MLregression}. The comparison of the classification
and regression is covered in Section~\ref{Sec:MLcomparison}. Observations relevant for the future studies
of systematic errors are addressed. 
The Summary, Section~\ref{Sec:Summary}, closes the paper.

In Appendix~\ref{App:DNN} more technical details on the {\it DNN} architecture are given together with arguments
supporting such a choice. In addition, we describe briefly the data preprocessing chain.

\section{Physics content of the problem}
\label{Sec:CPtheory}
The most general Higgs  boson Yukawa coupling to $\tau$ lepton pair,
expressed with the help of the
scalar--pseudoscalar parity mixing
angle $\phi^{CP}$ reads as
\begin{equation}
  {\cal L}_Y= N\;\bar{\tau}{\mathrm  h}(\cos\phi^{CP}+i\sin\phi^{CP}\gamma_{5})\tau,
\end{equation}
where $N$ denotes normalization, $\mathrm h$ Higgs field and $\bar\tau$, $\tau$ spinors of the $\tau^+$ and $\tau^-$.  
As we will see later, this simple analytic form translates itself into useful properties of observable distributions
convenient for our goal, determination of the $\phi^{CP}$. Recall of the definitions is thus justifiable, and helpful
to systematize properties and correlations of the observable quantities (features).

The matrix element squared for the scalar / pseudoscalar / mix parity Higgs, with  decay
into $\tau^+ \tau^-$ pairs can be expressed as
\begin{equation} \label{eq:matrix}
  |M|^2\sim 1 +  h_{+}^{i}  h_{-}^{j} R_{i,j}; \;\;\;\;\; i,j=\{x,y,z\}
\end{equation}
 where  $h_{\pm}$ denote polarimetric
vectors of $\tau$ decays (solely defined by $\tau$ decay matrix elements) and
$R_{i,j}$ density matrix of the $\tau$ lepton pair spin state.
In Ref.~\cite{Desch:2003rw} details of the frames used for $R_{i,j}$ and $h_{\pm}$
definition are given.
 The corresponding CP sensitive spin weight $wt$ has the form:
 \begin{equation}
   \label{eq:wt_master}
wt = 1-h_{{+}}^{z} h_{{-}}^{z}+ h_{{+}}^{\perp} R(2\phi^{CP})~h_{{-}}^{\perp}.
\end{equation}
The formula is valid for $h_\pm$ defined in $\tau^\pm$ rest-frames,
$h^{z}$ stands for longitudinal and  $h^{\perp}$ for  transverse components of $h$.
The $R(2\phi^{CP})$ denotes  the $2\phi^{CP}$ angle rotation matrix  around the  $z$ direction:
$R_{xx}= R_{yy}={\cos2\phi^{CP}}$, $R_{xy}=-R_{yx}={\sin2\phi^{CP}}$.
The $\tau^\pm$ decay polarimetric vectors $h_{+}^i$,  $h_{-}^j$, in the simplest case
of $\tau^{\pm} \to \pi^{\pm} \pi^0 \nu $ decay, read 
\begin{equation}
h^i_\pm =  {\cal N} \Bigl( 2(q\cdot p_{\nu})  q^i -q^2  p_{\nu}^i \Bigr), \;\;\;  
\end{equation}
where  $\tau$ decay products  $\pi^{\pm}$, $\pi^0$ and $\nu_{\tau}$   4-momenta are denoted respectively as
$ p_{\pi^{\pm}}$, $p_{\pi^0}, p_{\nu}$ 
%q--> p_{\pi^\pm} -p_{\pi^0}
and $q=p_{\pi^\pm} -p_{\pi^0}$.
Obviously, complete CP sensitivity can be extracted only if $p_{\nu}$ is known
(for $\tau^\pm \to \pi^\pm\pi^\pm\pi^\mp \nu$ formula is longer, dependence on modeling of the decay appear
too \cite{Barberio:2017ngd}, but is of no principle differences).

Note that the spin weight $wt$ is a simple first order trigonometric polynomial in a 
$2 \phi^{CP}$ angle. This observation  is valid for all $\tau$ decay channels.
For the clarity of the discussion on the {\it DNN} results, we introduce $\alpha^{CP}= 2\phi^{CP}$,
which spans over (0, 2\pi) range. The $\alpha^{CP}= 0, 2\pi$ corresponds to scalar state, the  $\alpha^{CP}= \pi$ to
pseudoscalar one. Spin weight can be expressed as

\begin{equation}
  wt = C_0 + C_1 \cdot \cos(\alpha^{CP}) + C_2 \cdot \sin(\alpha^{CP}),
  \label{eq:ABC_alpha}
\end{equation}
where
\begin{eqnarray}
  \label{eq:CPcoeff}
  C_0 &=& 1 -  h_{+}^{z} h_{-}^{z}, \nonumber \\
  C_1 &=& - h_{+}^{x} h_{-}^{x} +  h_{+}^{y} h_{-}^{y}, \\
  C_2 &=& - h_{+}^{x} h_{-}^{y} -  h_{+}^{y} h_{-}^{x},\nonumber 
\end{eqnarray}
depend on the $\tau$ decays only.

\begin{figure}
   \begin{center}
     {
   \includegraphics[width=5.0cm,angle=0]{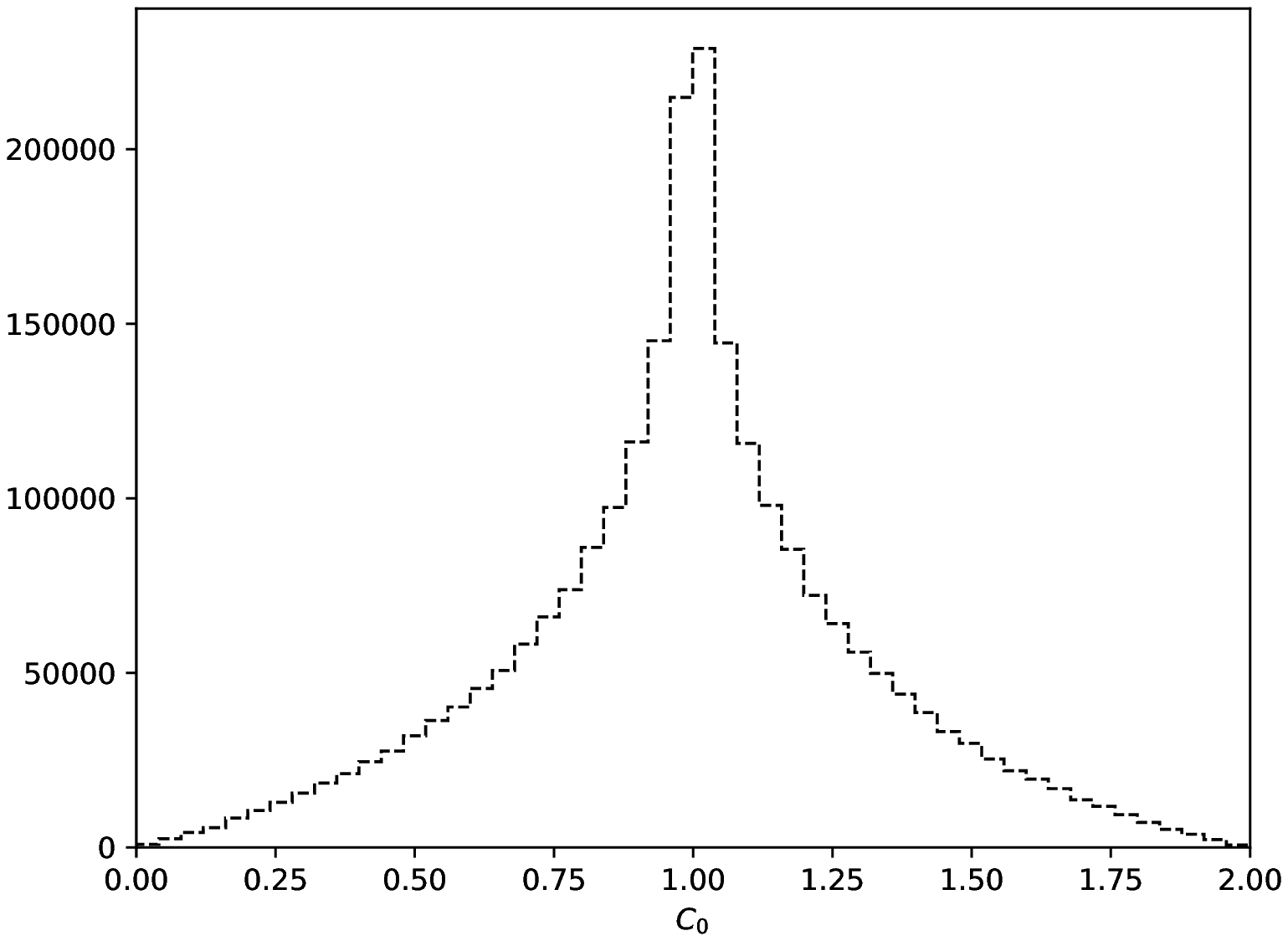}
   \includegraphics[width=5.0cm,angle=0]{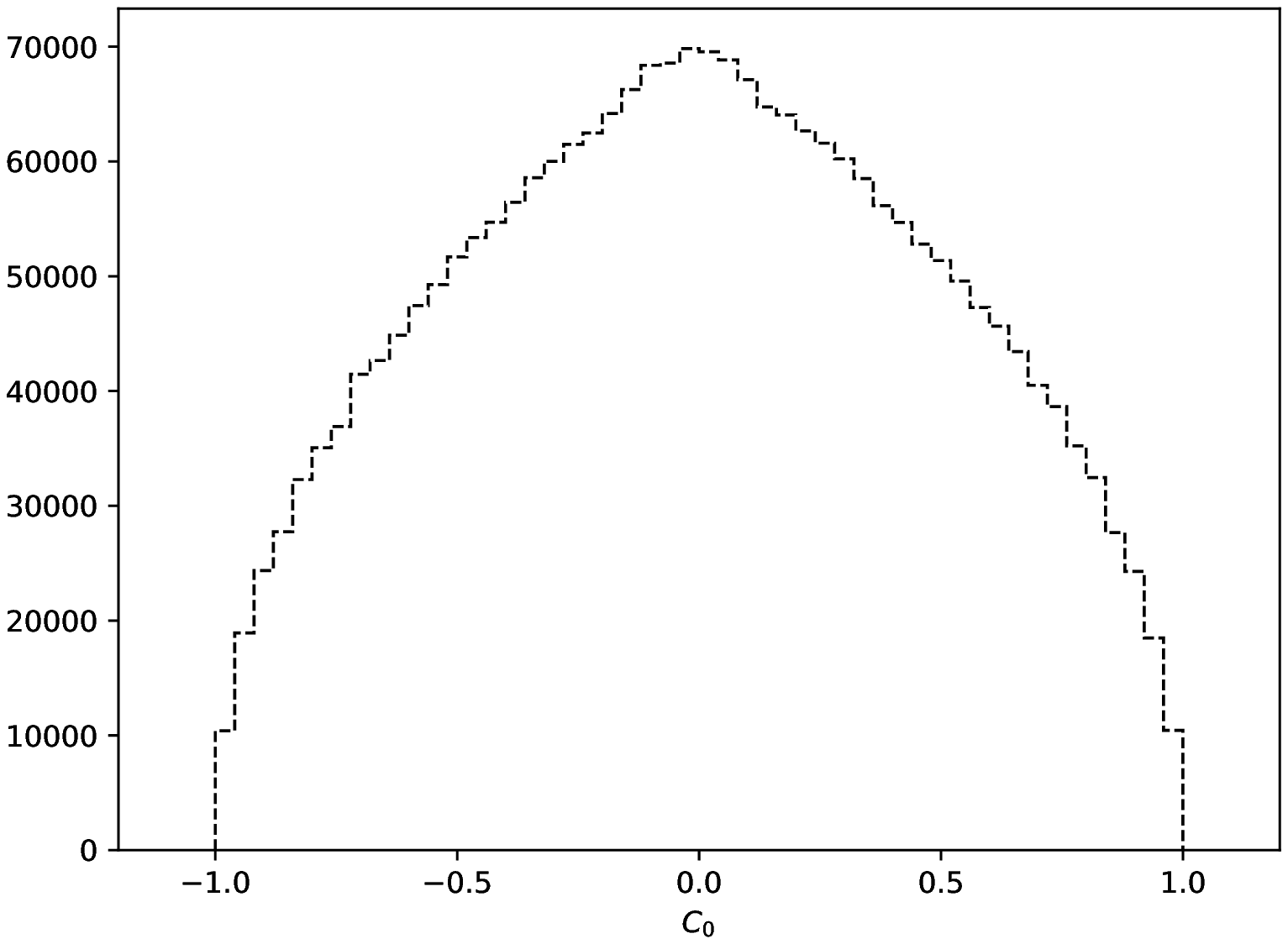}
   \includegraphics[width=5.0cm,angle=0]{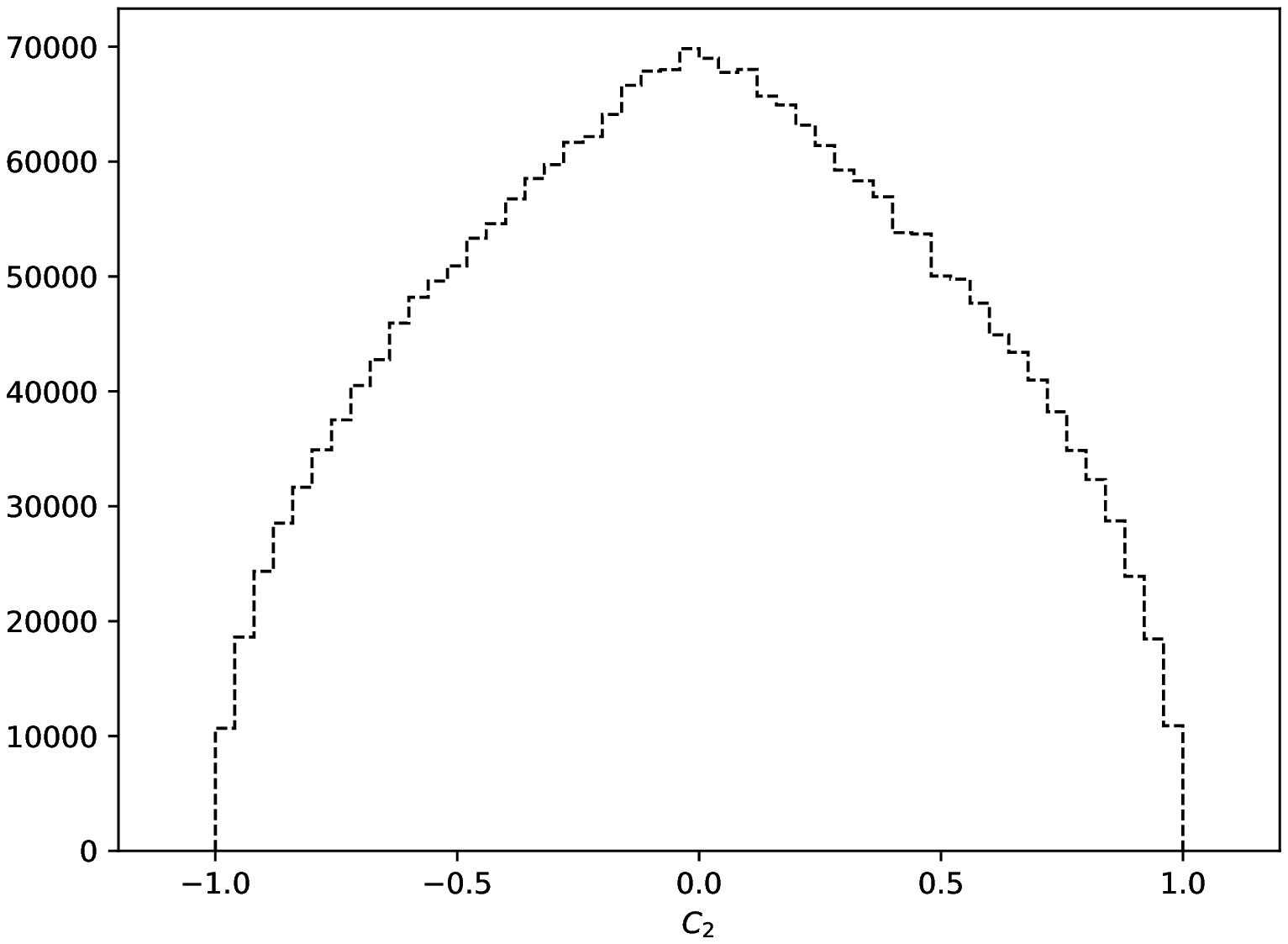}
     }
\end{center}
   \caption{Distributions of the formula~(\ref{eq:ABC_alpha})  $C_0, C_1, C_2$ coefficients, for the $H \to \tau \tau$ events sample.
   \label{fig:c012s}}
 \end{figure}

Distribution of the $C_0, C_1, C_2$ coefficients, for the sample of $H \to \tau \tau$ events used for our numerical results
is shown in Fig.~\ref{fig:c012s}. The $C_0$ spans (0, 2) range, while  $C_1$ and $C_2$ of (-1, 1) range
have a similar shape, quite different than the one of  $C_0$.

The amplitude of the $wt$ as function of  $\alpha^{CP}$ depends on the multiplication of the length of the transverse components
of the polarimetric vectors. The longitudinal component  $h_{+}^{z} h_{-}^{z}$ is defining shift with respect to zero of
the $wt$ mean value over a full (0, 2\pi) range. The maximum of the $wt$ distribution is reached for
$\alpha^{CP}$ = $\measuredangle (h_{+}^{T}, h_{-}^{T})$, the opening angle of transverse components of the polarimetric vectors.
%{\bf Sprawdzic analitycznie/numerycznie ten wzor raz jeszcze}.

The spin weight of formula (\ref{eq:ABC_alpha})
can be used to introduce transverse spin effects  into the event sample when for the generation
transverse spin effects were not taken into account at all.
The above statement is true, independently if longitudinal spin effects were included and which $\tau$ decay channels
complete cascade of  $H \to \tau \tau$ decay. The shape of weight dependence on the Higgs coupling to $\tau$
parity mixing angle is preserved.

\begin{figure}
   \begin{center}
     {
   \includegraphics[width=7.0cm,angle=0]{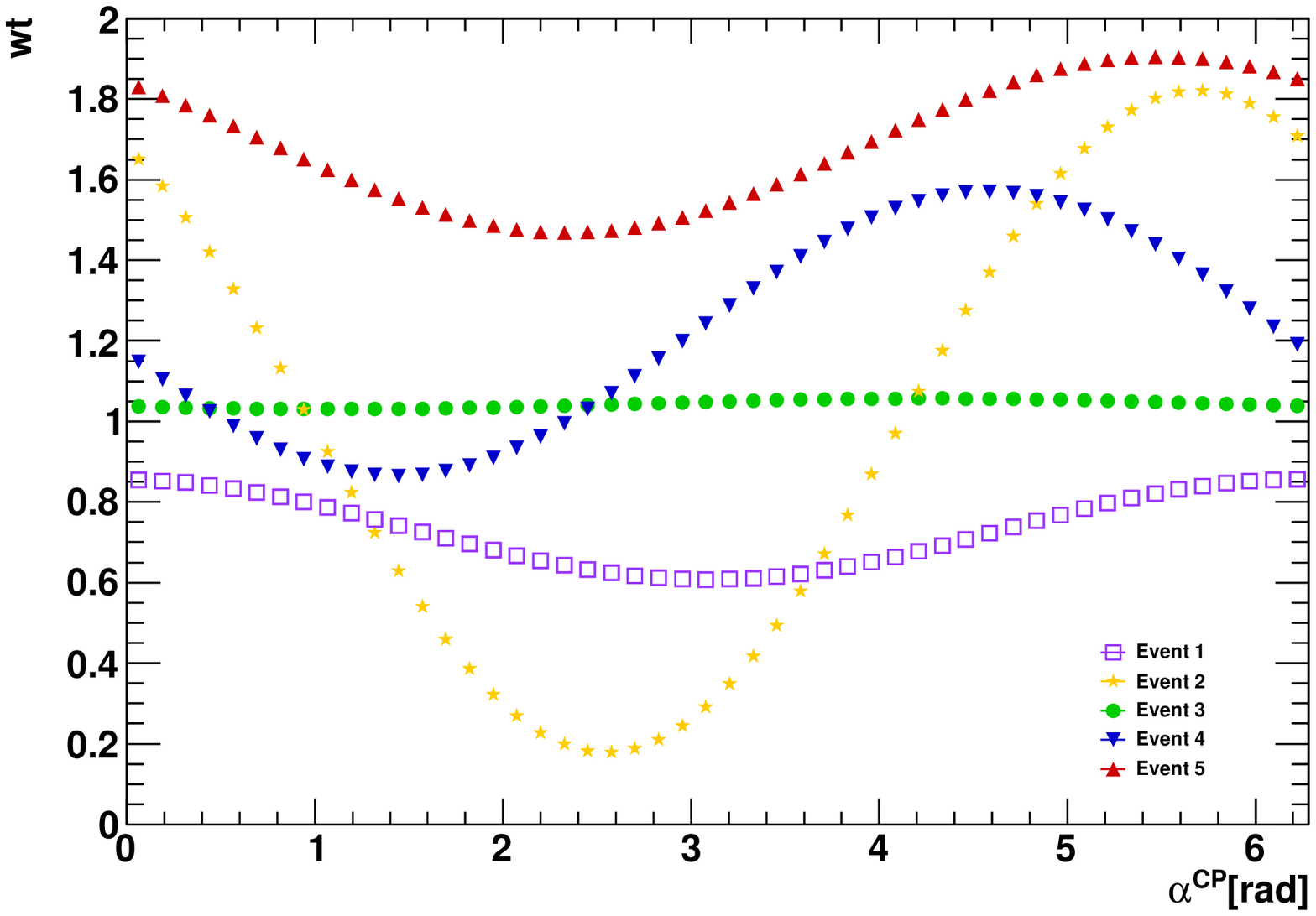}
   \includegraphics[width=7.0cm,angle=0]{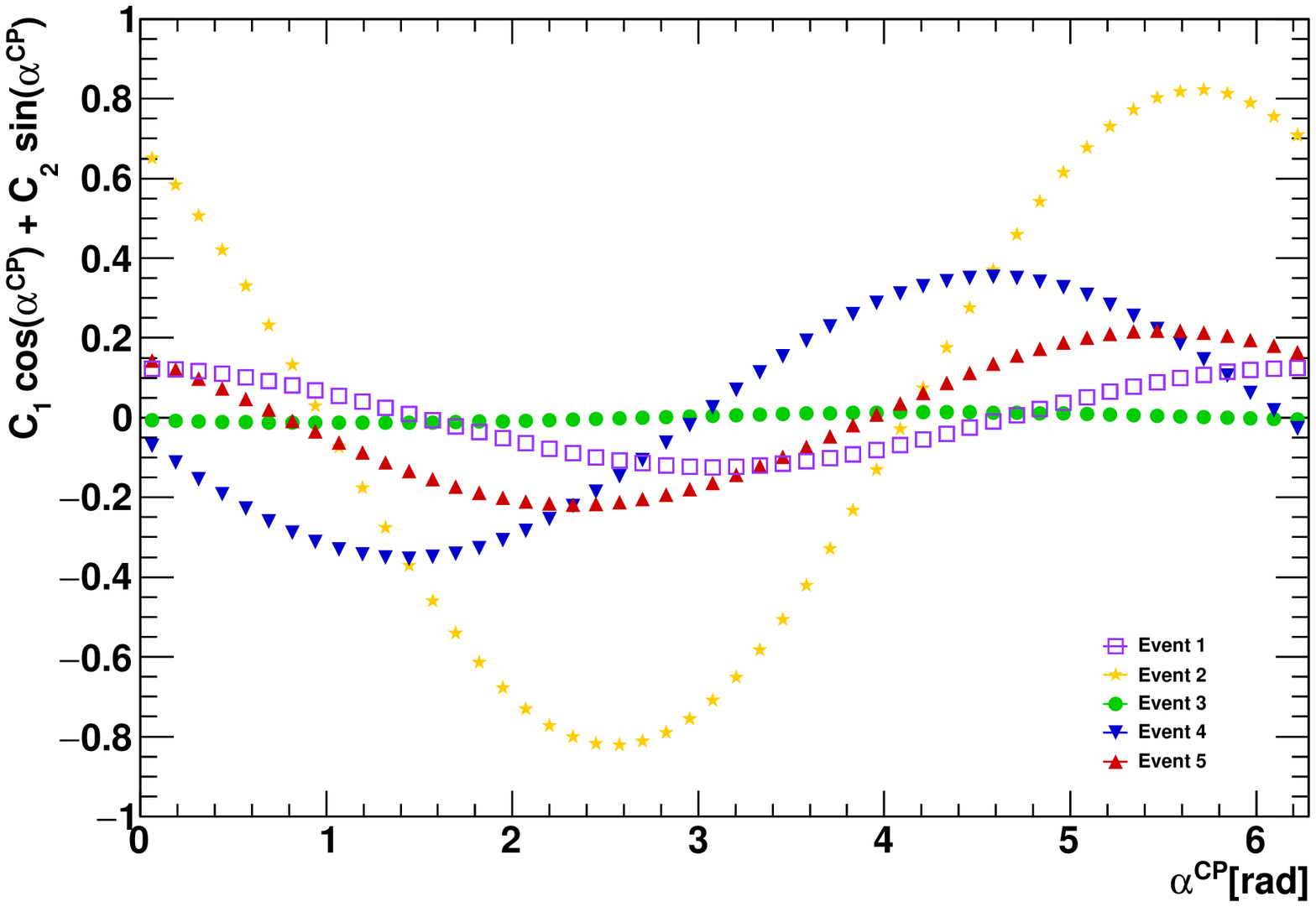}
     }
\end{center}
   \caption{The spin weight $wt$ (left plot) and only its $\alpha^{CP}$ dependent component (right plot) for
     five  $H \to \tau \tau$ events of Table~\ref{Tab:CPcoeff}. Note the vertical scale change between left and right plots.
   \label{fig:CPweight}}
 \end{figure}

In Fig.~\ref{fig:CPweight} we show distribution of spin weight $wt$ for five example $H \to \tau \tau$ events collected
in Table~\ref{Tab:CPcoeff}. For each event, depending on the polarimetric vectors,
single value of $\alpha^{CP}$ is preferred (by the largest weight). For a physics  model with $\alpha^{CP}$
the sample  will be more abundantly populated with events for which the angle between polarimetric vectors,
$\measuredangle (h_{+}^{T}, h_{-}^{T})$, is close to $\alpha^{CP}$. We show distributions when 
complete polarimetric vectors are used for spin weight $wt$ and when only hadronic parts
of polarimetric vectors are used. The second case is indicating easier to attain sensitivity part of observables.
The $\alpha^{CP}$ at which spin weight has its maximum is then a bit shifted. Table~\ref{Tab:CPcoeff} specifies values of the
polarimetric vectors and the resulting coefficients $C_i$ calculated from formulas~(\ref{eq:CPcoeff}) and
for events of Fig.~\ref{fig:CPweight}. It also explicitly gives $\measuredangle (h_{+}^{T}, h_{-}^{T})$ calculated from
complete polarimetric vectors and (in brackets) from their hadronic parts only.

\begin{table}
 \vspace{2mm}
  \caption{
    Polarimetric vectors,  resulting $C_i$ coefficients of formulas (\ref{eq:CPcoeff}) and
    angle $\measuredangle (h_{+}^{T}, h_{-}^{T})$ between
    transverse components of polarimetic vectors for five  example events of $H \to \tau^+ \tau^-, \tau^\pm \to \rho^\pm \nu_\tau$.
    In brackets, angle of only hadronic part of polarimetric vector is given.
    \label{Tab:CPcoeff}}
      \begin{center}
    \begin{tabular}{|l|c|c||c|c|c||c|}
    \hline
    Events  & Polarimetric vectors & $|h_{+}^{T}||h_{-}^{T}|$ & $C_0$ & $C_1$ & $C_2$ &  $\measuredangle (h_{+}^{T}, h_{-}^{T})$ [rad] \\
            &                      &                       &     &     &     &  (hadronic part only)   \\ 
    \hline
    Event 1 &  $h_{+}^{x,y,z}$ = (0.7547 -0.2232 -0.6167) & 0.7519 & 0.8179 & 0.7517 & 0.0183 &  6.2586 \\
    &  $h_{-}^{x,y,z}$ = (-0.9093 -0.2931 -0.2953) & & & &                                    &  (6.1738)   \\
    \hline
    Event 2 &  $h_{+}^{x,y,z}$ = (0.8617 0.0485  0.5050)   & 0.8535   & 1.0751 & 0.5518 & -0.6511 &  5.4134\\
    &  $h_{-}^{x,y,z}$ = (-0.5959 0.7892 -0.1487) & & & &                                          & (5.6307) \\
    \hline
    Event 3 &  $h_{+}^{x,y,z}$ = (0.3402 0.9377 -0.0682) &   0.8339  & 0.9626 & -0.1619 & -0.8180 & 5.2130 \\
    &  $h_{-}^{x,y,z}$ = (0.8262 0.1272 -0.5487) & & & &                                          & (4.1923) \\
    \hline
    Event 4 &  $h_{+}^{x,y,z}$ = (-0.6964 0.6204 -0.3605) & 0.4138   & 0.6769 & -0.0919 & -0.4035 & 4.4883\\
    &  $h_{-}^{x,y,z}$ = (0.2142 -0.3885 -0.8962) & & & &                                          & (4.5127) \\
    \hline
    Event 5 &  $h_{+}^{x,y,z}$ = (0.1115 -0.4989 -0.8595) &  0.1201   & 1.8354 & 0.0317 & -0.1158 & 4.9793 \\
    &  $h_{-}^{x,y,z}$ = (-0.2347 -0.01108 0.9720) & & & &                                        & (5.4300) \\
    \hline
 \hline
    \end{tabular}
  \end{center}
\end{table}

 \section{Monte Carlo samples and feature lists}
 \label{Sec:Features}

 For compatibility with our previous publications~\cite{Jozefowicz:2016kvz, Lasocha:2018jcb}, we use the same generated event
 samples, namely Monte Carlo events of the Standard Model, 125~GeV Higgs boson, produced in pp collision at 13~TeV
 centre-of-mass energy, generated with {\tt Pythia 8.2}~\cite{Sjostrand:2014zea} and with spin correlations introduced
 with {\tt TauSpinner}~\cite{Przedzinski:2014pla} package. For $\tau$ lepton decays we use {\tt Tauolapp}
 library~\cite{Davidson:2010rw}. All spin and parity effects are implemented with the help of weight
 $wt$~\cite{Czyczula:2012ny, Przedzinski:2018ett}. The sample is generated without spin effects, and
 the spin weights $wt_i$ for few different values of CP mixing angle $\alpha^{CP}_i$ are stored.
 Spin weight, formula~(\ref{eq:wt_master}), is calculated using $R_{i,j}$ density matrix and  polarimetric vectors $h_\pm$.

\begin{figure}
   \begin{center}
     {
       \includegraphics[width=7.0cm,angle=0]{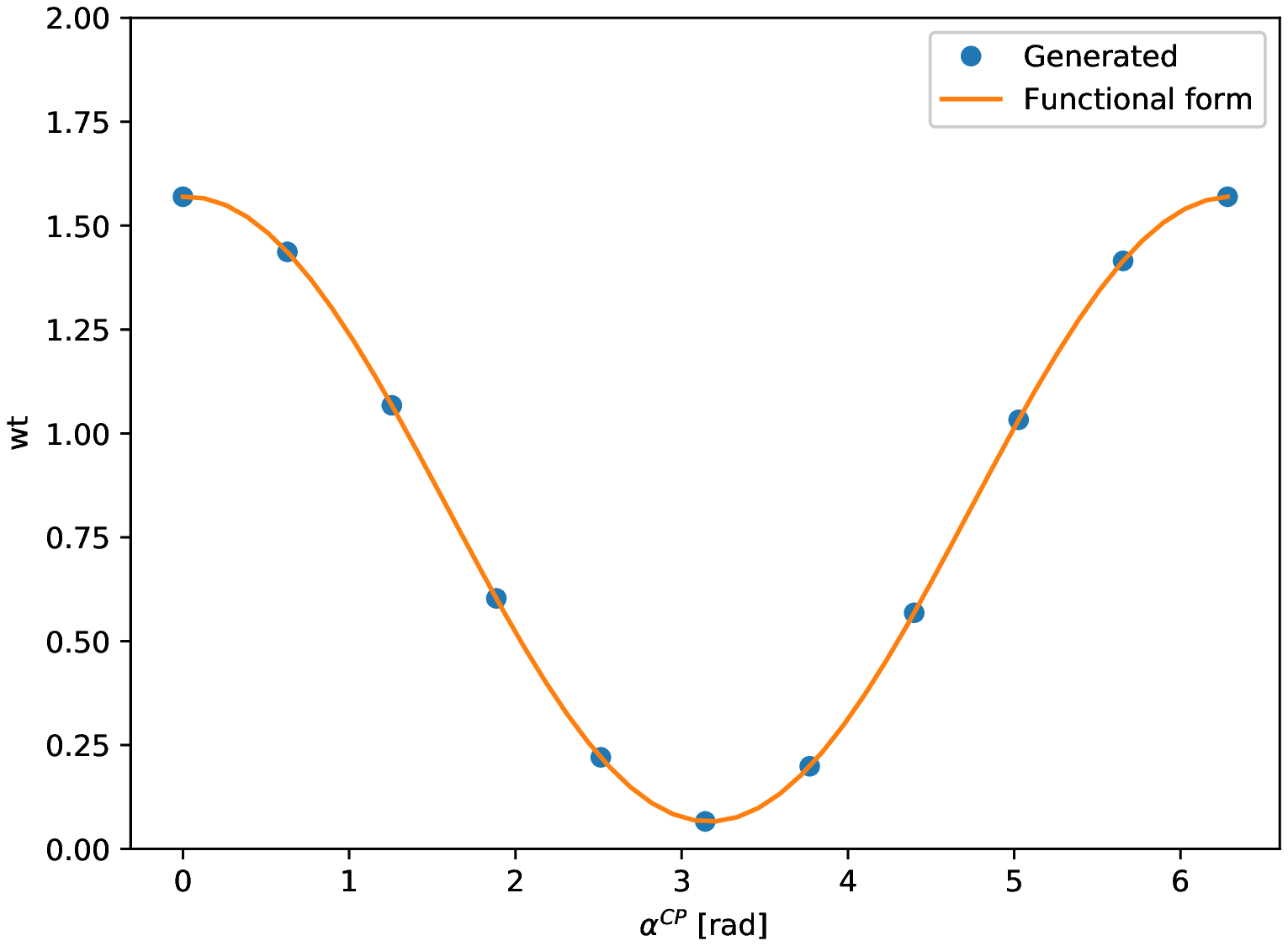}
       \includegraphics[width=7.0cm,angle=0]{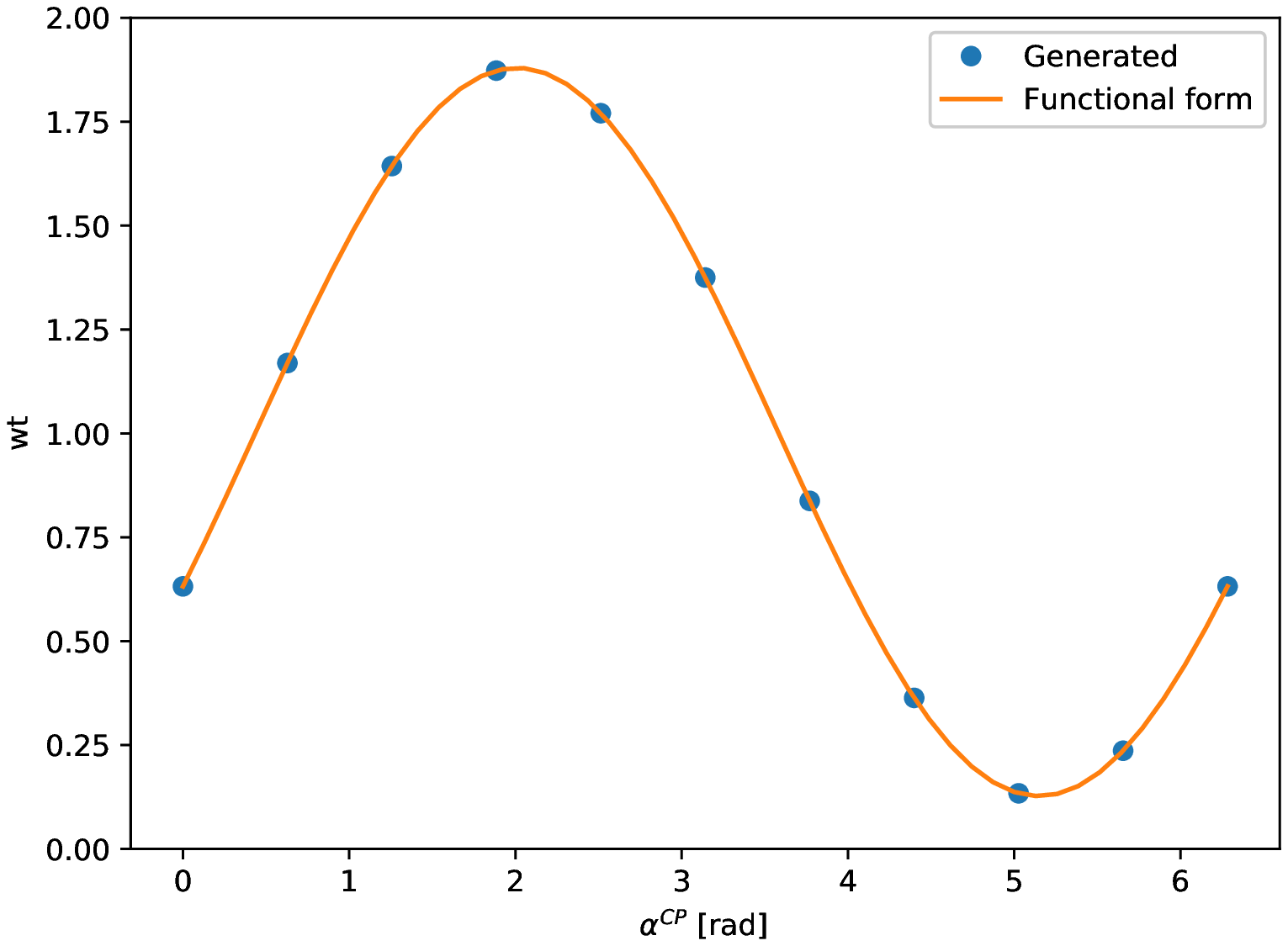}
     }
\end{center}
   \caption{Cross-check distributions of the spin weight $wt$ calculated at generation (blue points) and from functional form of Eq.~(\ref{eq:ABC_alpha})
     (orange line), as a function of CP mixing parameter $\alpha^{CP}$. For left and right plots two different example events were used.
     Coefficients $C_i$ are reconstructed from eq.~(\ref{eq:ABC_alpha}) and $wt$ is taken at three different  $\alpha^{CP}$.
   \label{fig:ABCtest}}
 \end{figure}

Later, for a given event it is possible to calculate coefficients $C_0, C_1, C_2$, using three $\alpha^{CP}$
and linear equation~(\ref{eq:ABC_alpha}).
Fig.~\ref{fig:ABCtest} shows the cross-check how well this procedure works. The functional form (orange line) and evaluated
spin weights (blue dots) for two example events are shown. The $C_0, C_1, C_2$ coefficients for the functional form are
calculated solving Eq.~(\ref{eq:ABC_alpha}) for $wt$ stored in the generated event samples at three values of $\alpha^{CP}$.

 In this paper we present results for the case when both $\tau$'s decay $\tau^{\pm} \to \rho^{\pm} \nu_{\tau}$ and
 about $5 \cdot 10^6$ simulated Higgs events are used. To partly emulate detector conditions, 
 a minimal set of cuts is used. We require that the transverse momenta of the visible decay products combined,
 for each $\tau$, are larger than 20 GeV. It is also required that the transverse momentum of each $\pi^\pm$ is larger than 1 GeV.

 The emphasis of the paper is to explore different ML approaches to the problem, and we discuss only the case of the {\tt Variant-All}
 feature list from paper~\cite{Lasocha:2018jcb}. It contains the four-momenta of {\it all} decay products of $\tau$ leptons defined
 in the rest frame of intermediate resonance pairs, and with sum of hadronic decay products aligned with $z$-axis are.
 This represents an ideal benchmark case scenario, for performance monitoring.

\section{Binary classification}
\label{Sec:binaryclas}

The use of the {\it DNN} for binary classification have been discussed in our previous
papers~\cite{Jozefowicz:2016kvz, Lasocha:2018jcb}.
The focus was on discriminating between CP-scalar ($\mathscr{H}_0$ hypothesis) and CP-pseudoscalar ($\mathscr{H}_{1}$ hypothesis).

Now we apply the same procedure but with alternative hypothesis ($\mathscr{H}_{\alpha^{CP}}$) representing the scalar-pseudoscalar
mixed state of mixing parameter $\alpha^{CP}$.
To quantify performance for Higgs CP state classification the weighted Area Under Curve (AUC)~\cite{roc-1,Fawcett2006} is used again.
For each simulated event we know also Bayes optimal probability that it is sampled from $\mathscr{H}_0$ or
$\mathscr{H}_{\alpha^{CP}}$ hypothesis, see more detailed description in Appendix~\ref{App:DNN}.
This forms the so called {\it oracle predictions}, i.e. ultimate discrimination for this problem.
We calculate oracle predictions and evaluate the results of {\it DNN}.
This is a straightforward extension of the method used in~\cite{Jozefowicz:2016kvz, Lasocha:2018jcb}.
That is why, simple attempt on future discussion of systematic error may follow that suggested in~\cite{Lasocha:2018jcb}:
variations within expected range of detector response can be easily introduced and biases studied. 

\begin{figure}
   \begin{center}
     {
       \includegraphics[width=7cm,angle=0]{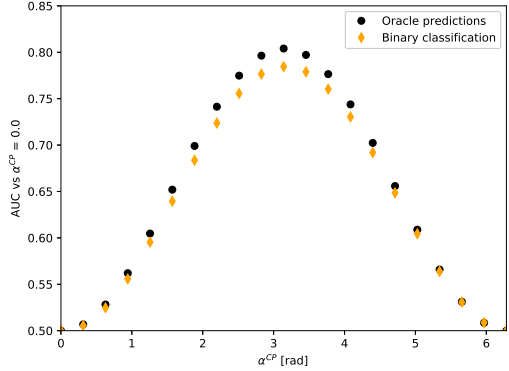}
      }
\end{center}
   \caption{ The AUC score for binary classification  between $\mathscr{H}_0$  and $\mathscr{H}_{\alpha^{CP}}$ hypotheses and
     corresponding oracle predictions.
        \label{figApp:DNN_binary}}
 \end{figure}

The oracle predictions for discriminating between $\mathscr{H}_0$  and $\mathscr{H}_{\alpha^{CP}}$ hypotheses is increasing with
$\alpha^{CP}$ and reach  AUC=0.78 for $\alpha^{CP} = \pi$.
The performance of {\it DNN} is following similar pattern, reaching maximum at $\alpha^{CP} = \pi$ (pure pseudo-scalar case).
It decreases for smaller or larger  $\alpha^{CP}$, where admixture of the scalar component appear.
In case of complete feature list, it is almost achieving the performance of oracle predictions.
In Fig.~\ref{figApp:DNN_binary}, the AUC values are plotted for full $\alpha^{CP}$ range.
The distributions are (almost) symmetric around  $\alpha^{CP} = \pi$.
Note that  the functional form of spin weight $wt$, Eq.~(\ref{eq:ABC_alpha}),
encapsulating sensitivity to  $\alpha^{CP}$ is not symmetric, see Fig.~\ref{fig:ABCtest}.
In Table~\ref{Tab:DNN_binary} we show numerical results for few $\alpha^{CP}$.

\begin{table}
 \vspace{2mm}
  \begin{center}
  \caption{
    The AUC scores for discriminating between Higgs CP states. Results from oracle predictions
    and binary classification for discriminating between  $\mathscr{H}_0$ hypothesis that Higgs CP is a scalar
    ( CP-mixing angle $\alpha^{CP}$ = 0.0 or 2\pi) and $\mathscr{H}_{\alpha^{CP}}$ hypothesis,
    when Higgs CP is of a parity mixed state, are shown.
    CP-mixing angle $\alpha^{CP}= \pi$ corresponds to pseudo-scalar case.
    \label{Tab:DNN_binary}}
 \vspace{2mm}
    \begin{tabular}{|l|c|c|}
    \hline
    CP-mixing angle $\alpha^{CP}$   & Oracle predictions & Binary \\
    (units of $\pi$)  &            &              classification         \\
    \hline
     0.2 & 0.528 & 0.525   \\
     0.4 & 0.605 & 0.595   \\
     0.6 & 0.699 & 0.684   \\
     0.8 & 0.775 & 0.756   \\
     1.0 & 0.804 & 0.784   \\
    \hline
    \end{tabular}
  \end{center}
\end{table}

\section{Multiclass classification}
\label{Sec:ML-multiclass}

The binary classification discussed in previous Section is easy to generalize to the multiclass case.
The {\it DNN} is learning to provide per-event probabilities to associate with each class.
Single class represents either discrete point or a specific range in 1-dimensional parameter space. 
We explore three approaches, each providing complementary physics information, but all allowing
to quantify, on the per-event basis, which is the preferred mixing angle of the studied Higgs sample:

\begin{itemize}
\item
The {\it DNN} classifier is learning per-event spin weight as a function of mixing angle $\alpha^{CP}$.
The range of mixing angle $(0, 2\pi)$ is discretised into equally spaced points called classes.
This approach is described in Section~\ref{Sec:Classification:wt},
and used for the figures labeled with:  {\tt Classification:wt}.
\item
The {\it DNN} classifier is learning per-event coefficients $C_0, C_1, C_2$.
The allowed range of coefficients is split into several equal size ranges (classes), single
class represents a range for a coefficient value. The {\it DNN} is trained for each coefficient separately.
This approach is described in Section~\ref{Sec:Classification:C012} and used for the figures labeled with:
{\tt Classification:$C_0, C_1, C_2$}. 
\item
The {\it DNN} classifier is learning per-event most probable mixing angle $\alpha^{CP}_{max}$, i.e. 
value of  $\alpha^{CP}$ at which spin weight is maximal. The range of mixing angle $(0, 2\pi)$ is split into several
equally spaced points (classes).
This approach is described in Section~\ref{Sec:Classification:alphaCPmax} and used for the figures labeled with:
{\tt Classification:$\alpha^{CP}_{max}$}.
\end{itemize}
  
We monitor performance of the learning process in a standard manner, with the loss function on the training
and validation sets. Respective distributions are shown in Fig.~\ref{figApp:DNN_loss} of Appendix~\ref{App:DNN}. 
Note that the loss function, the {\tt tf.nn.softmax\_cross\_entropy\_with\_logits}
of the {\tt Tensorflow}, allows to predict probabilities of the class labels, and not the actual value of the observable
at a given class. In case of predicting spin weight distribution, only the normalized to unity shape is predicted.
In case of predicting values of $C_i$ coefficients or $\alpha^{CP}_{max}$, vector of probabilities is returned, and
the one-hot encoding transformation selecting most probable class is then applied to retrieve actual predicted value of the parameter.

\subsection{Learning spin weight $wt$}
\label{Sec:Classification:wt}

The  {\it DNN} classifier is trained with per-event feature list and as a label normalized to unity $N_{class}$-dimensional
vector of spin weights \footnote{The $wt_i$ remains in the (0,4) range, as explained in~\cite{Przedzinski:2018ett}.}
$wt_i^{norm} = wt_i /\sum_{i=1}^{i=N_{class}} wt_i$
is given, each component of $wt^{norm}(\alpha^{CP})$ vector corresponds to the i-th discrete value of mixing
angle $\alpha^{CP}_i$. $N_{class}$ denotes number of points to which range $(0, 2 \pi)$ was discretised. 
The number of classes is kept odd,
to assure that $\alpha^{CP}= 0, \pi, 2 \pi$, corresponding respectively to scalar/pseudoscalar/scalar cases,
are always represented as a separate class. Training of {\it DNN} is performed with $N_{class}$ varying from 3 to 51.
This is to understand the tradeoff between the better approximation given by high number of classes and smaller
complexity of the low-class system.

\begin{figure}
   \begin{center}
     {
      \includegraphics[width=7.0cm,angle=0]{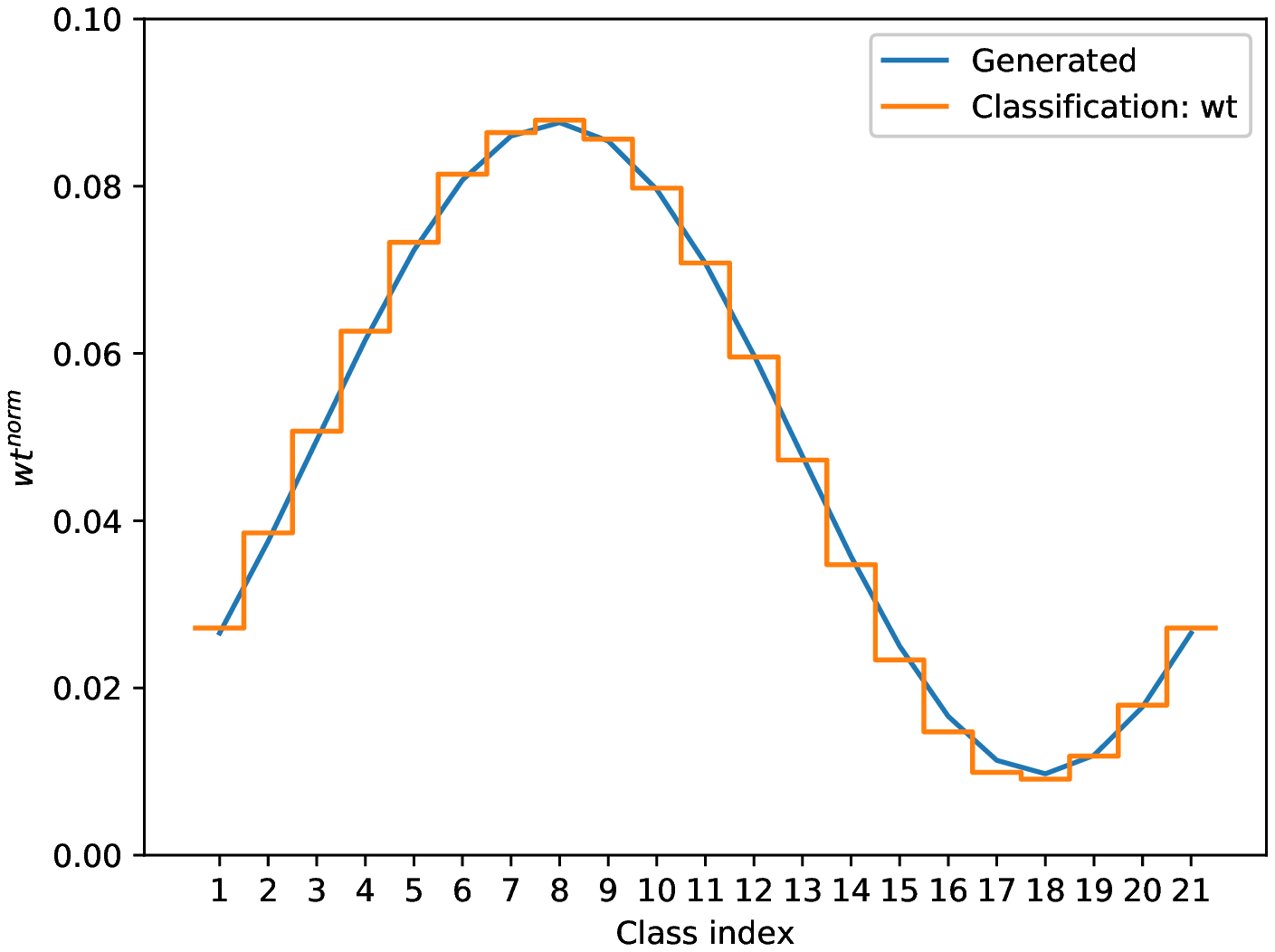}
      \includegraphics[width=7.0cm,angle=0]{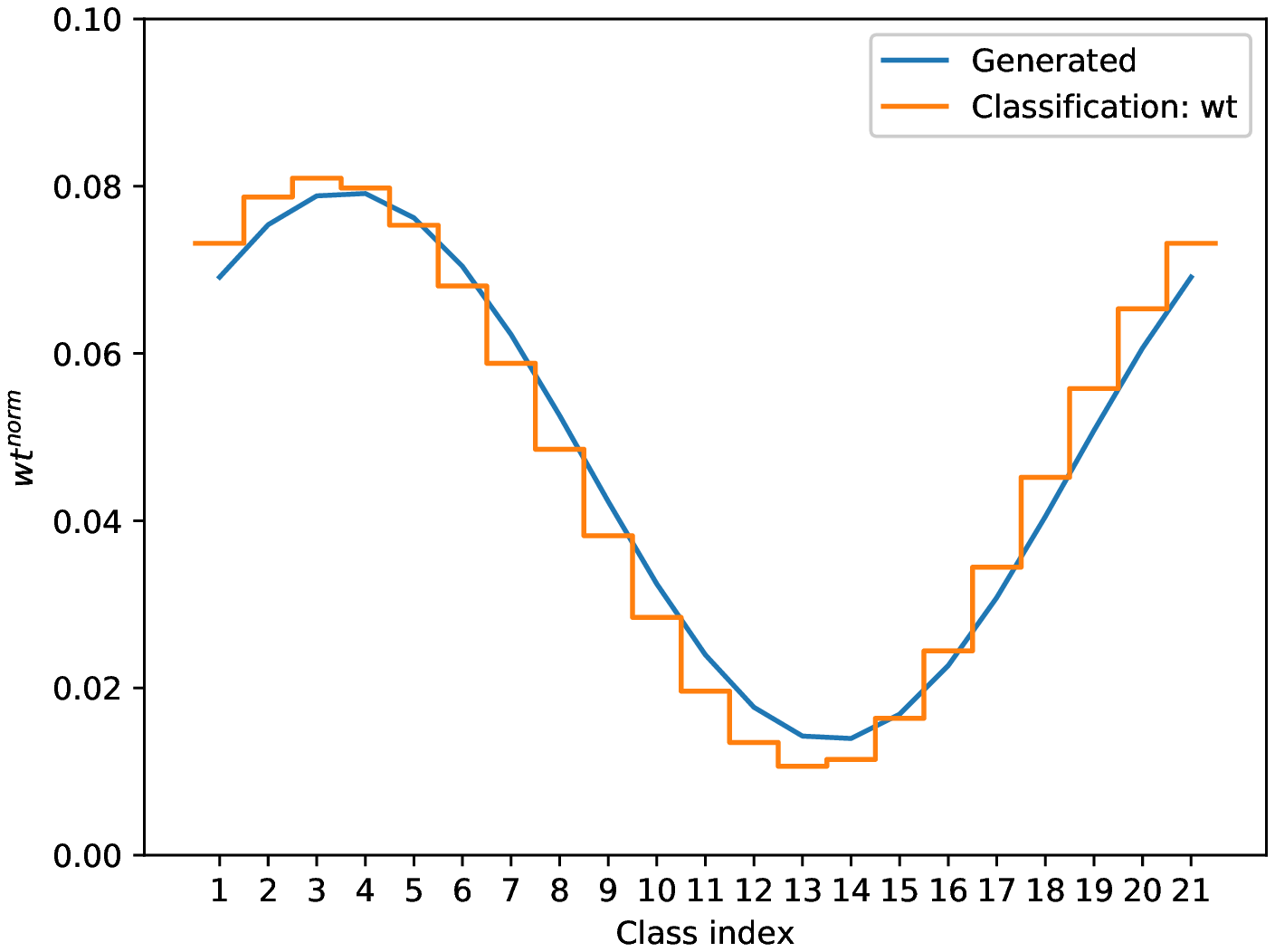}
     }
\end{center}
   \caption{Normalized to probability  spin weight $wt^{norm}$, predicted  (orange steps) and true (blue line),
     as a function of $\alpha^{CP}_i$ for two example events (left and right plots).
     {\it DNN} was trained with $N_{class} = 21$ spanning range $(0, 2 \pi)$.
   \label{fig:DNN_soft_nc_predwt}}
 \end{figure}

We quantify the {\it DNN} performance for classification problem in the context of physics
relevant criteria.
The first question is how well {\it DNN} is able to reproduce per-event shape of the spin weight $wt^{norm}$.
For two example events, true and predicted spin weight $wt^{norm}$ distribution with $N_{class} = 21$ is shown in
Fig.~\ref{fig:DNN_soft_nc_predwt}
as a function of either continuous mixing parameter $\alpha^{CP}_i$ or class index i (representing discretised
mixing parameter $\alpha^{CP}_i$). Blue line denote true weights while orange steps denote weights predicted by
{\it DNN} classifier. 
In overall, predicted weights follow smoothly true shape of linear  $\cos (\alpha^{CP})$ and
$\sin (\alpha^{CP})$ combination. This is encouraging, because the loss function is not correlating explicitly nearby
classes. The {\it DNN} is discovering this pattern in the process of learning.
%Although such information could be used with adding some penalty term.
 
 To quantify those observations, performance of {\it DNN} is monitored on the statistical basis with $l_2$ norm.
 The $l_2$ norm is defined as a square root of the integral of squared difference between predicted $ p_{k}$ and true $wt^{norm}_{k}$
 over the whole interval $(0, 2 \pi)$. It then averaged over the number of events $N_{evt}$. Although $p_{k}$ and $wt^{norm}_{k}$ are
 functions of $\alpha^{CP}$, we shall usually skip the argument for the notation brevity.

 \begin{equation}
 l_2  =  \sum_{k=1}^{N_{evt}} \frac{\sqrt{  \int_{0}^{2 \pi} \left(wt^{norm}_{k}(\alpha^{CP}) - p_{k}(\alpha^{CP}) \right)^2 d \alpha^{CP}}}{N_{evt}}.
\label{eq:l2_def}
\end{equation}

\begin{figure}
   \begin{center}
     {
   \includegraphics[width=7.0cm,angle=0]{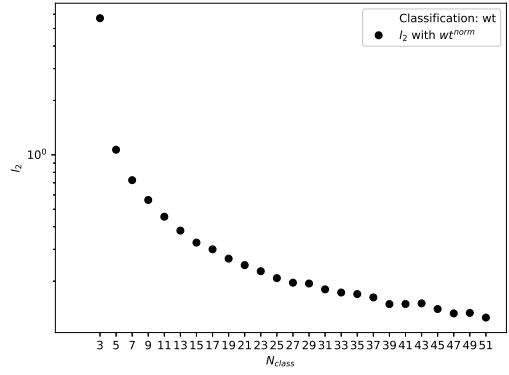}
     }
\end{center}
   \caption{The $l_2$ norm, quantifying difference between true and predicted spin weight $wt^{norm}$, as a function of
     class multiplicity $N_{class}$.
   \label{fig:DNN_soft_nc_l2}}
 \end{figure}

 The $p_{k}$ corresponds to the k-th event and is represented as a step function, with step levels given by a $N_{class}$-dimensional
 output of DNN. For true weights, represented as continuous function~(\ref{eq:ABC_alpha}), we scale them in such a way
 that $\int_{0}^{2 \pi} wt^{norm} d \alpha^{CP} = 1$, to enable the comparison. 
Distribution of  $l_2$ norm is shown in Figure \ref{fig:DNN_soft_nc_l2}, as a function of class multiplicity $N_{class}$.
With increasing number of classes, $l_2$ decreases. The slope remains very steep up to $N_{class} = 21$, and seems to flatten
around $N_{class} = 51$. These two values of $N_{class}$ we've chosen as representative for the rest of the paper.

 From physics perspectives, learning the shape of $wt$ distribution as function of $\alpha^{CP}$,
 is equivalent to learning components of the polarimetric vectors. But, because only the shape, not the
 normalization, is available the $C_i$ coefficients cannot be fully retrieved from formula~(\ref{eq:ABC_alpha}).
 It is not necessary the aim anyway. The physics interest is more to learn $\alpha^{CP}$
 which is preferred by events of the analyzed sample, i.e. value at which $wt$ distribution has its maximum.
 This corresponds to determining CP mixing angle of the analyzed sample.
 
 The second criterium is the difference between most probable predicted class and most probable true class, denoted as $\Delta_{class}$.
 When calculating difference between class indices, periodicity of the functional form~(\ref{eq:ABC_alpha})
 is taken into account. Class indices represent discrete values of $\alpha^{CP}$, in range $(0, 2\pi)$.
 The distance between the first and the last class is zero. We take the distance which corresponds to the smaller
 angle difference and we take the sign according to clock-wise orientation vs class index at which
 true $wt$ has its maximum.
 
 Let's $idp_{max}$ denote the index of most probable predicted class, $idc_{max}$ be index of true most probable
 class.
 The distance $|\Delta_{class}|$ is defined as:
\begin{equation}
    |\Delta_{class}| = min( (|idp_{max}-idc_{max}|), ((N_{class}-1)-|(idp_{max}-idc_{max})|) ),
    \label{eq:defDelt_c0}
\end{equation}
and the sign is attributed
\begin{equation}
    \Delta_{class} = sign(idp_{max}-idc_{max}) \ |\Delta_{class}|,
    \label{eq:defDelt_c1}
 \end{equation}
 if $(|idp_{max}-idc_{max}|) < ((N_{class}-1)-|(idp_{max}-idc_{max})|)$, or 
 
\begin{equation}
    \Delta_{class} = sign(idc_{max} - idp_{max}) \ |\Delta_{class}|,
    \label{eq:defDelt_c2}
 \end{equation}
 otherwise.

\begin{figure}
   \begin{center}
     {
       \includegraphics[width=7.0cm,angle=0]{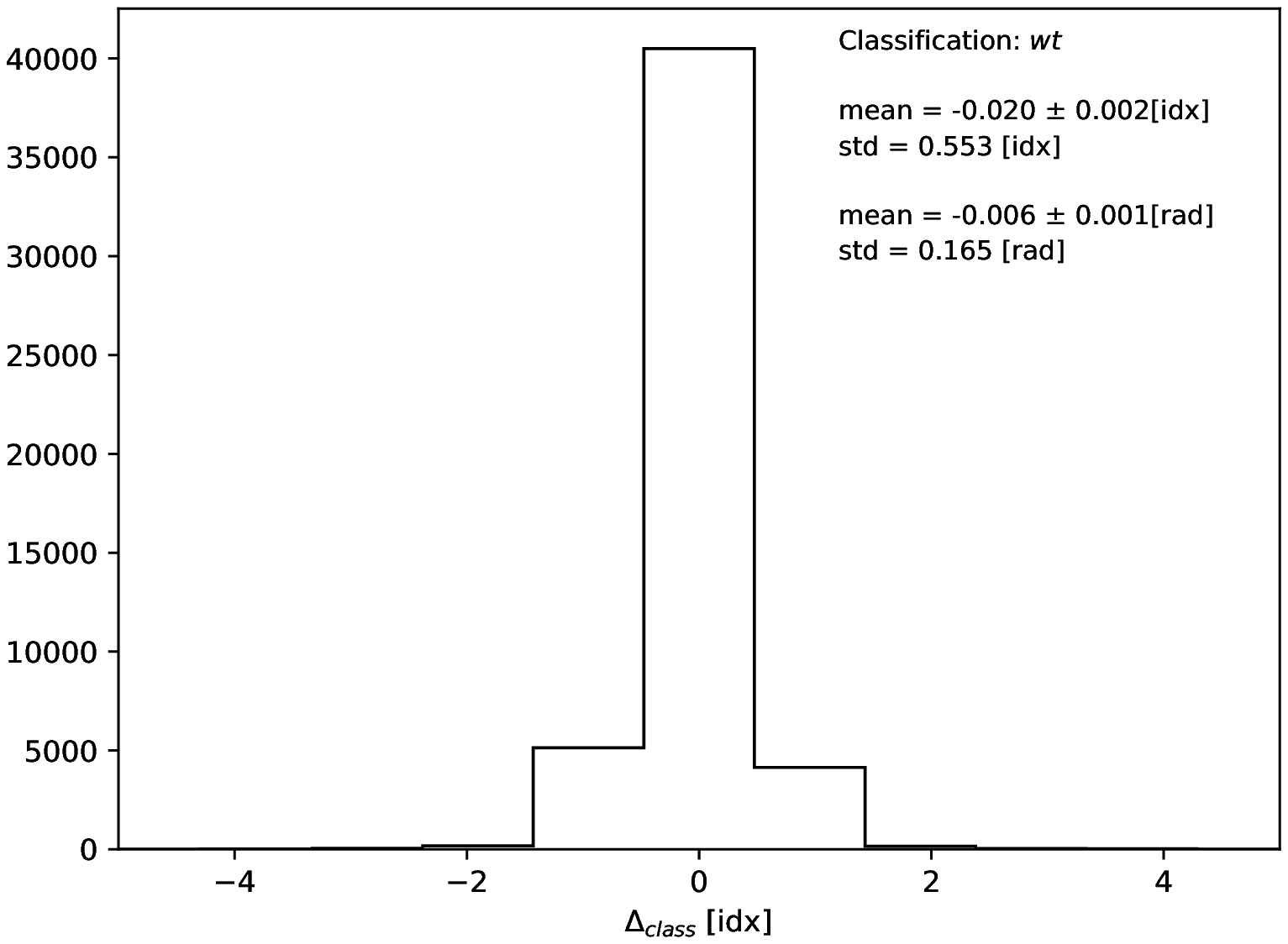}
       \includegraphics[width=7.0cm,angle=0]{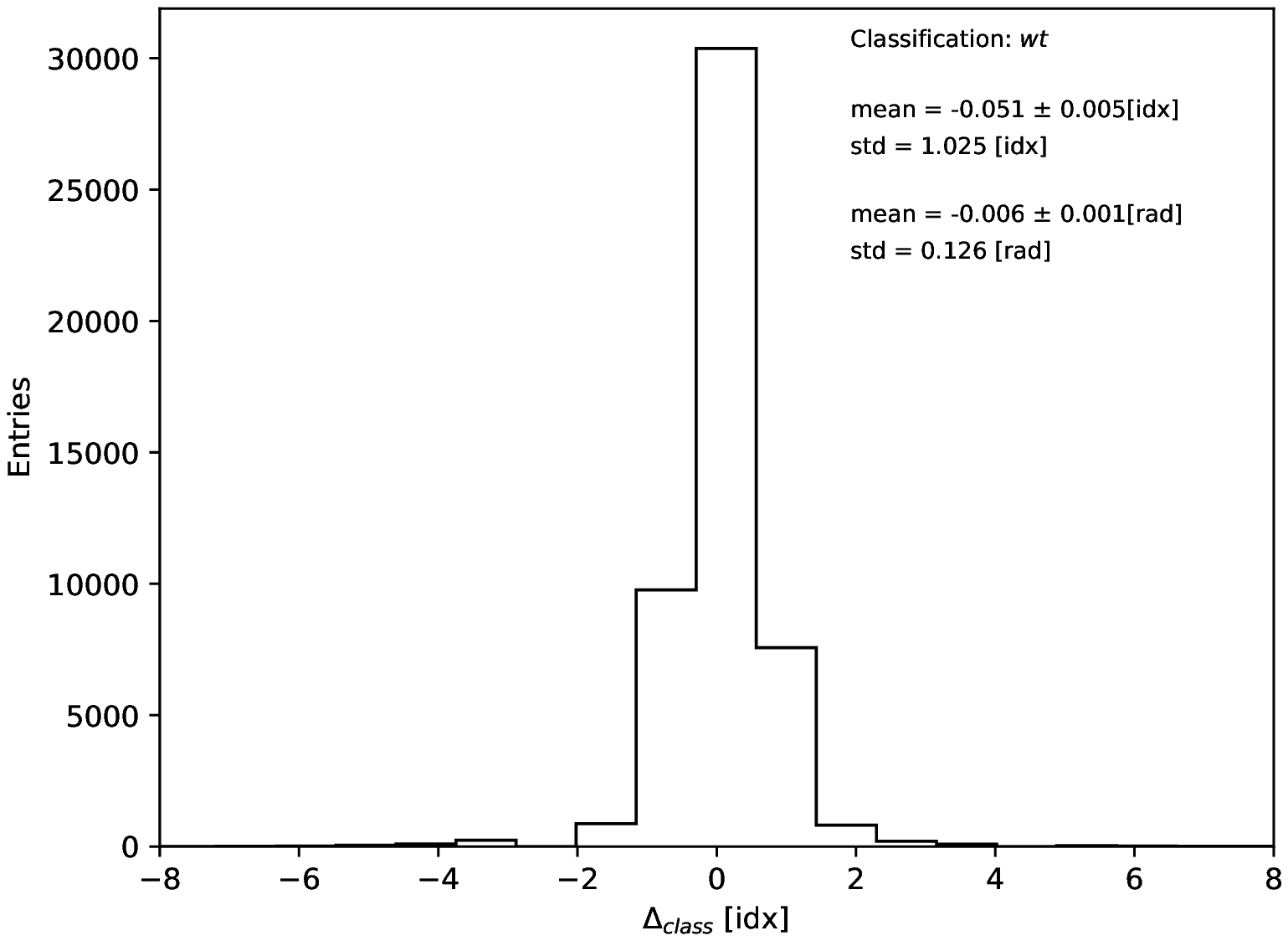}
     }
\end{center}
   \caption{Distribution of $\Delta^{max}_{class}$ between predicted most probable class and true most probable
     class for  $N_{class}$ = 21 and 51 respectively.
     The {\tt mean} and {\tt std} are calculated in units of class index [idx] or units of radians [rad].
   \label{fig:DNN_soft_deltaclass_nc}}
 \end{figure}

 In Fig.~\ref{fig:DNN_soft_deltaclass_nc}  distributions of $\Delta_{class}$ for $N_{class}$ = 21 and 51 respectively are shown.
 The shapes are Gaussian-like and centered around zero.
The mean <$\Delta_{class}$>~=~-0.006~[rad] in both cases
and this we can interpret as the bias of the method. The standard deviation of per-event distribution is
$\sigma_{\Delta_{class}}$ = 0.165 [rad] for $N_{class}$ = 21 and $\sigma_{\Delta_{class}}$~=~0.126~[rad] for $N_{class}$ = 51.
As we can see, the performance has not improved significantly with  $N_{class}$ exceeding 21.

 The {\it DNN} classifier which is predicting normalized spin weight $wt^{norm}$, provides enough information to identify the most
 probable mixing angle $\alpha^{CP}_{max}$ with high precision. The information is not sufficient though to reconstruct complete
 set of $C_i$ coefficients and the polarimetric vectors.

\subsection{Learning $C_0, C_1, C_2$ coefficients}
\label{Sec:Classification:C012}

The second approach is to learn
formula~(\ref{eq:ABC_alpha}) coefficients $C_0, C_1, C_2$ for the spin weight $wt$.
They can be then used to predict not only normalized $wt^{norm}$, but also original $wt$. Coefficients $C_0, C_1, C_2$
represent physical observables,
products of longitudinal and transverse components of polarimetric vectors, as shown in formulas~(\ref{eq:CPcoeff}). 

The classification technique using {\it DNN} is configured to learn each of the $C_i$ with separate training.
The allowed range is well known, the $C_0$ spans the range (0.0, 2.0) and $C_1, C_2$
the range  (-1.0, 1.0), see Fig.~\ref{fig:c012s}. The allowed range is binned into $N_{class}$ and as a label,
the $N_{class}$-dimensional vector with one-hot encoded value of the $C_i$ parameter is associated with each event.
Therefore in this case, a single class represents range of the $C_i$ coefficient.
During training, the  {\it DNN} is learning per-event association between feature list and the class labels.
The output is a probability $N_{class}$-dimensional vector, which is then converted to one-hot encoded representation,
i.e. the most probable class is chosen as a predicted value of the $C_i$ coefficient. 

\begin{figure}
   \begin{center}
     {
       \includegraphics[width=5.0cm,angle=0]{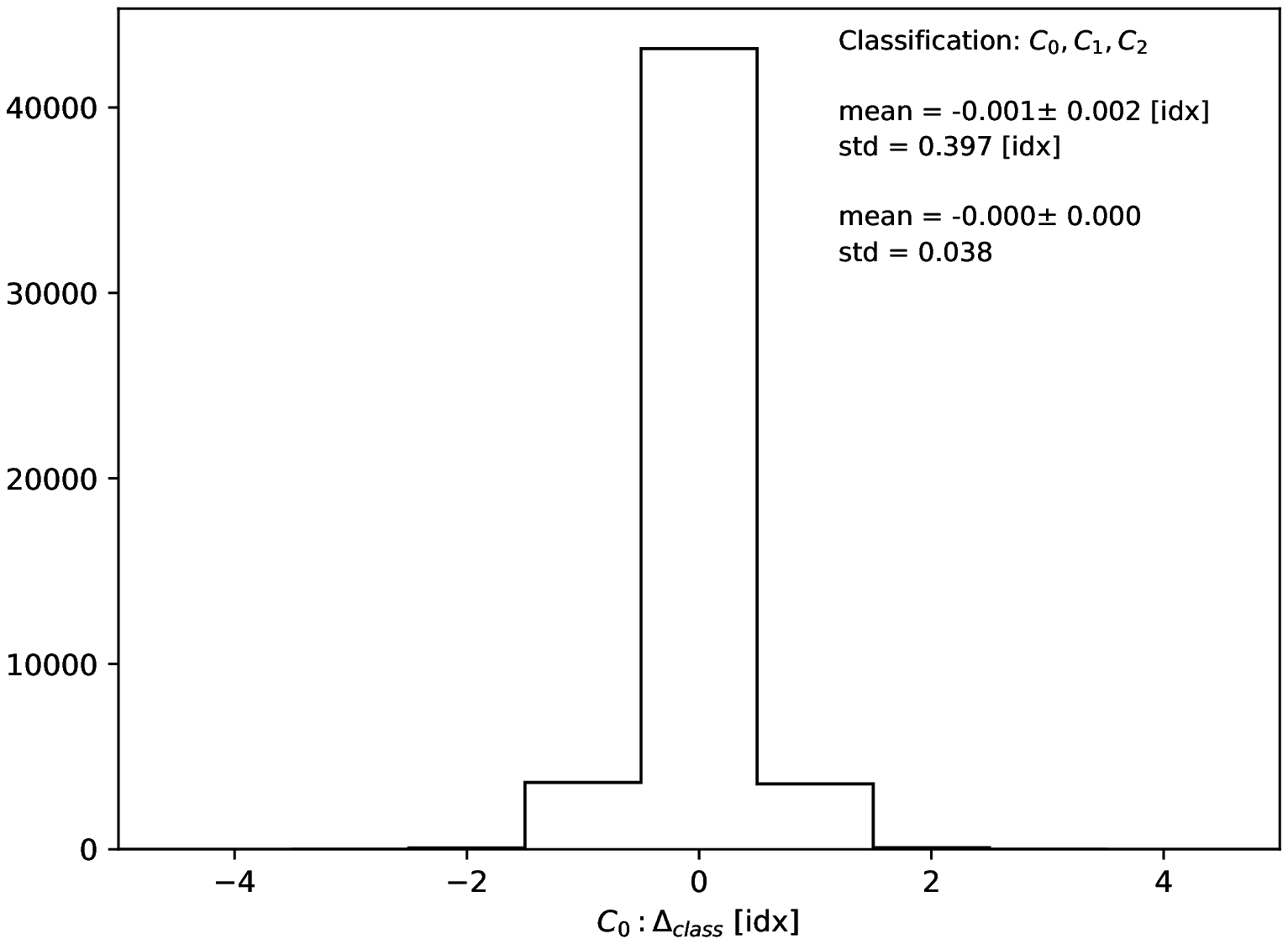}
       \includegraphics[width=5.0cm,angle=0]{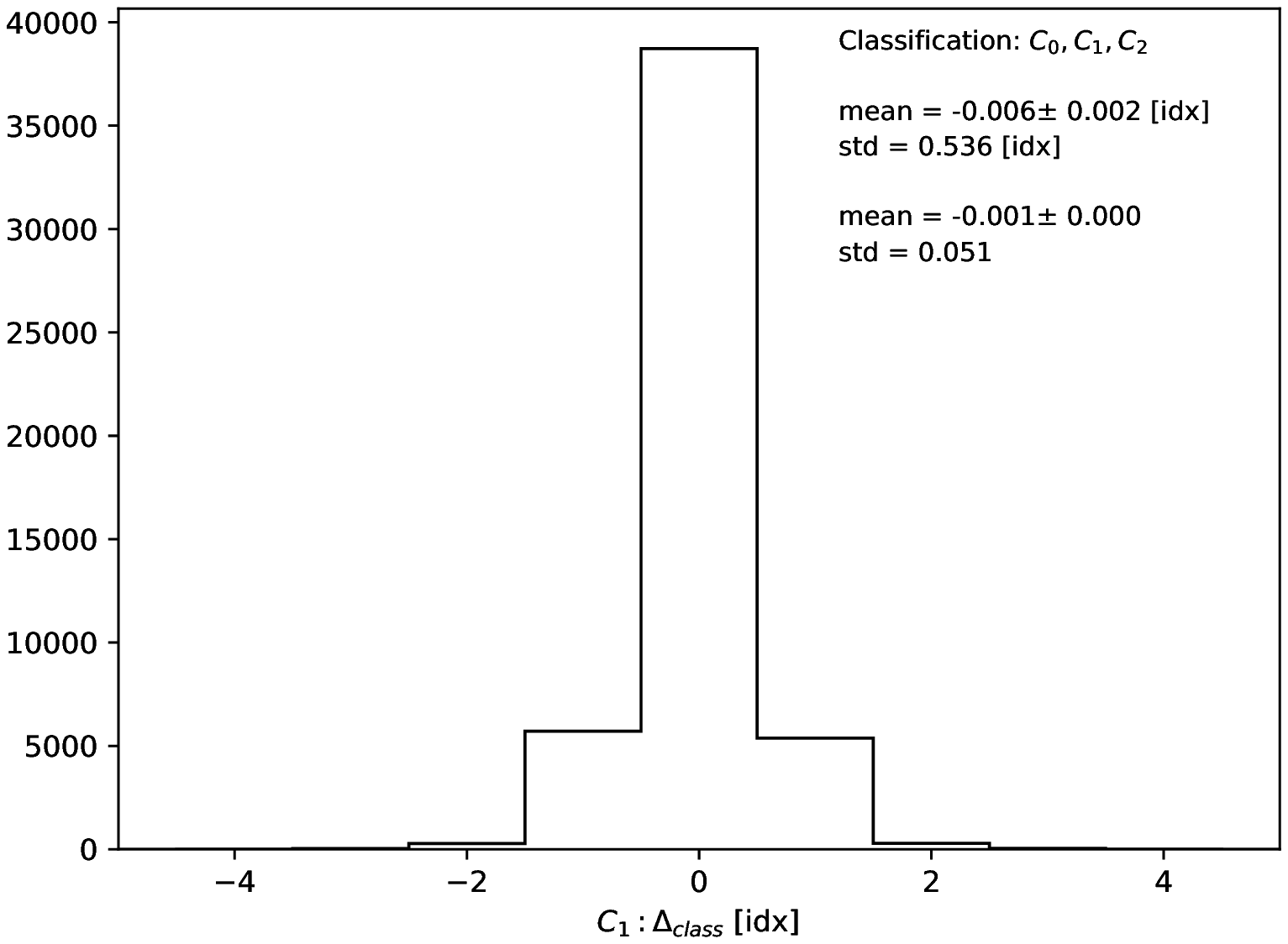}
       \includegraphics[width=5.0cm,angle=0]{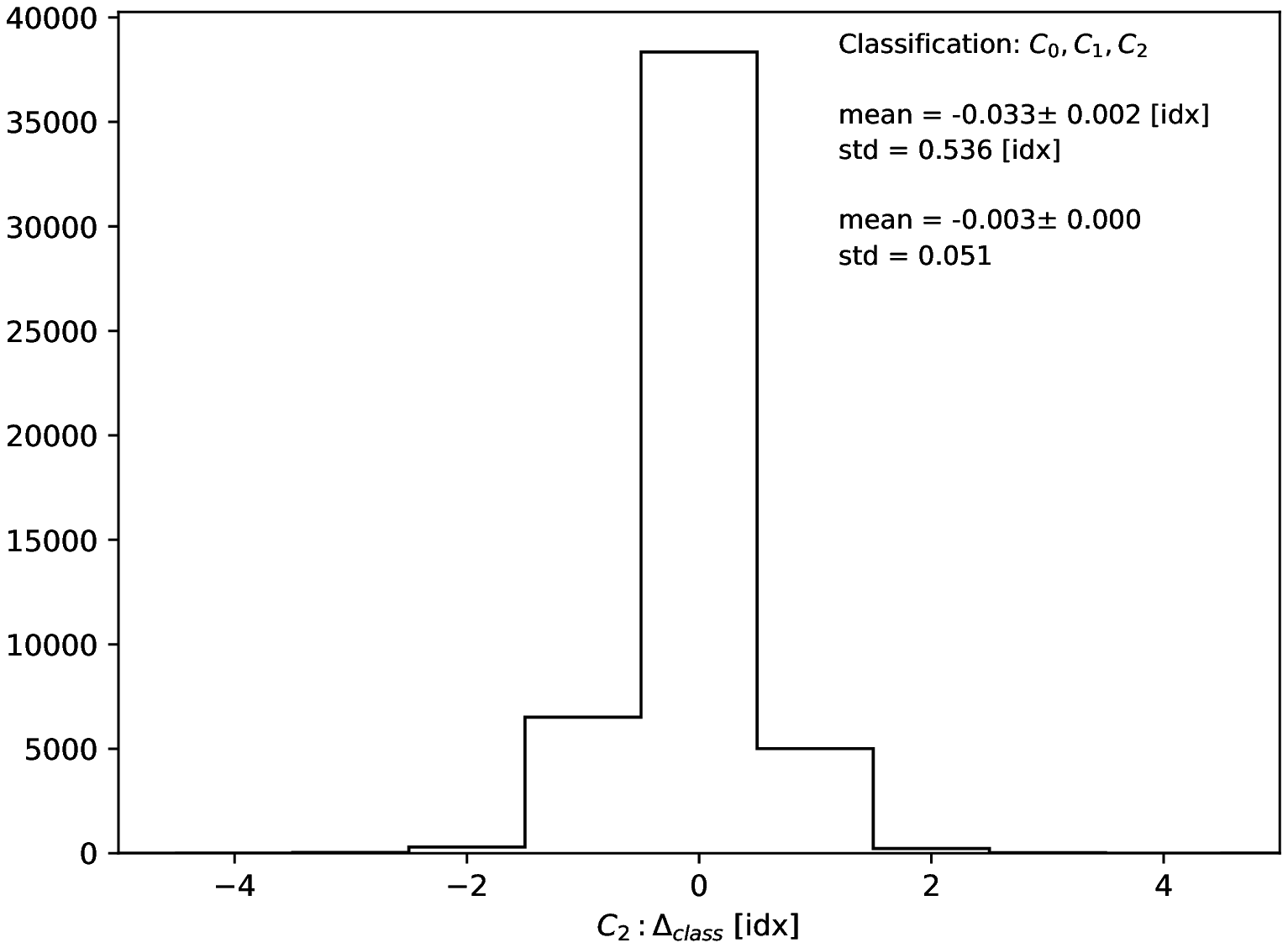}
     }
\end{center}
   \caption{Difference between true and predicted  coefficients $C_0, C_1, C_2$ of formula (\ref{eq:ABC_alpha}).
     For {\it DNN} training the granularity of  $N_{class}$= 21 was used.
   \label{figApp:DNN_soft_ABC}}
 \end{figure}

Distributions of the  difference between true and predicted $C_i$ coefficients are shown in Figs.~\ref{figApp:DNN_soft_ABC}.
In that case, as there is no periodicity involved, $\Delta_{class} = idp - idc $ where
$idp$, $idc$ denote respectively true and predicted class index. Mean of $\Delta C_i$ is close
to zero and standard deviation is of 0.038-0.051, which is less than 5\% of the range.
Precision with which $C_i$ coefficients are predicted is clearly limited by the $N_{class}$. 

\begin{figure}
   \begin{center}
     {
       \includegraphics[width=7.0cm,angle=0]{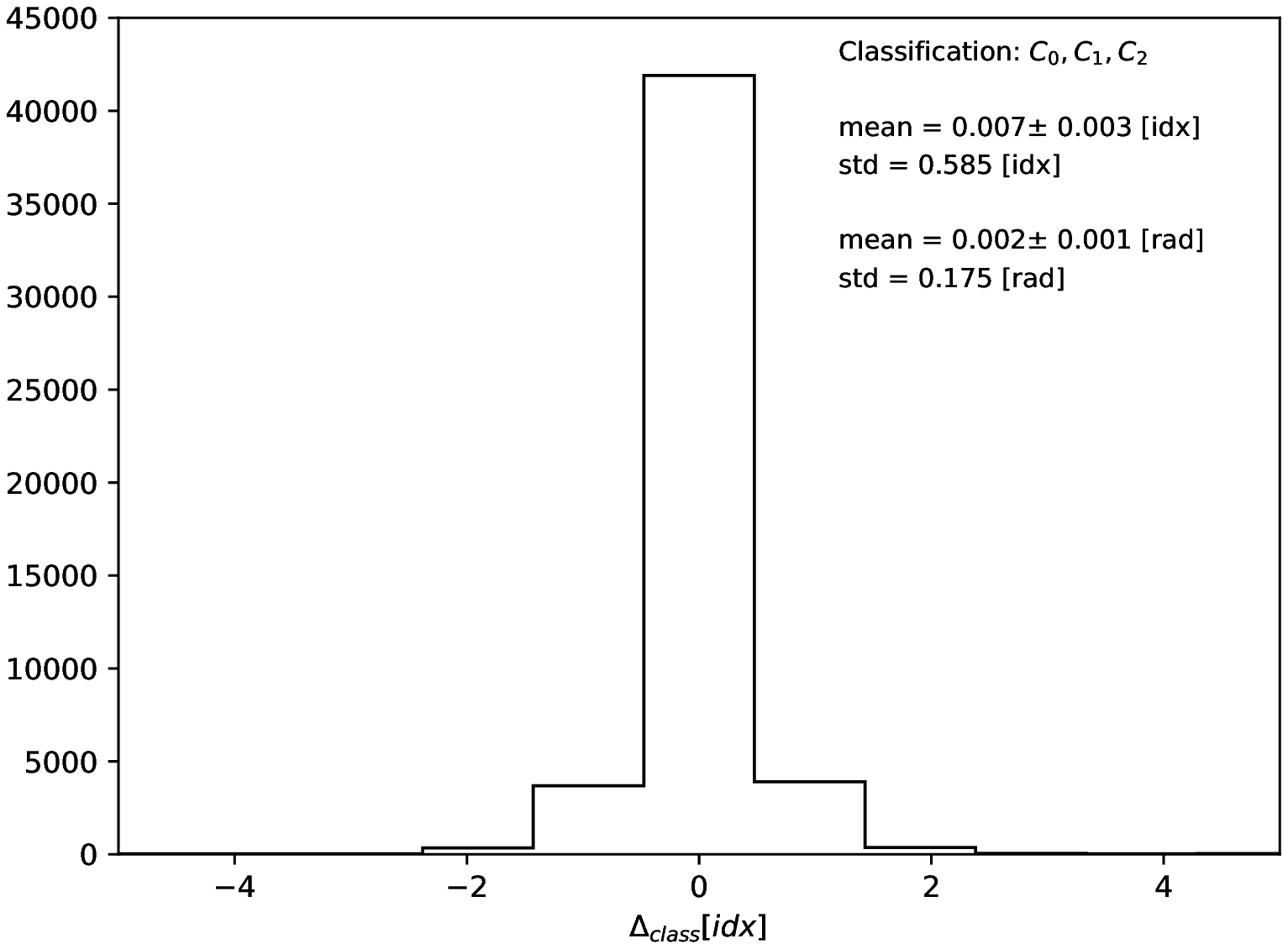}
       \includegraphics[width=7.0cm,angle=0]{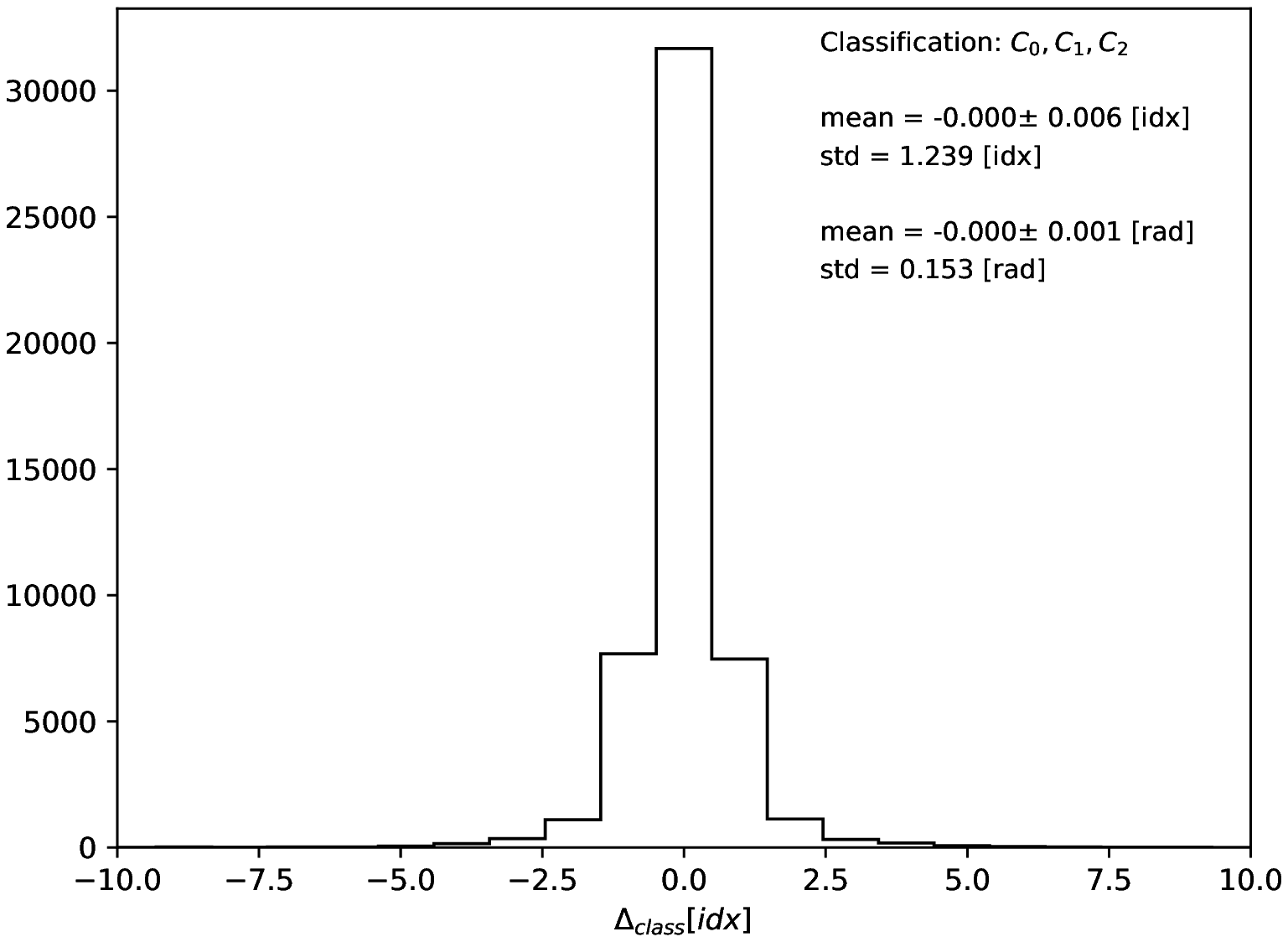}
     }
\end{center}
   \caption{The difference between true and predicted most probable mixing angle $\alpha^{CP}_{max}$, calculated
     using formula (\ref{eq:ABC_alpha}) 
     and coefficients $C_0, C_1, C_2$ learned with classification method.
     The granularity of $\alpha^{CP}_{max}$,   $N_{class}$= 21 and 51 was used respectively for left and right-hand plot.
   \label{figApp:DNN_soft_ABC_alphamax}}
 \end{figure}

We use the true and predicted $C_0, C_1, C_2$ coefficients to calculate $wt$ distribution of~(\ref{eq:ABC_alpha}).
It is then discretised with $N_{class}$ points (the $N_{class}$ could be different than the one used for learning coefficients),
and the $\alpha^{CP}_{max}$ is determined from the class of maximal weight. The difference between true and predicted
$\alpha^{CP}_{max}$ is shown in Fig.~\ref{figApp:DNN_soft_ABC_alphamax} for  $N_{class}$ = 21 and 51.
The Gaussian-like shape of those distributions, centered around zero, clearly demonstrated that method works as expected.
The mean and standard deviation of the distributions are close to those obtained with {\tt Classification:wt} approach,
of Fig.~\ref{fig:DNN_soft_deltaclass_nc}.

\begin{figure}
   \begin{center}
     {
       \includegraphics[width=5.0cm,angle=0]{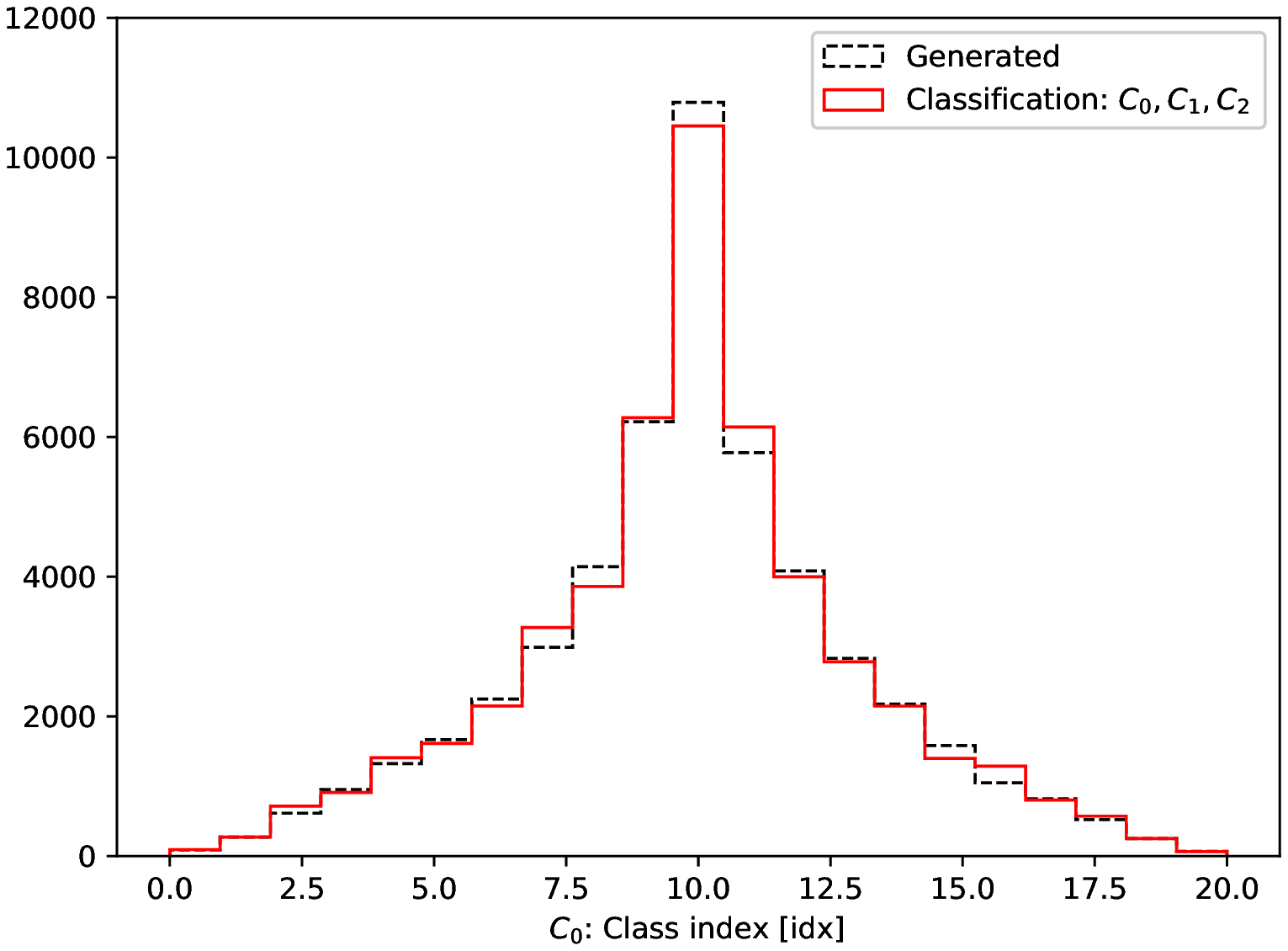}
       \includegraphics[width=5.0cm,angle=0]{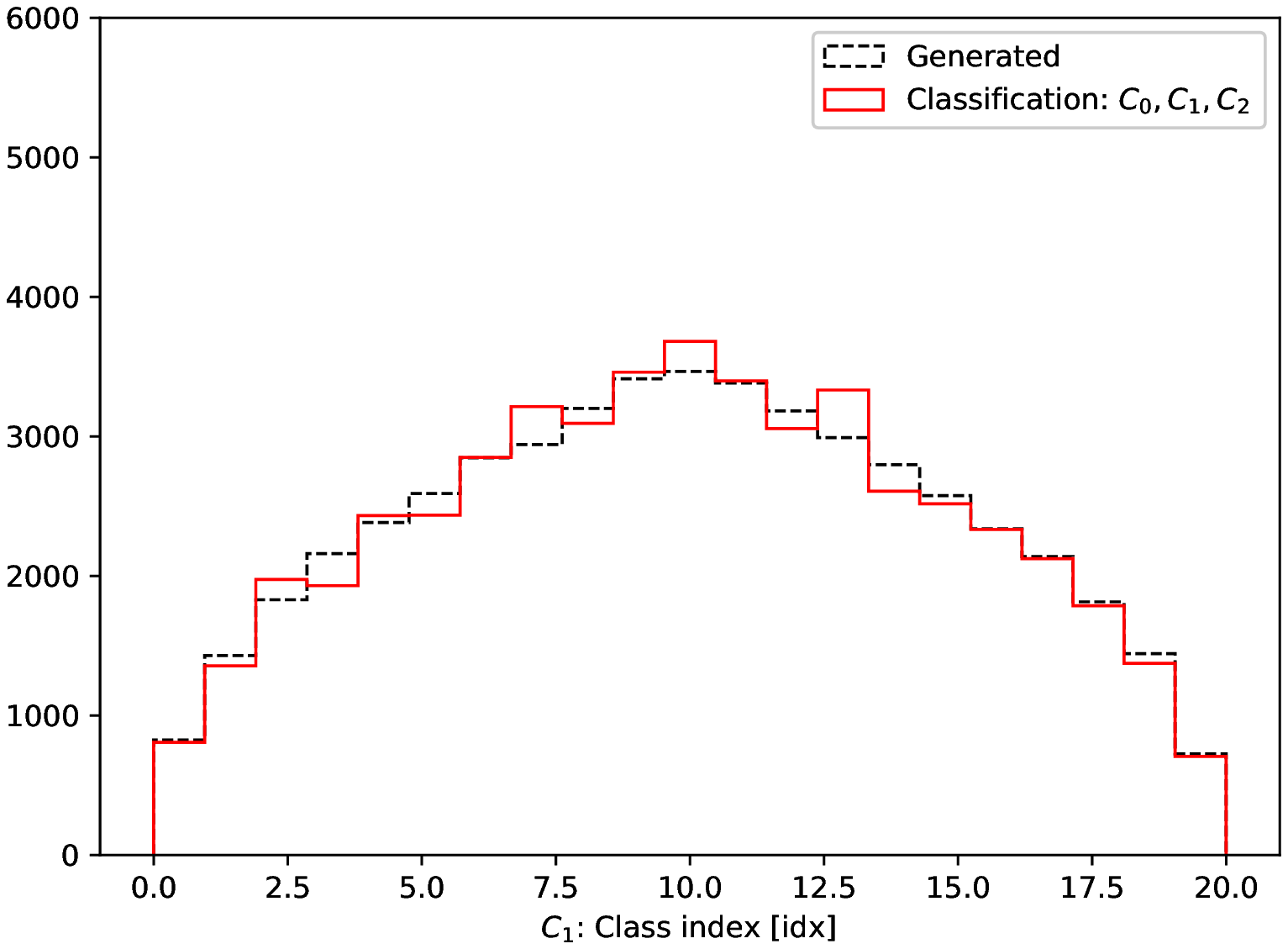}
       \includegraphics[width=5.0cm,angle=0]{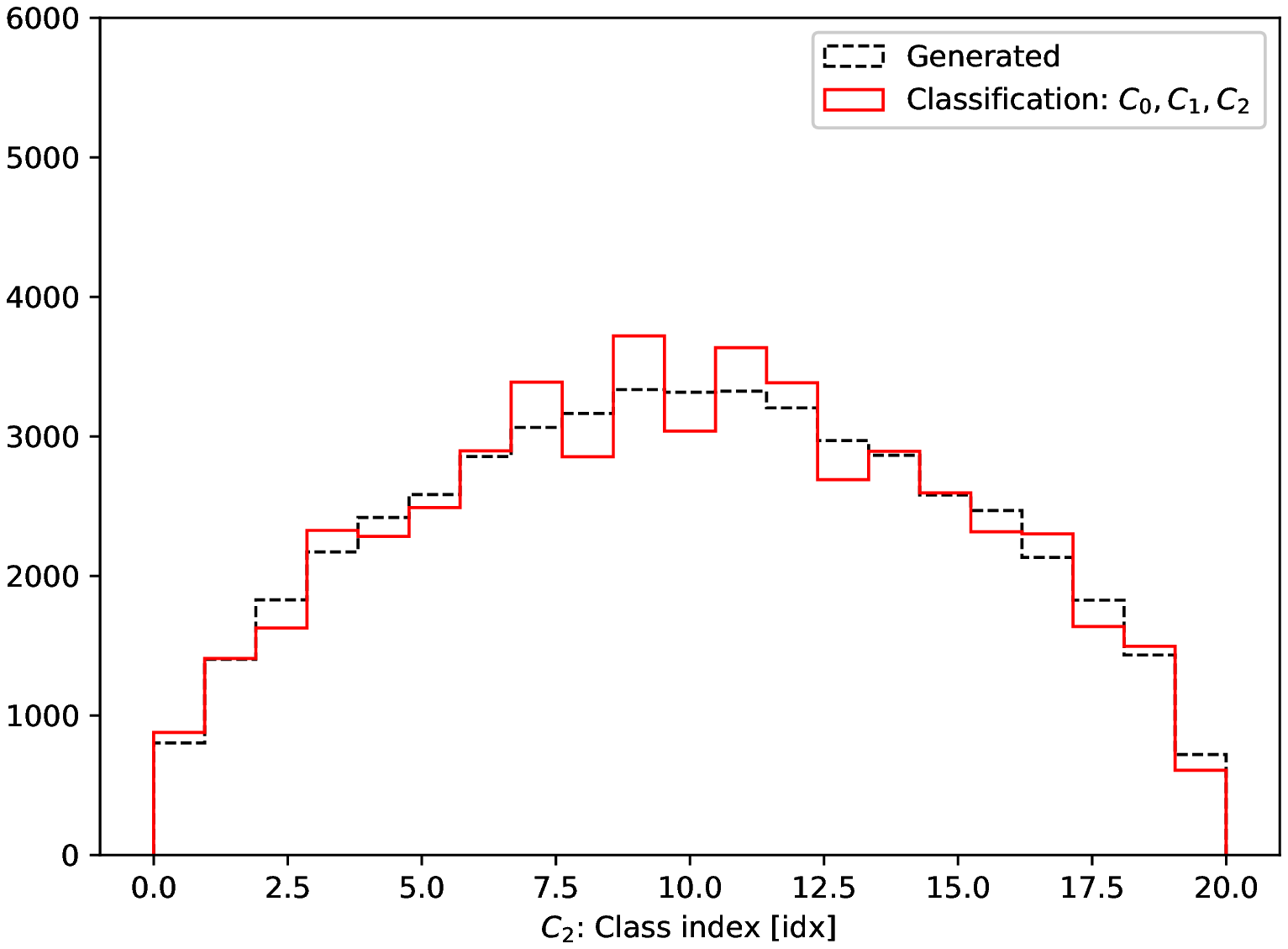}
     }
\end{center}
   \caption{Distributions of true and predicted coefficients $C_0, C_1, C_2$ of formula (\ref{eq:ABC_alpha}).
     For {\it DNN} training the granularity of  $N_{class}$= 21 was used.
   \label{fig:DNN_soft_CO12}}
 \end{figure}

 Finally, as sanity check we have compared the true distributions of $C_0, C_1, C_2$ with the predicted ones.
 As we can see in Fig.~\ref{fig:DNN_soft_CO12}, both distributions match very well for all $C_i$.

 \subsection{Learning the $\alpha^{CP}_{max}$}
 \label{Sec:Classification:alphaCPmax}

 The third approach is to directly learn per-event most  preferred mixing angle, $\alpha^{CP}_{max}$.
 The allowed range (0, 2\pi) is again binned into $N_{class}$ classes, where single bin represents discrete $\alpha^{CP}$.
 For training, for every event we take the one-hot encoded vector of $N_{class}$-dimension as a label.
 The  {\it DNN} is returning $N_{class}$-dimensional vector of probabilities, which is then transformed into a single number,
 that is the class of the highest probability $\alpha^{CP}_{max}$.
 With this approach, neither spin weight nor $C_i$ coefficients are predicted. 
 
\begin{figure}
   \begin{center}
     {
       \includegraphics[width=7.0cm,angle=0]{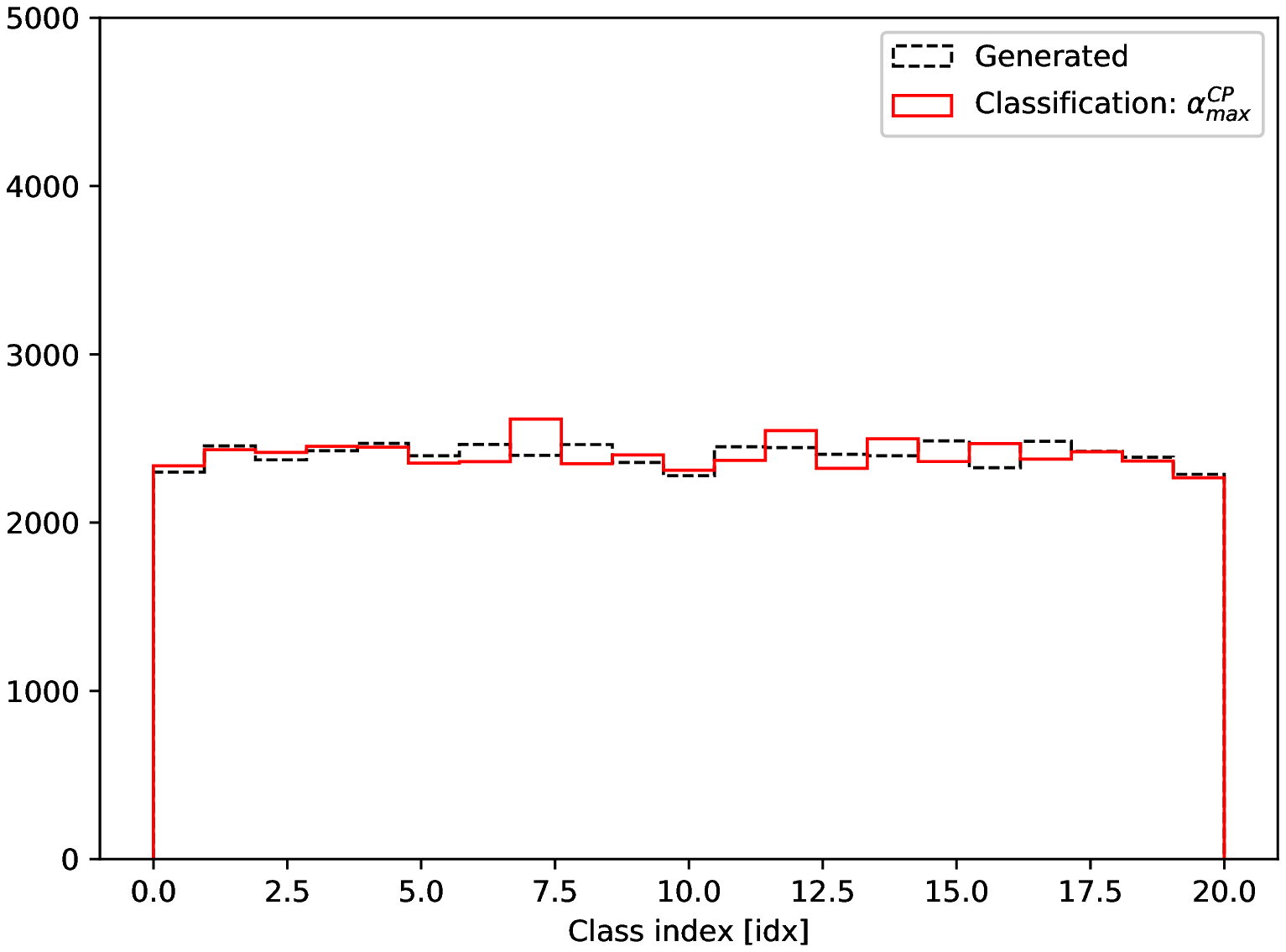}
       \includegraphics[width=7.0cm,angle=0]{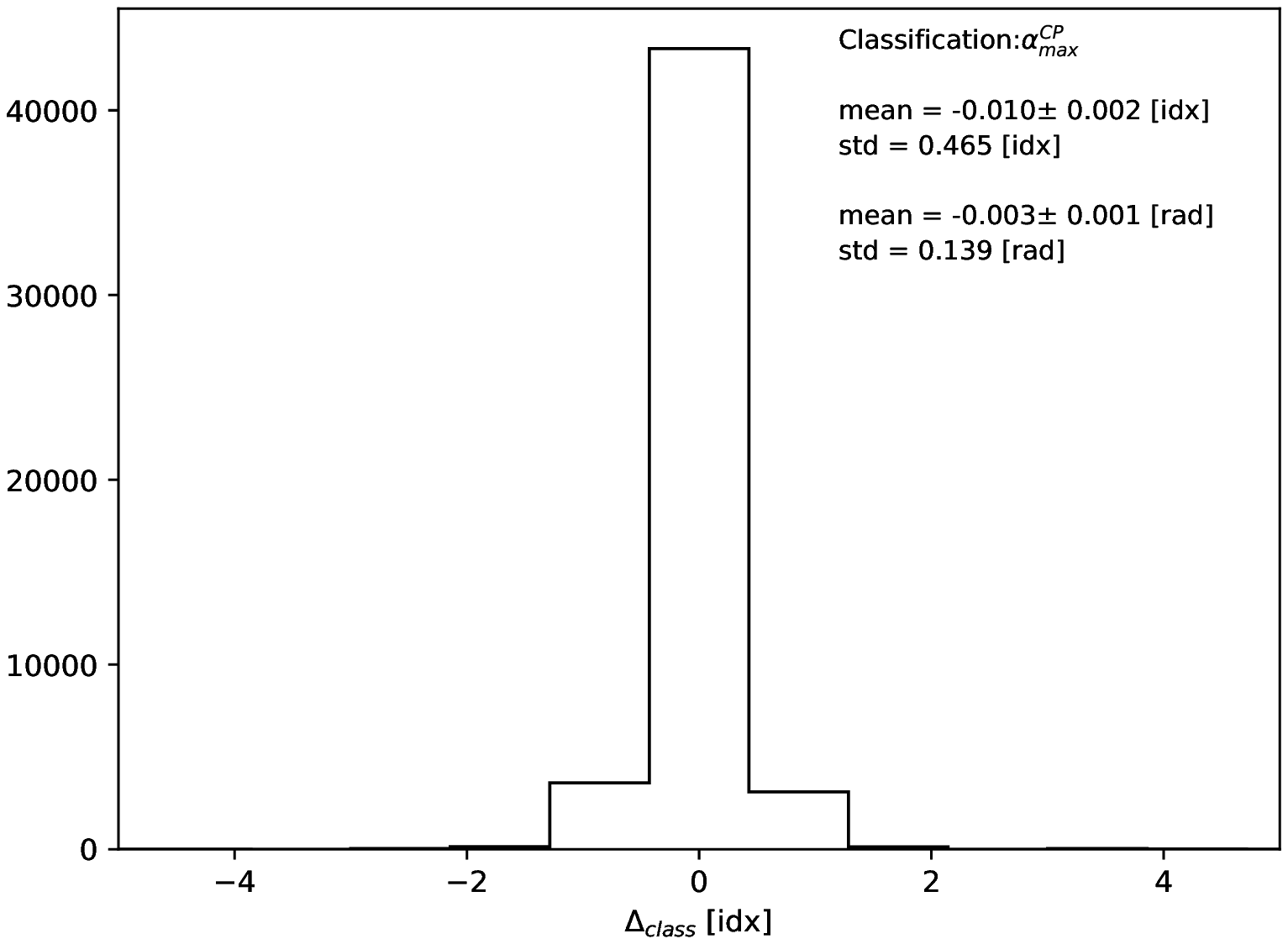}
     }
\end{center}
   \caption{Distributions (left-hand plot) of true and predicted most preferred mixing angle $\alpha^{CP}$.
     The distribution of per-event difference of the two is shown on the right-hand plot. 
     The granularity of $N_{class}$= 21 was used for training {\it DNN}.
   \label{figApp:DNN_soft_argmax}}
 \end{figure}

 As the event sample is generated without any CP mixture favoured, the distribution of the $\alpha^{CP}_{max}$
 is expected to be uniform, and such sanity check is demonstrated in the left plot of Fig.~\ref{figApp:DNN_soft_argmax}.
 The {\it DNN} is well reproducing this behaviour. The $\Delta \alpha^{CP}_{max}$, the difference between true and predicted value
 of the  $\alpha^{CP}_{max}$ is shown in the right plot of Fig.~\ref{figApp:DNN_soft_argmax}. In the case of $N_{class}$ = 21, it has a Gaussian-like shape
 with the mean $<\Delta \alpha^{CP}_{max}>$ = 0.003 $\pm$ 0.001 [rad] and standard deviation 0.139~[rad].
 Results are again comparable with the ones obtained with the previously discussed approaches.

\section{Regression}
\label{Sec:MLregression}

The ML regression is not so commonly used in the high energy physics analyses. The main feature is, that contrary to the classification case,
we get a continuous parameter (or set of parameters) as a {\it DNN} output. We explore three approaches, defined similarly as in Section~\ref{Sec:ML-multiclass}

\begin{itemize}
\item
The {\it DNN} is learning to predict per-event spin weight as a function of mixing angle $\alpha^{CP}$.
The range of mixing angle $(0, 2\pi)$ is split into discrete points of $\alpha^{CP}$ at which value of
spin weight is learned.
This approach is described in Section~\ref{Sec:regr_wt} and used for the figures labeled with:
{\tt Regression:wt}.
\item
The {\it DNN} is learning to predict per-event value of the coefficients $C_0, C_1, C_2$ of the functional
form~(\ref{eq:ABC_alpha}). The {\it DNN} is trained for all coefficients simultaneously.
This approach is described in Section~\ref{Sec:regr_C012} and used for the figures labeled with:
{\tt Regression:$C_0, C_1, C_2$}.
\item
The {\it DNN} is learning to predict per-event most probable mixing angle $\alpha^{CP}_{max}$, i.e. where
$\alpha^{CP}$ spin weight has maximum. 
This approach is described in Section~\ref{Sec:regr_alphaCPmax} and used for the figures labeled with:
{\tt Regression:$\alpha^{CP}_{max}$}.
\end{itemize}
  
We continue with {\tt Tensorflow} package, but now with
{\it tf.losses.mean\_squared\_error} function as a {\it loss} in the training procedure
of Section~\ref{Sec:regr_wt},~\ref{Sec:regr_C012} and self-defined function in the training
procedure of Section~\ref{Sec:regr_alphaCPmax}. Mentioned self-defined function is discussed in the appendix.

\subsection{Learning spin weight $wt$}
\label{Sec:regr_wt}

Similarly as in the classification case, the {\it DNN} regression is trained on an
input information consisting of per-event feature list.  As a training output we provide a vector of the spin
weight $wt_i$ for the discrete values of $\alpha^{CP}$. Training is performed for different granularities of
$\alpha^{CP}$ discretisation, to monitor performance sensitivity. 
Again in this case we use odd number of equally spaced points $\alpha_i^{CP}$, so the $\alpha^{CP}=0, \pi, 2 \pi$
coincide with a single point.
It is worth noting, that in case of regression, both shape and normalization of the $wt$ are learned by the {\it DNN}.

\begin{figure}
   \begin{center}
     {
      \includegraphics[width=7.0cm,angle=0]{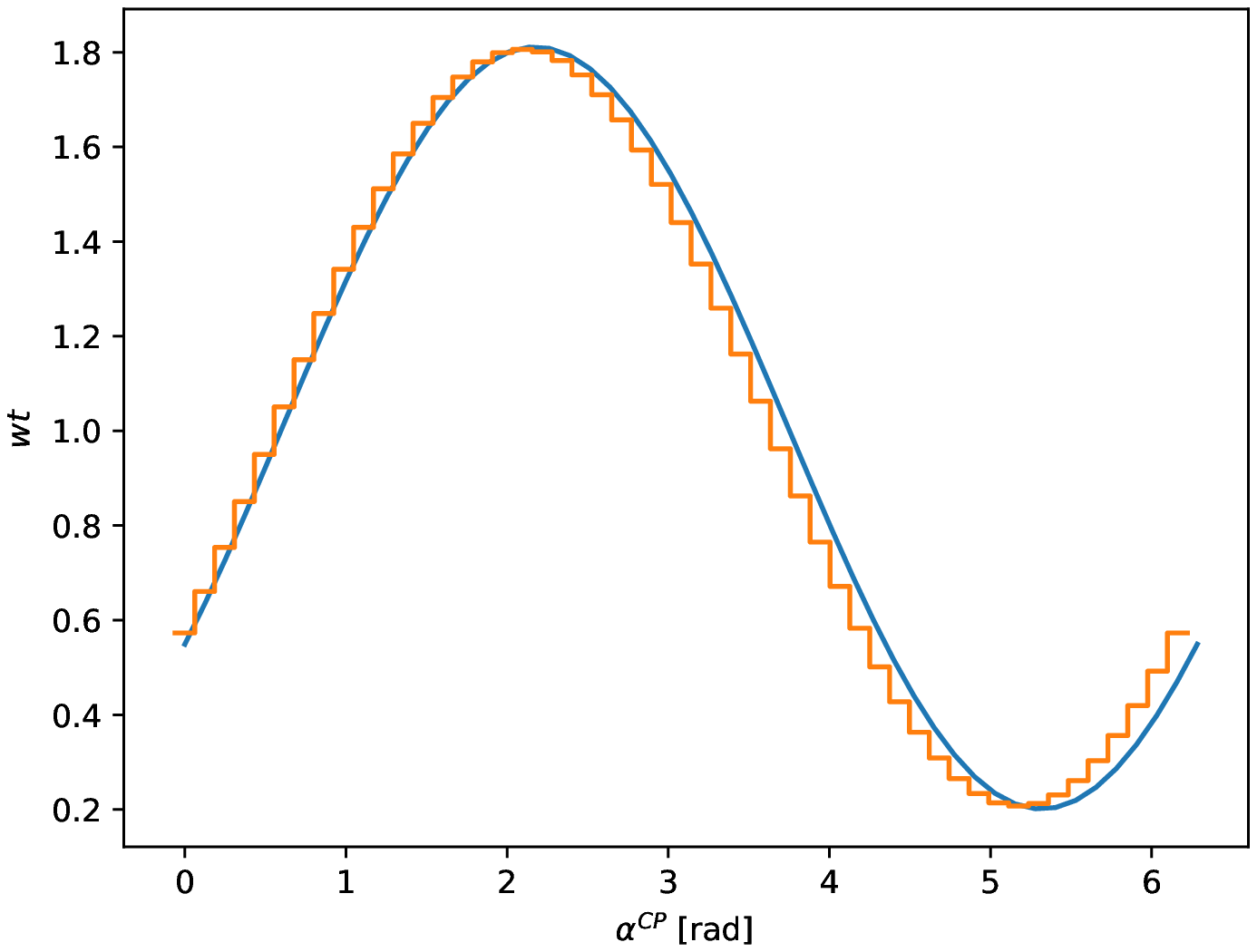}
      \includegraphics[width=7.0cm,angle=0]{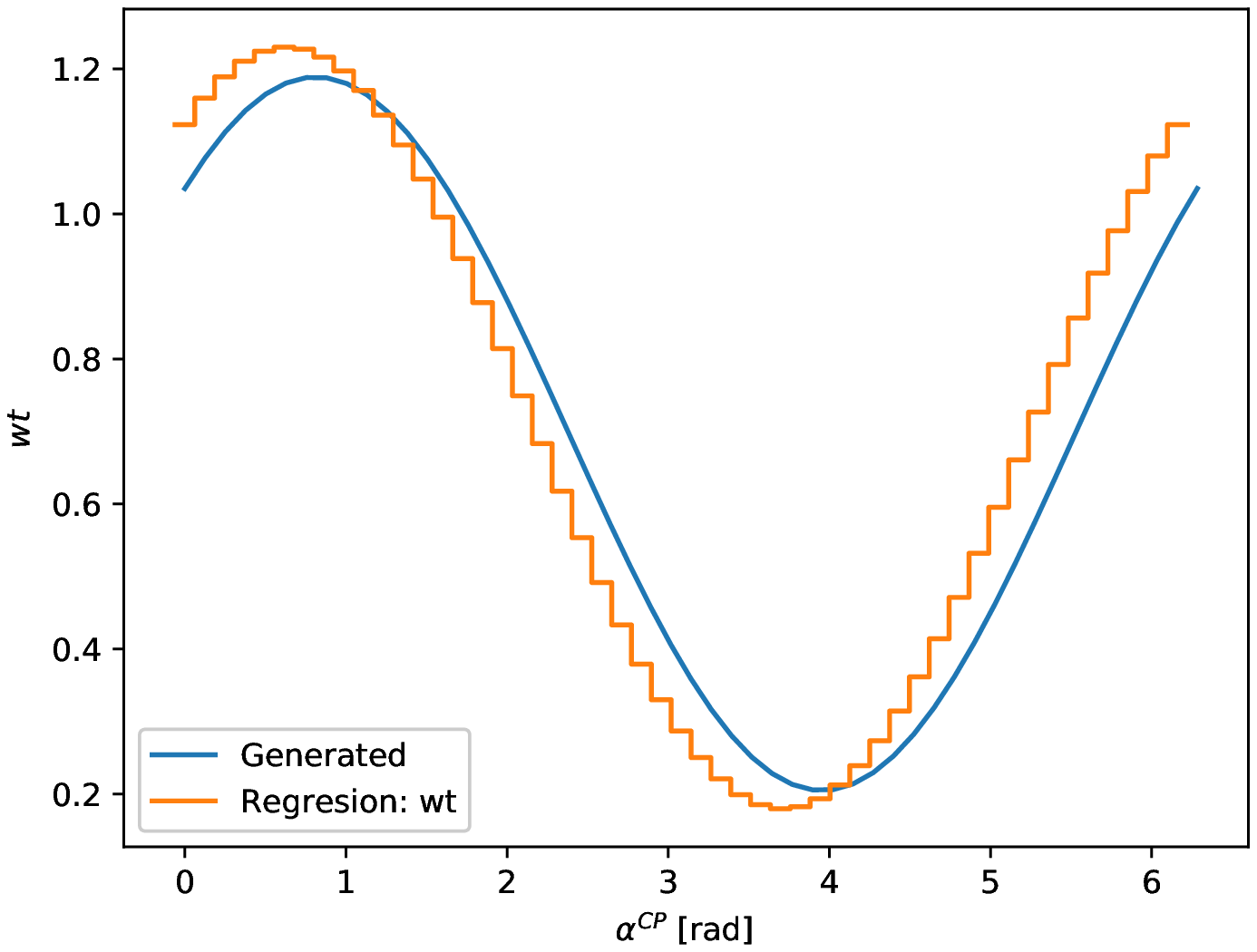}
     }
\end{center}
   \caption{Example plots with {\it DNN} regression results:
     the  spin weight $wt$, predicted  (orange steps) and true (blue line),
     as a function of $\alpha^{CP}_i$ for two example events (left and right plots).
     {\it DNN} was trained with $N_{class} = 51$ spanning range $(0, 2 \pi)$.
   \label{fig:DNN_regr_nc_predwt}}
 \end{figure}

For two example events in Fig.~\ref{fig:DNN_regr_nc_predwt}, true continuous spin weight $wt$ distribution
as well as step-function prediction is shown
as a function of mixing parameter $\alpha^{CP}$.
In overall, predicted weights follow smoothly expected shape of linear $\cos (\alpha^{CP})$ and
$\sin (\alpha^{CP})$ combination, even if no attempt to regularize for such smooth behaviour was made. 
 
\begin{figure}
   \begin{center}
     {
   \includegraphics[width=7.0cm,angle=0]{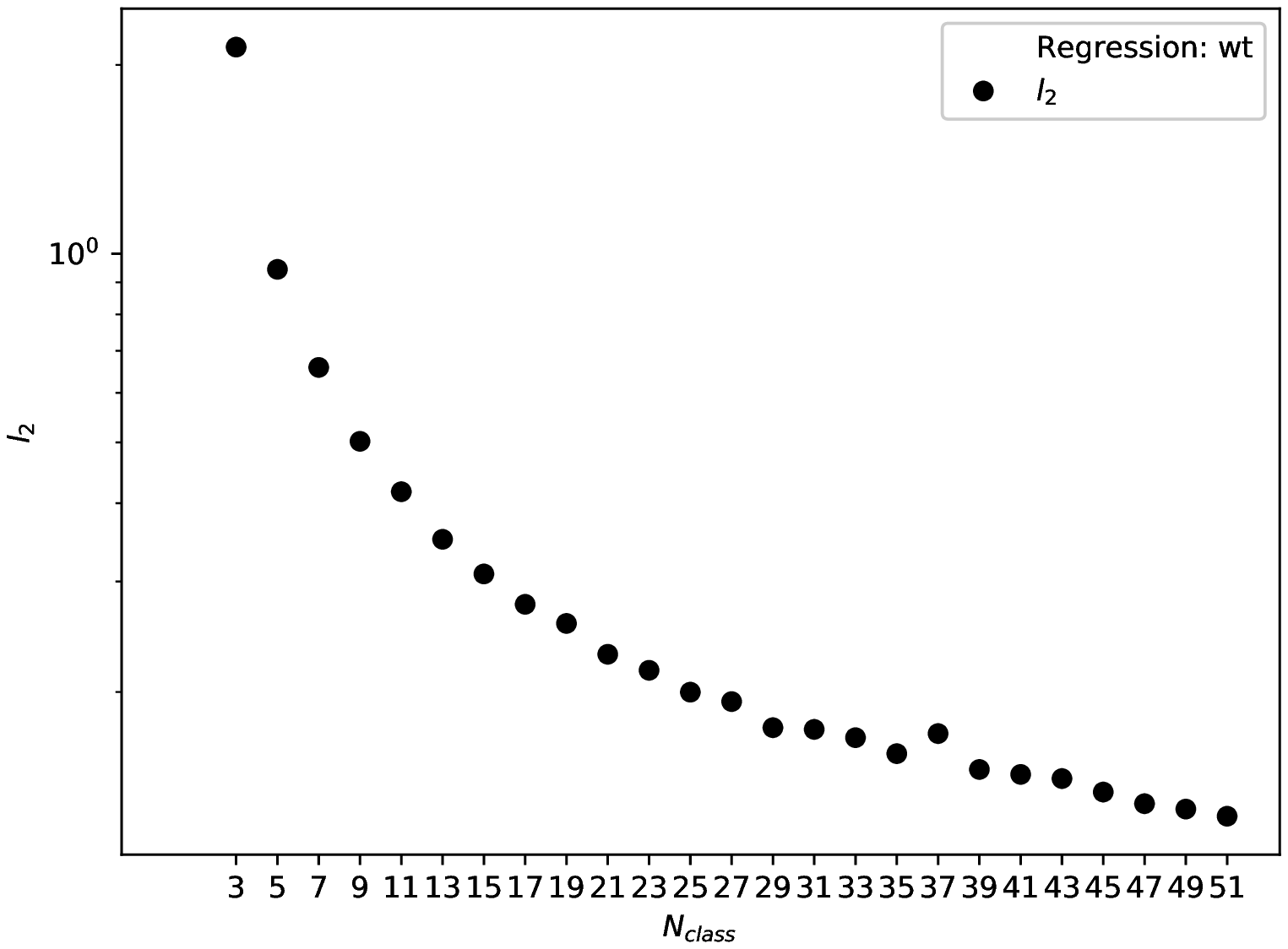}
   \includegraphics[width=7.0cm,angle=0]{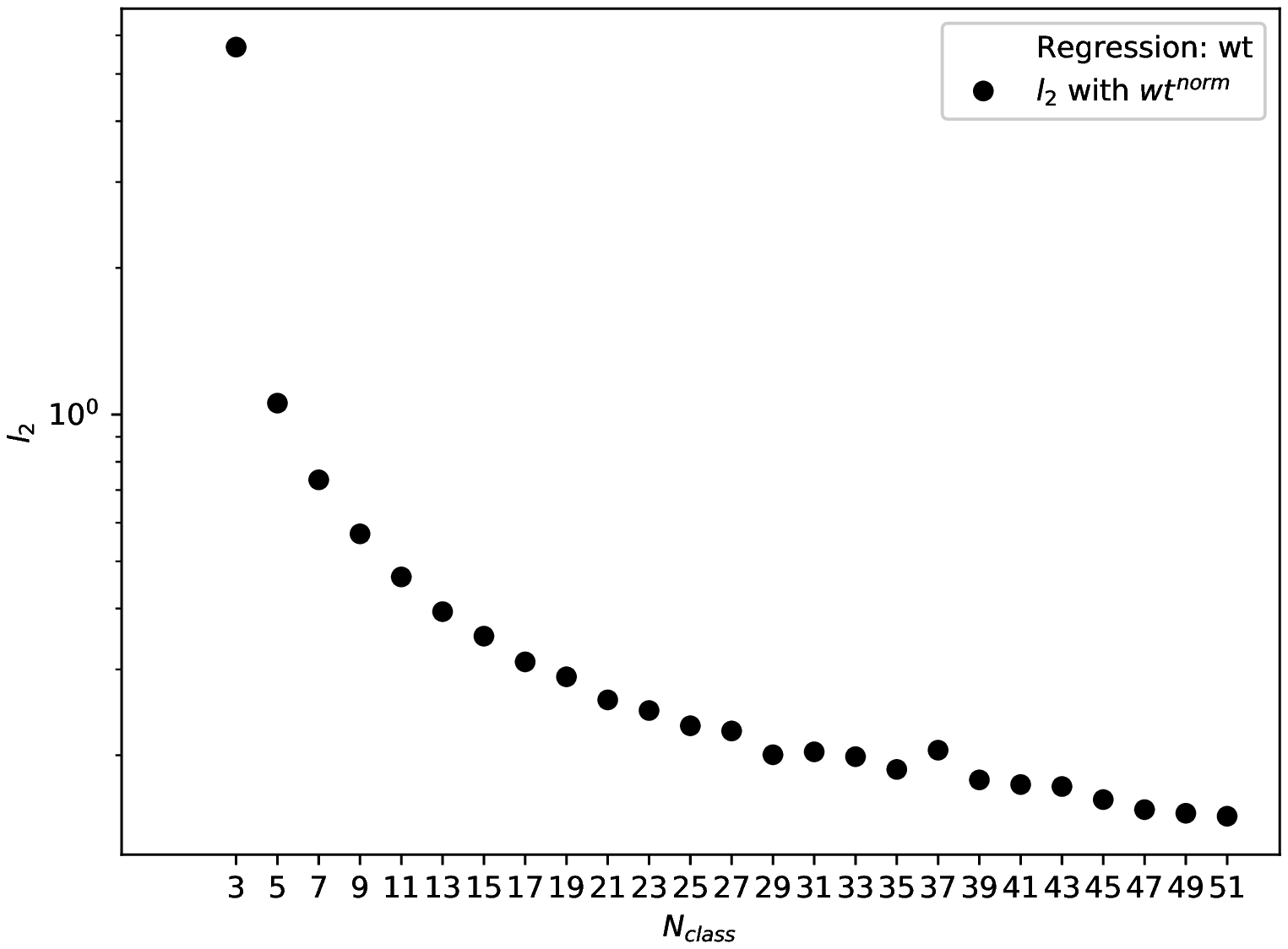}
     }
\end{center}
   \caption{The $l_2$ norm for predicted spin weight $wt$ (left) and  $wt^{norm}$ (right) as a function of $N_{class}$.
   \label{fig:DNN_regr_nc_l2}}
 \end{figure}

 Distributions of  $l_2$ norm, defined in the same way as in the classification case, 
 as a function of $N_{class}$ (granularity for discretising $\alpha^{CP}$) is shown in Figure \ref{fig:DNN_regr_nc_l2}.
 For more compatibility with the classification case of Section~\ref{Sec:Classification:wt} we present
 results for original $wt$, as well as normalized to unity $wt^{norm}$. The results are comparable, with a
 visible flattening of $l_2$ for higher values of $N_{class}$.

 \begin{figure}
   \begin{center}
     {
       \includegraphics[width=7.0cm,angle=0]{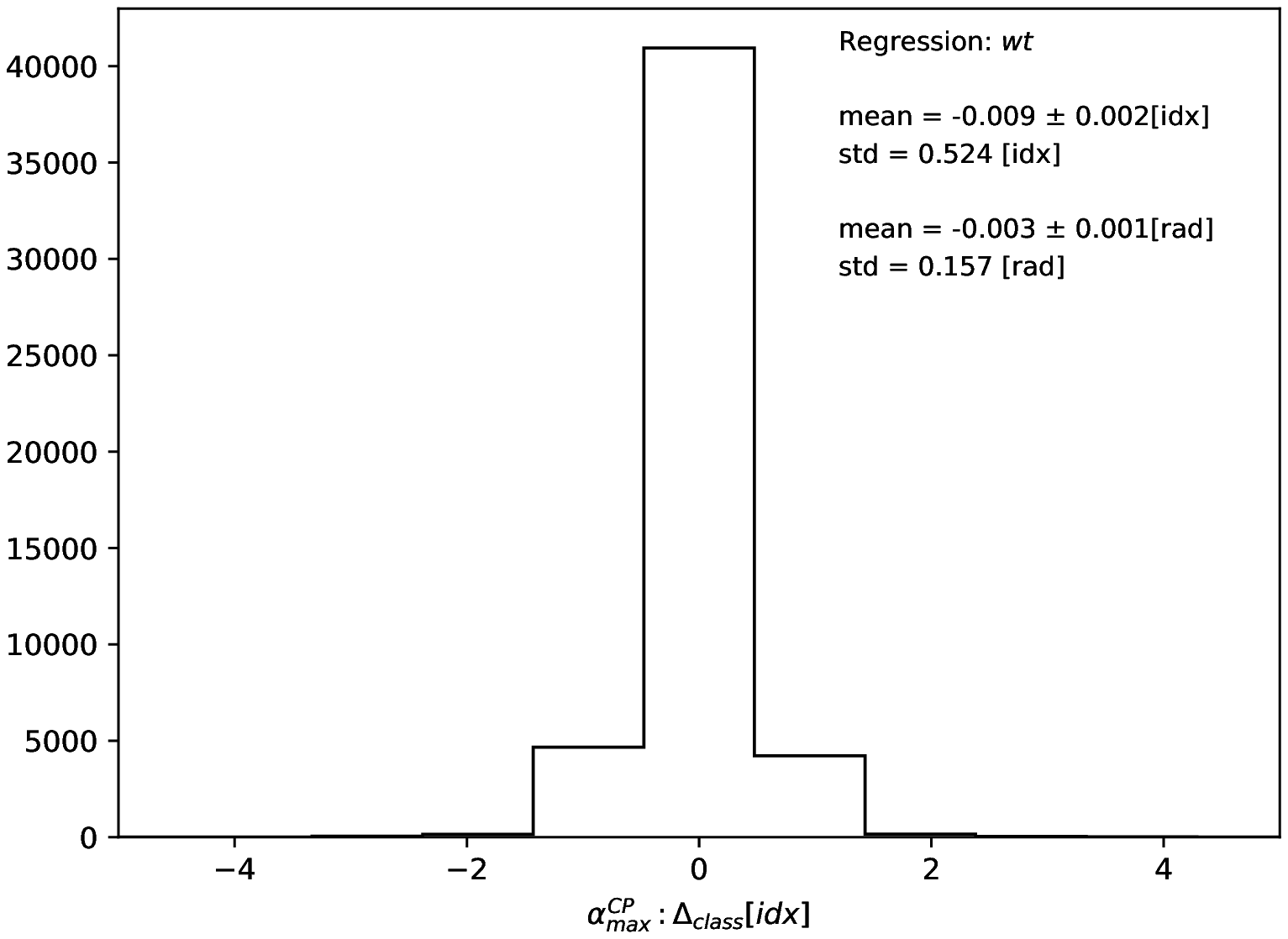}
       \includegraphics[width=7.0cm,angle=0]{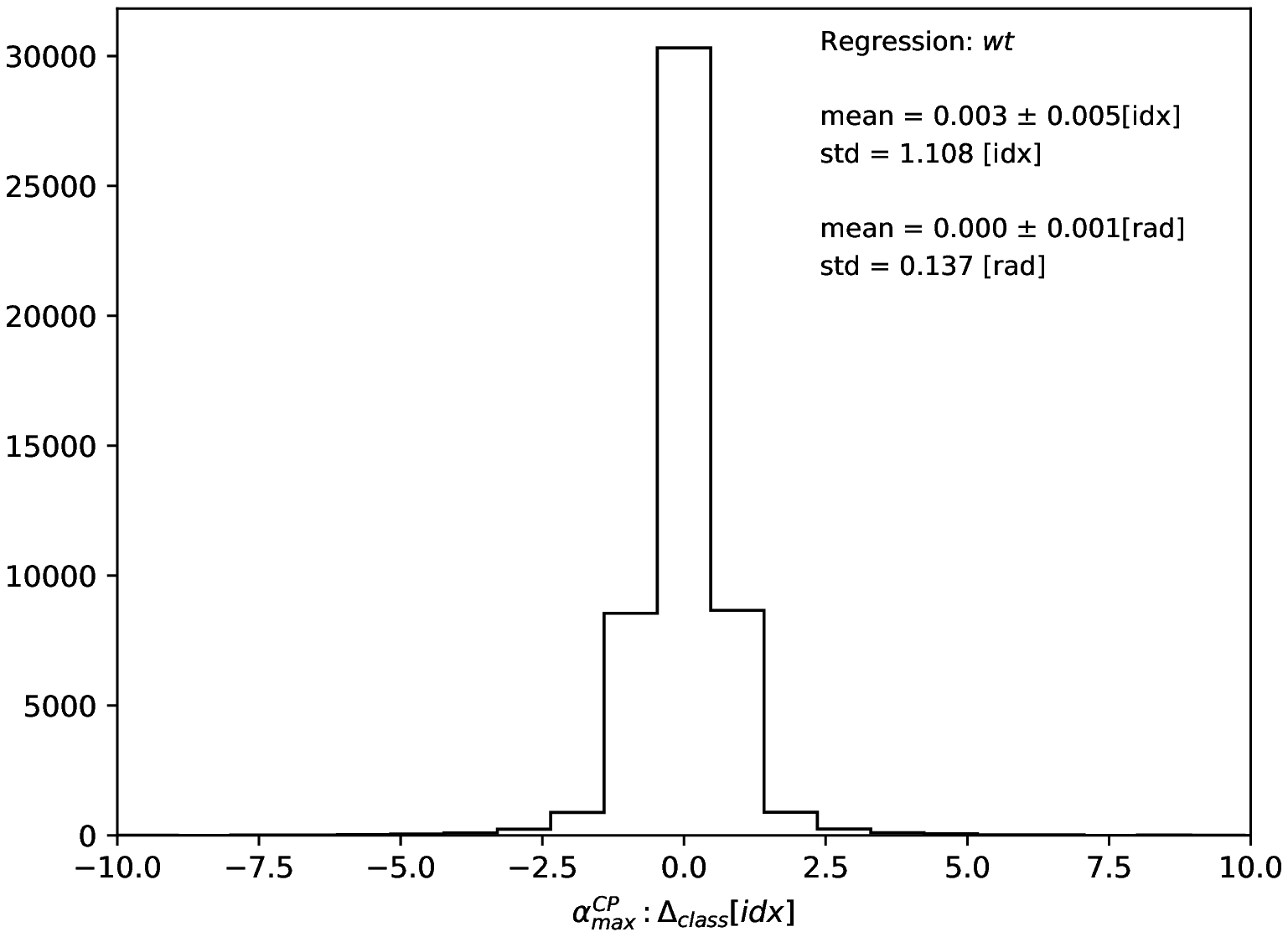}
     }
\end{center}
   \caption{Distribution of $\Delta_{class}$ between most probable predicted class and true most probable
     class. The $N_{class}$ = 21 and 51 are used for respectively left and right plot.
     The {\tt mean} and {\tt std} standard deviation are calculated in units of class index [idx] and
     units of radians [rad].
   \label{fig:DNN_regr_deltaclass_nc}}
 \end{figure}

 In Fig.~\ref{fig:DNN_regr_deltaclass_nc}  distributions of $\Delta_{class}$ for $N_{class}$ = 21 and 51 
 used to train {\it DNN} regression are respectively shown. The shape is Gaussian-like and as expected
 centered around  $\Delta_{class}$=0.

\subsection{Learning $C_0, C_1, C_2$ coefficients }
\label{Sec:regr_C012}

\begin{figure}
   \begin{center}
     {
       \includegraphics[width=5.0cm,angle=0]{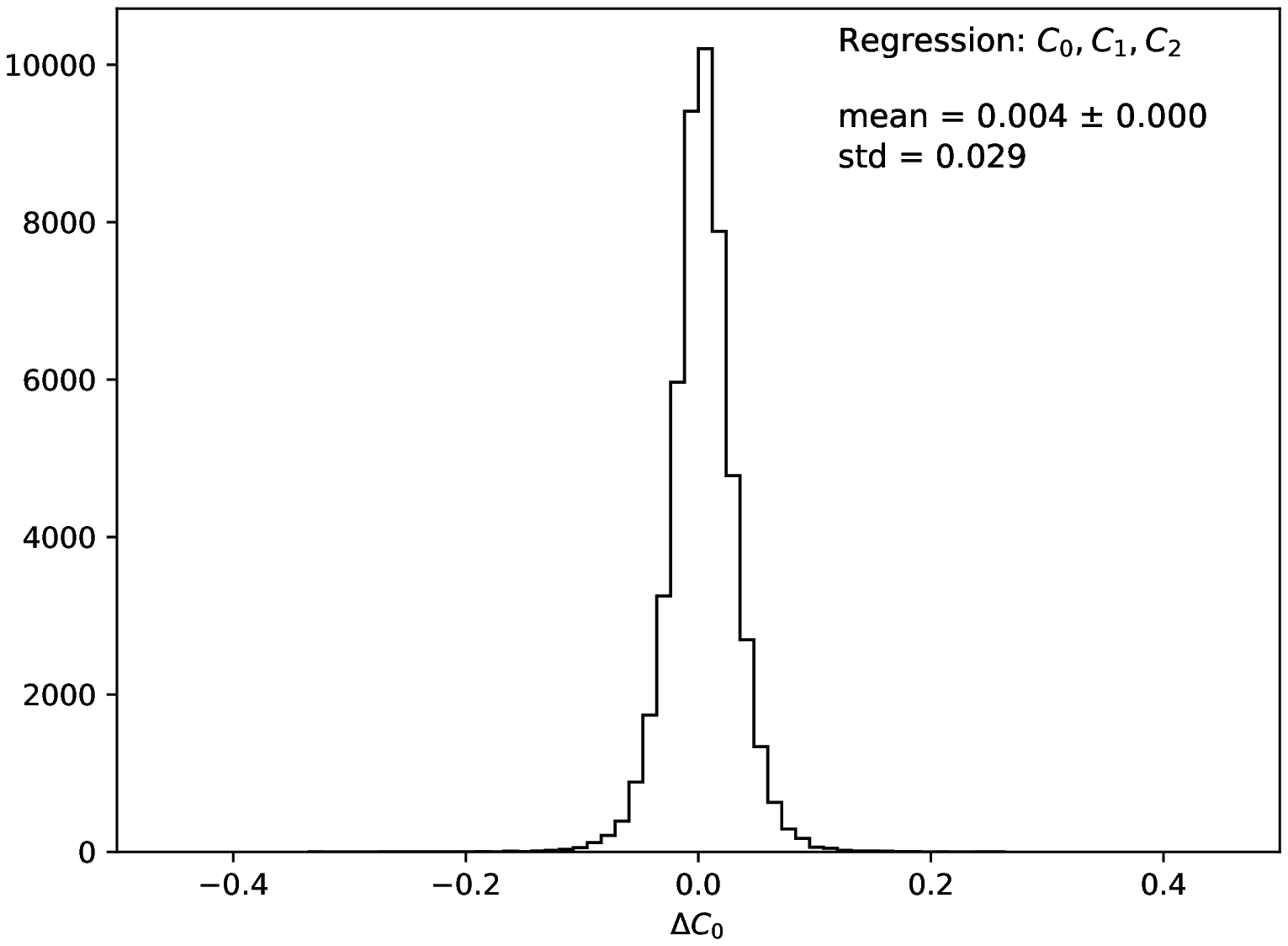}
       \includegraphics[width=5.0cm,angle=0]{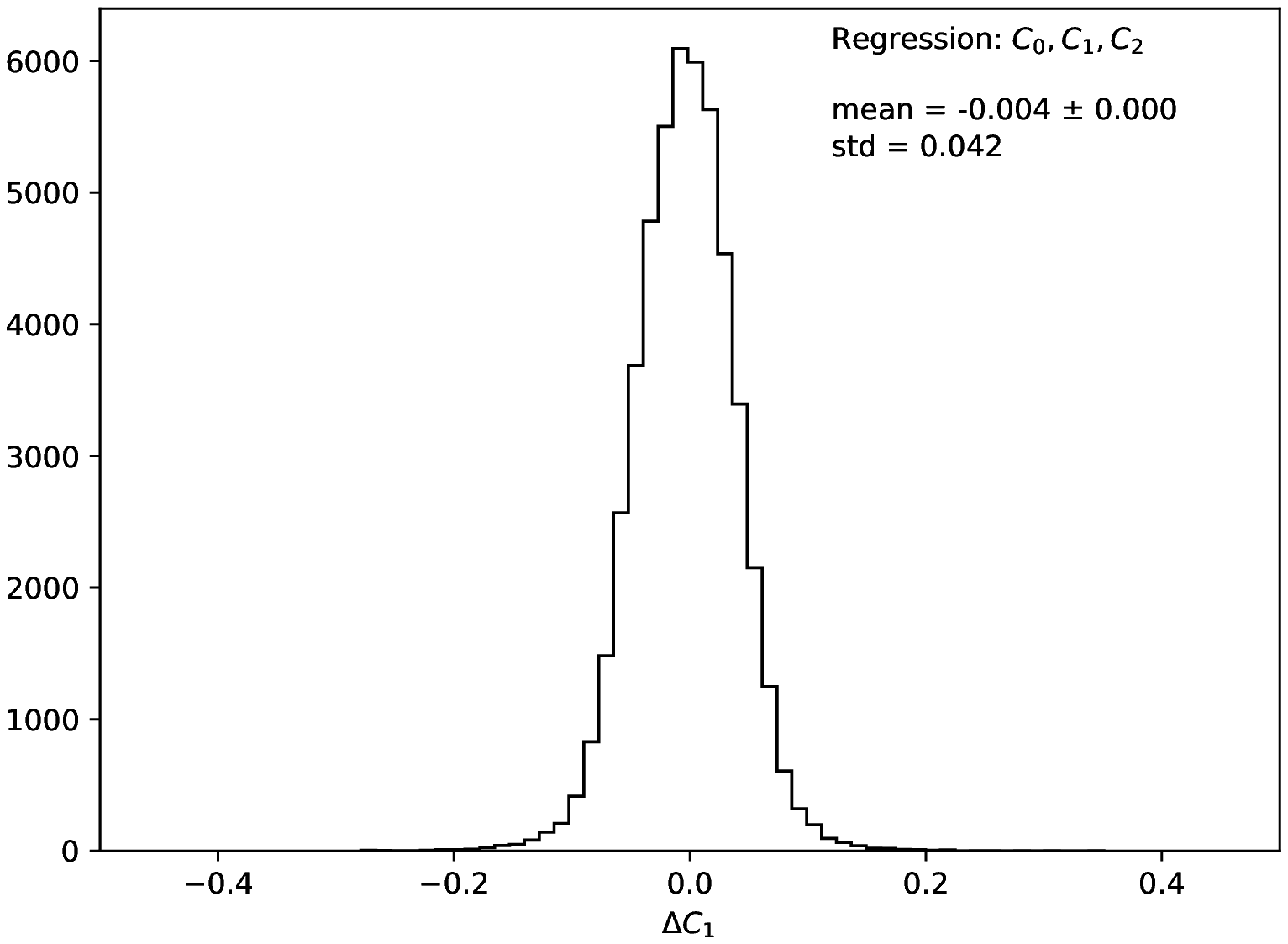}
       \includegraphics[width=5.0cm,angle=0]{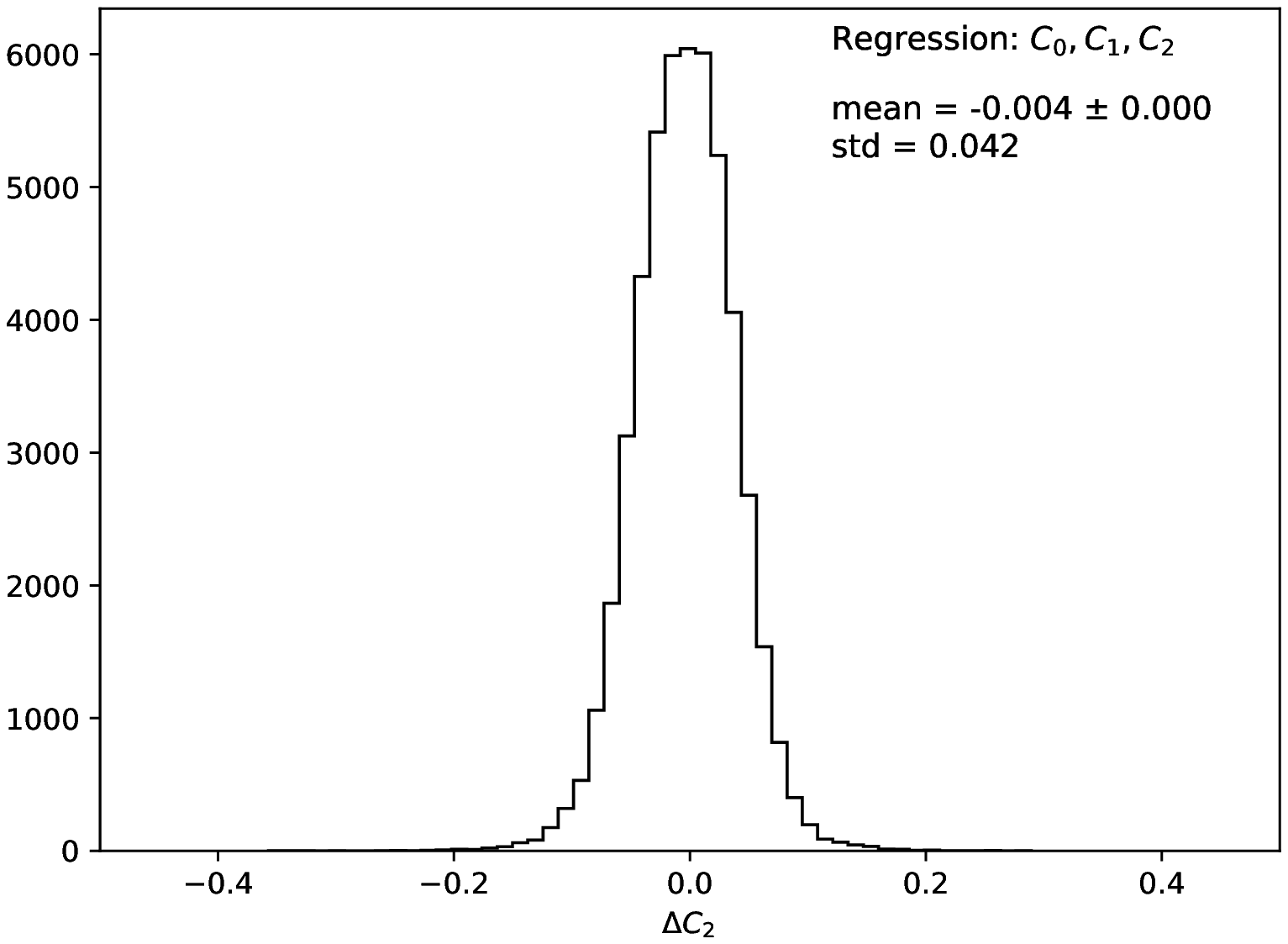}
     }
\end{center}
   \caption{Difference between true and predicted coefficients $C_0, C_1, C_2$ of formula (\ref{eq:ABC_alpha}).
   \label{fig:DNN_regr_ABC}}
 \end{figure}

Regression approach allows us to predict $C_0, C_1, C_2$ coefficients directly, without any need of discretization.
The differences between true and predicted ones are shown in Figs.~\ref{fig:DNN_regr_ABC}. On average, all three
coefficients are predicted reasonably well. Consistent are the statistical summaries of $\Delta C_i$: means remain
in the range $\pm 0.004$ and standard deviations in range (0.029-0.042).
Coefficients $C_i$  are then  used to calculate predicted spin weight $wt$ of formula~(\ref{eq:ABC_alpha}).

We have investigated also, how well predicted $C_0, C_1, C_2$ can be used to estimate the most preferred mixing angle,
$\alpha^{CP}_{max}$. For consistency, we evaluate it using the same criteria as for classification approaches.
This is achieved by using coefficients  $C_0, C_1, C_2$ to calculate
spin weight $wt$, and then turning it into discrete predictions for $wt$ and $wt^{norm}$ in the $N_{class}$ points.
As in Section~\ref{Sec:ML-multiclass} for classification approach, we use  $\Delta_{class}$,
defined by formulas~(\ref{eq:defDelt_c0})~-(\ref{eq:defDelt_c2}).

\begin{figure}
   \begin{center}
     {
       \includegraphics[width=7.0cm,angle=0]{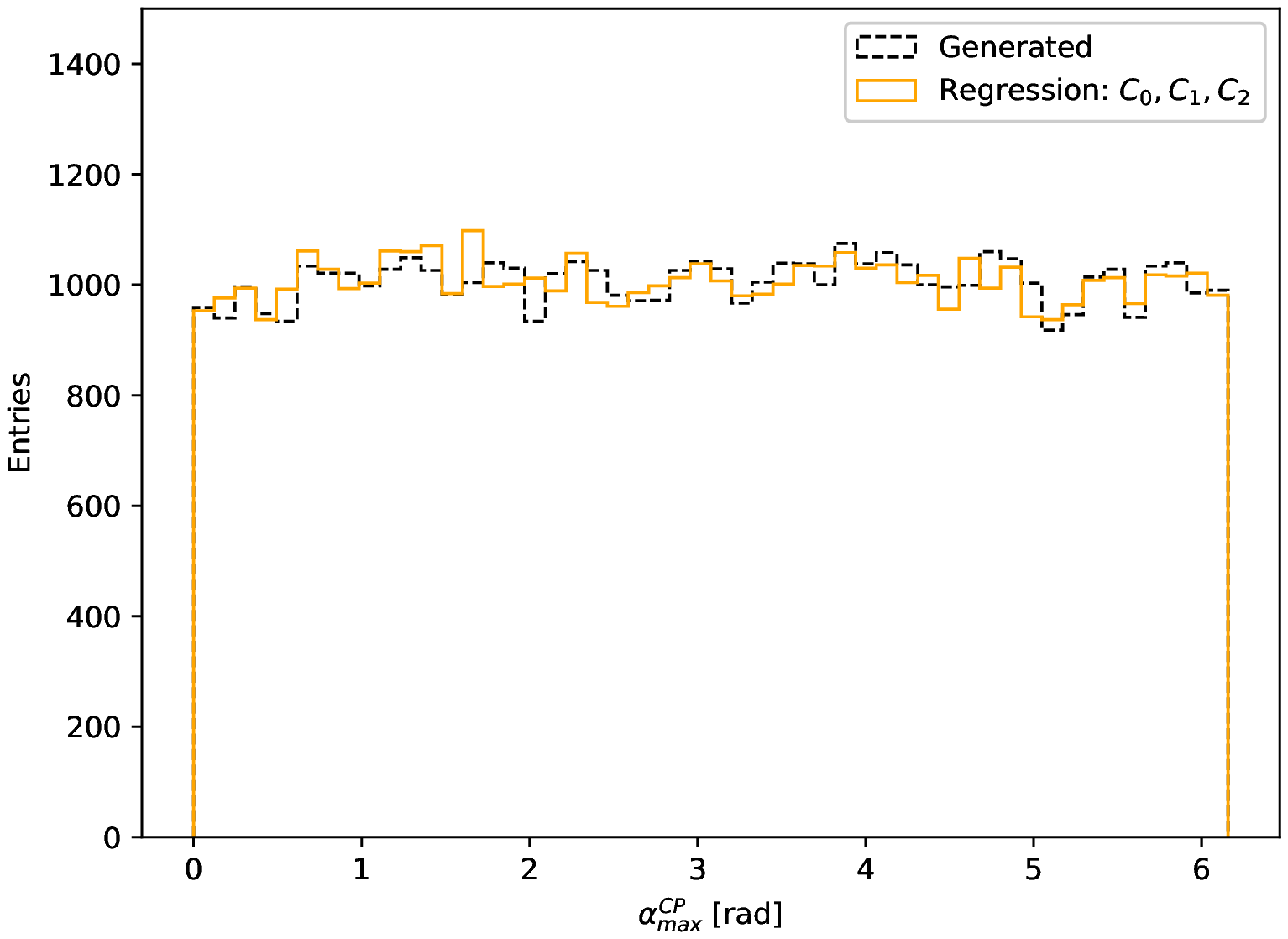}
       \includegraphics[width=7.0cm,angle=0]{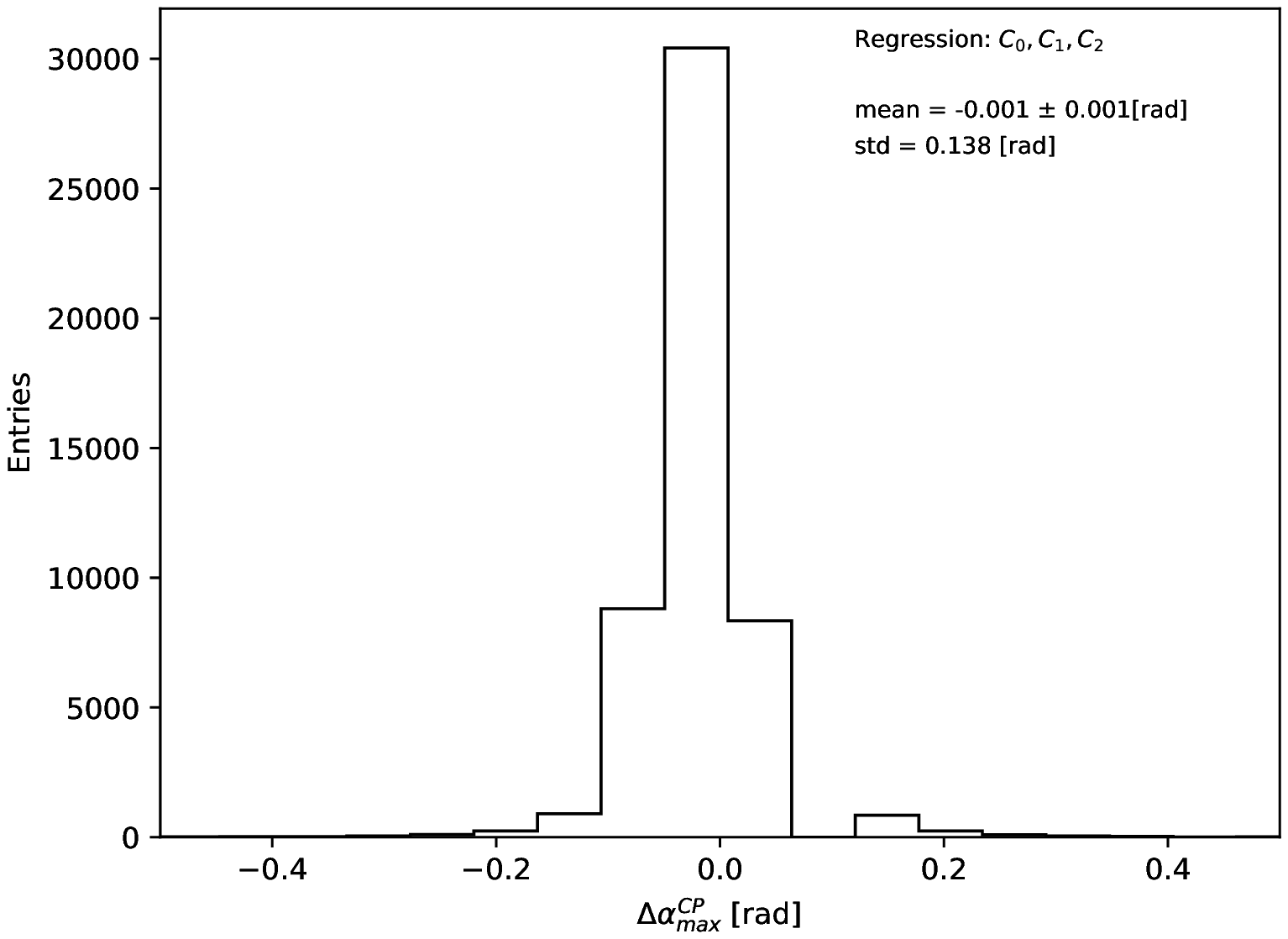}
     }
\end{center}
   \caption{Distributions (left plot) of true (black dashed line) and predicted (orange line) most preferred mixing angle $\alpha^{CP}$.
     The prediction was based on coefficients $C_0, C_1, C_2$.
     The distribution of per-event  difference of the two is shown on the right plot.
   \label{fig:DNN_Argmax_regr}}
 \end{figure}

The distributions of the true and predicted most probable class, $\alpha^{CP}_{max}$ and their difference
are shown in Figs.~\ref{fig:DNN_Argmax_regr} for the $N_{class}$ = 51. We expect the distributions to be flat as
sample was generated without any polarization correlation (carrier of CP effects) included, and this sanity
check seems to be positive.  
The difference between true and predicted $\alpha^{CP}_{max}$ forms a narrow peak with the
mean value $<\Delta \alpha^{CP}_{max}>~=~-0.001 \pm 0.001$ [rad] and standard deviation 0.138 [rad].

Finally, as a sanity check, we have compared the true overall distribution of $C_0, C_1, C_2$ with the predicted one.
As we can see in Fig.~\ref{fig:DNN_soft_CO12}, both distributions match very well.

\begin{figure}
   \begin{center}
     {
       \includegraphics[width=5.0cm,angle=0]{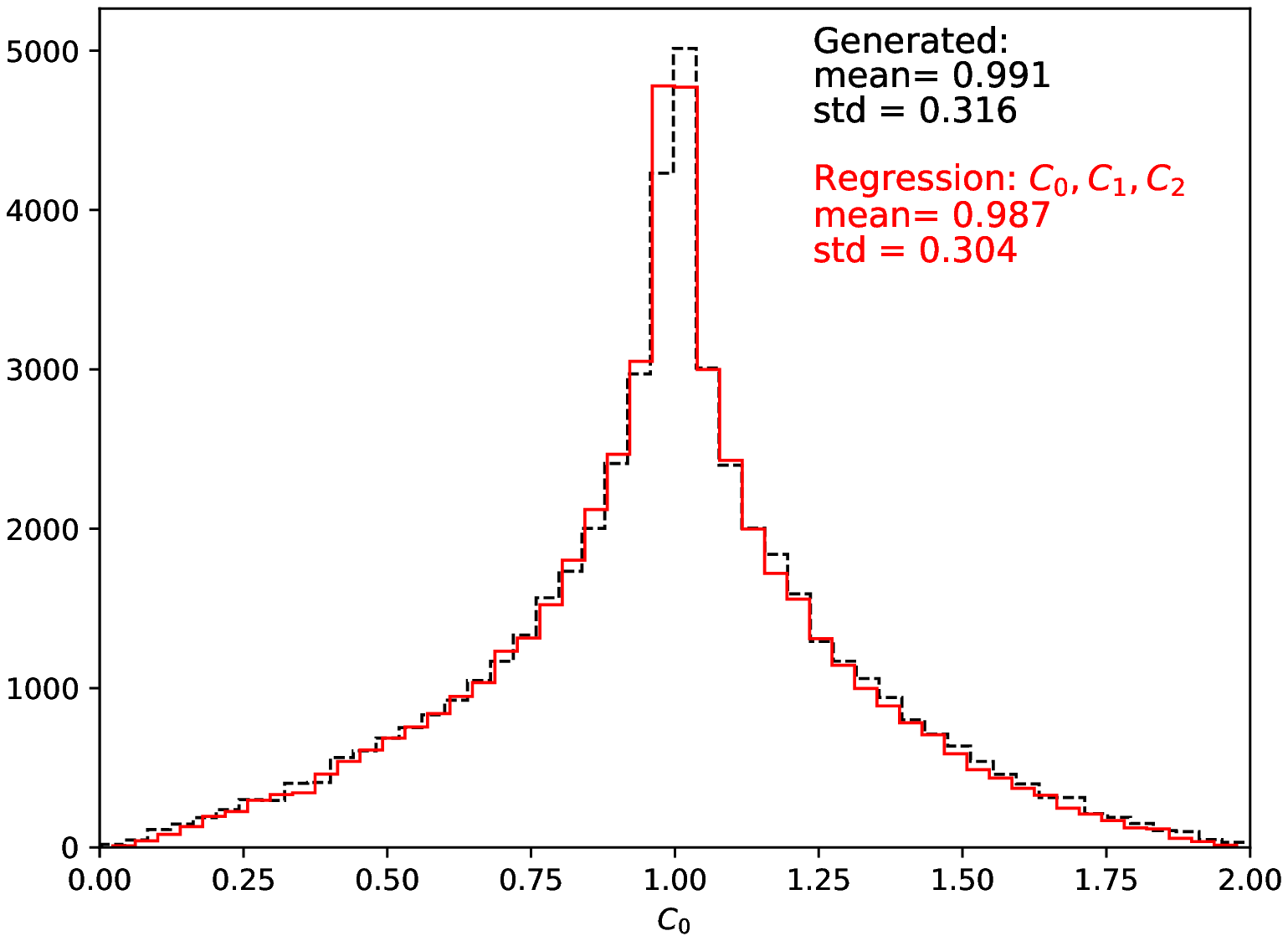}
       \includegraphics[width=5.0cm,angle=0]{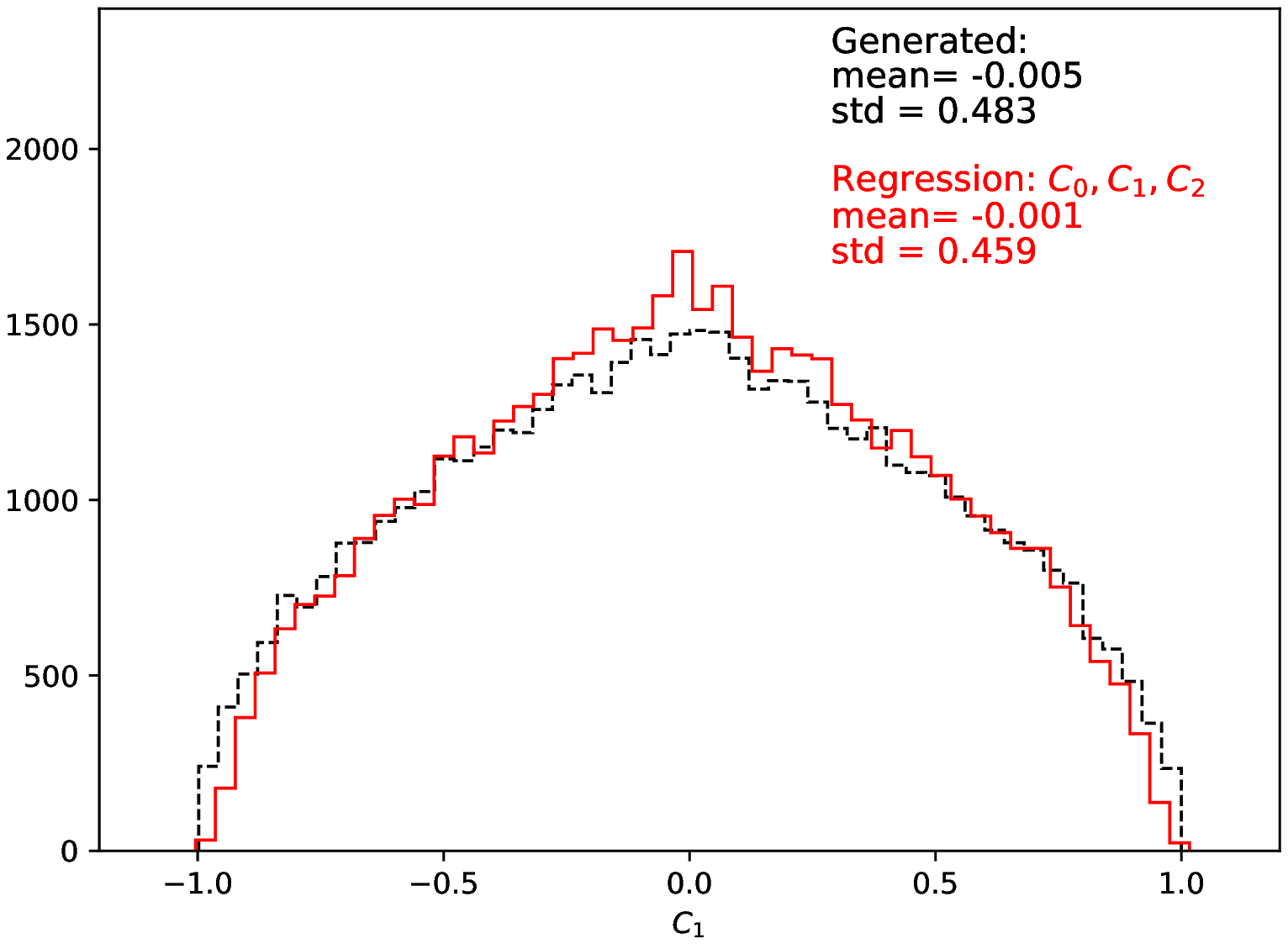}
       \includegraphics[width=5.0cm,angle=0]{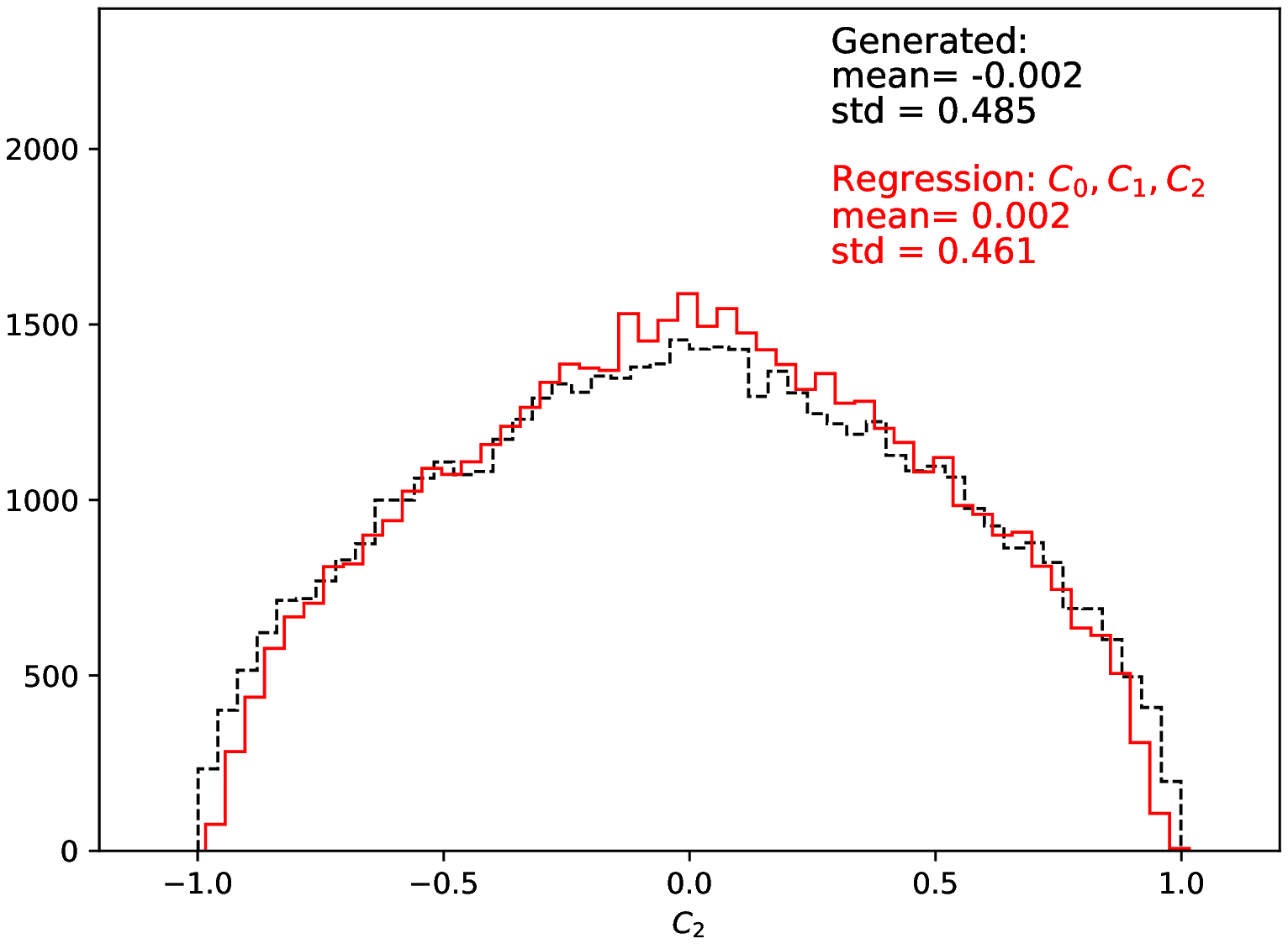}
     }
\end{center}
   \caption{Distributions of true and predicted coefficients $C_0, C_1, C_2$ of formula (\ref{eq:ABC_alpha}).
   \label{fig:DNN_regr_CO12}}
 \end{figure}

 \subsection{Learning the $\alpha^{CP}_{max}$}
\label{Sec:regr_alphaCPmax}

As was in the previous subsection, the implementation of the regression method allows a direct, non-discrete estimation of continuous parameters.
This is also desired with the most preferred mixing angle $\alpha^{CP}_{max}$.

\begin{figure}
   \begin{center}
     {
       \includegraphics[width=7.0cm,angle=0]{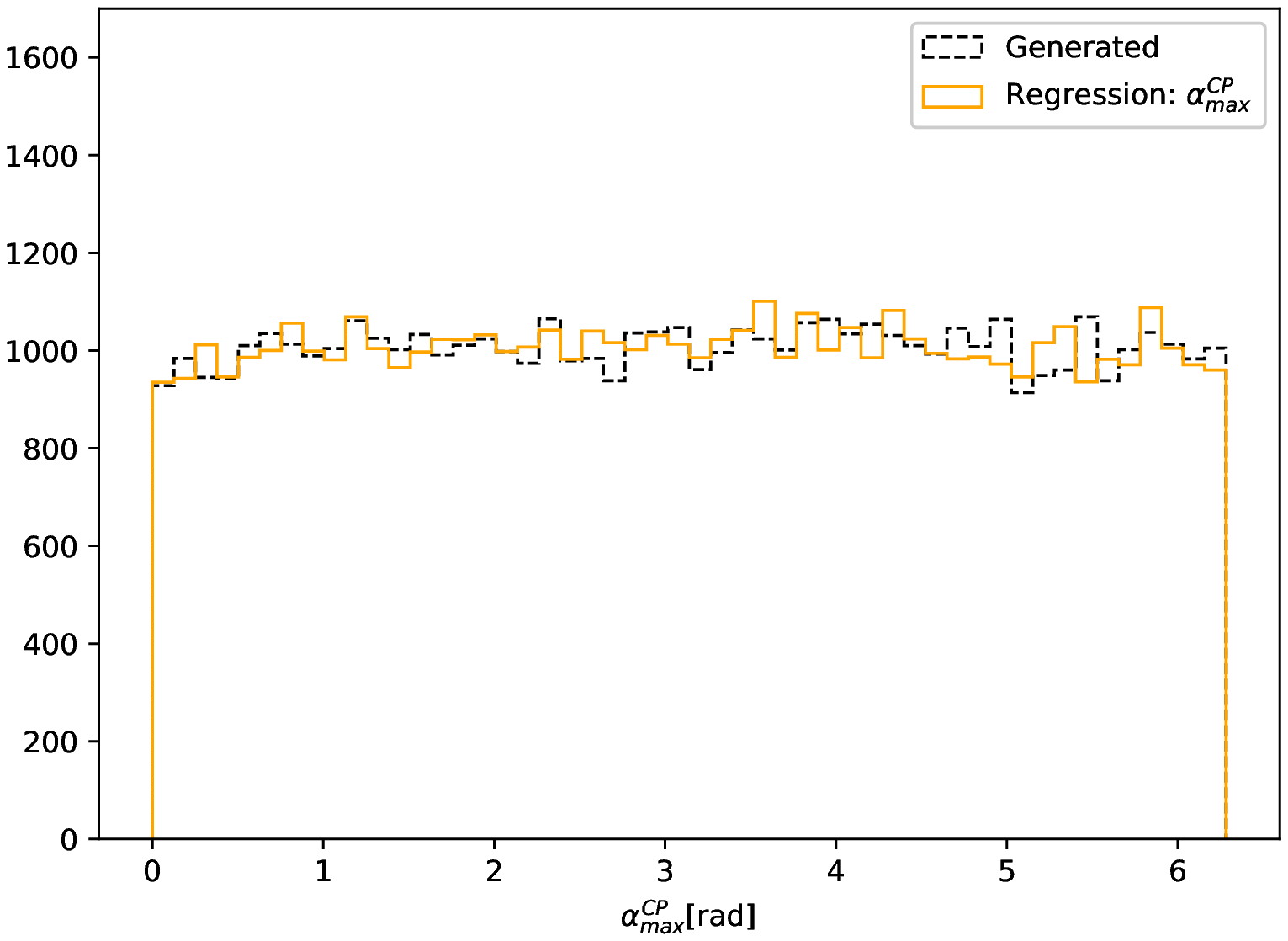}
       \includegraphics[width=7.0cm,angle=0]{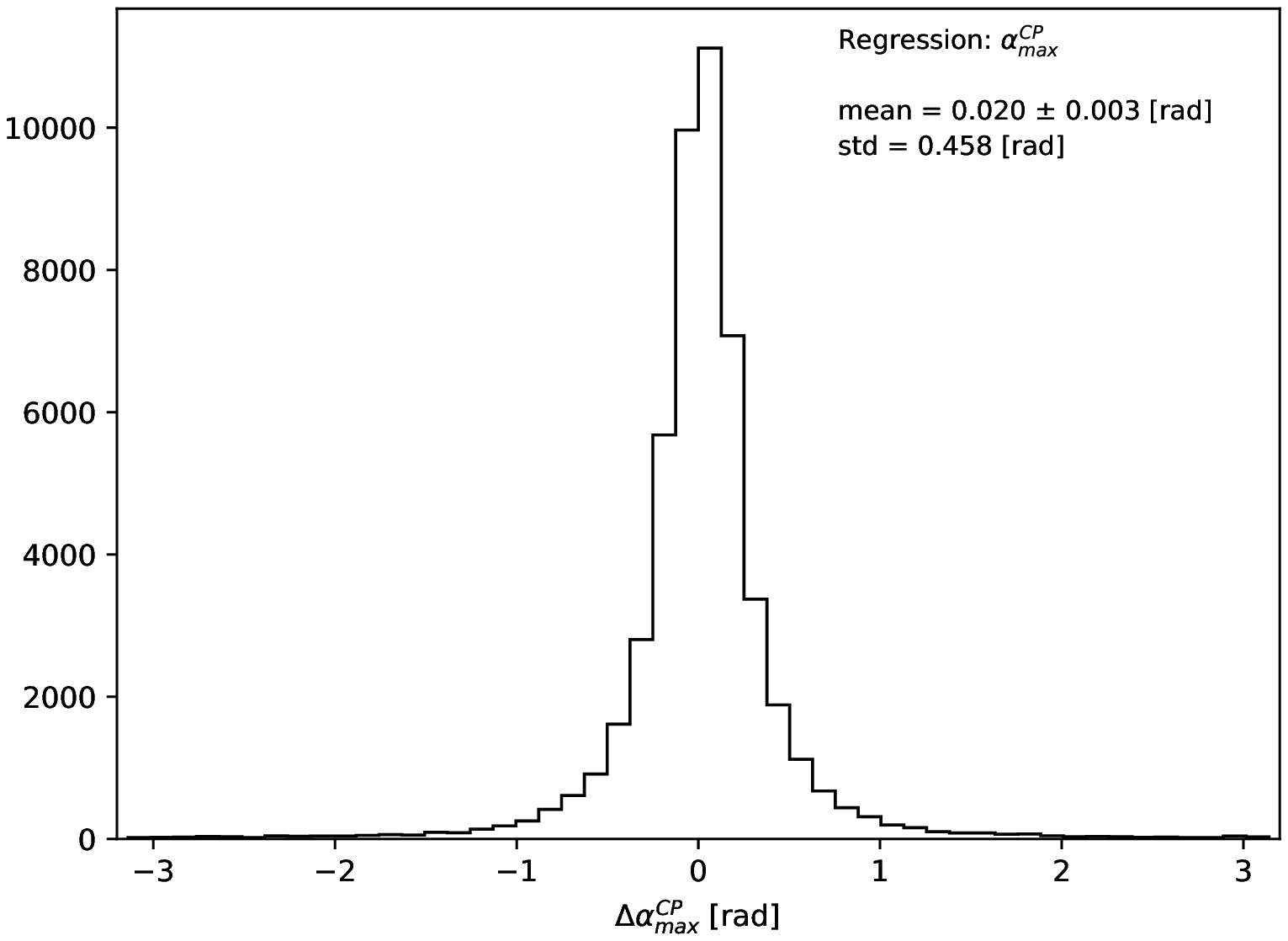}
     }
\end{center}
   \caption{Distributions (left plot) of true (black dashed line) and predicted (orange line) most preferred mixing angle $\alpha^{CP}$.
     The distribution of per-event  difference of the two is shown on the right plot.
   \label{fig:DNN_regr_argmax}}
 \end{figure}

The distributions of the true and predicted most probable class, $\alpha^{CP}_{max}$ and their difference
are shown in Figs.~\ref{fig:DNN_regr_argmax} for the $N_{class}$ = 51. We expect the distributions to be flat as sample was generated
without any polarization correlation (carrier of CP effects) included, and this sanity check seems to be positive.  
As the used event sample is generated without any polarization, the distribution of the $\alpha^{CP}_{max}$ is expected to be uniform,
see the left plot of Fig.~\ref{fig:DNN_regr_argmax}.
The {\it DNN} is reproducing this feature well.
The difference between true and predicted $\alpha^{CP}_{max}$ forms a narrow peak with the
mean $<\Delta \alpha^{CP}_{max}> = 0.020 \pm 0.003$ [rad] and standard deviation 0.458 [rad].

\section{Classification or regression: comparison and complementarity}
\label{Sec:MLcomparison}

In this Section we shortly compare classification and regression approaches.
In Table~\ref{Tab:c012sCoeff} we collect the mean and standard deviation
for difference between true and predicted with classification
and regression methods $C_i$.
There is no clear winner, both methods give predictions of similar precision,
with only $C_0$ being better predicted with regression.

In Table~\ref{Tab:DeltalphaCPmax} we compare the difference between
true and predicted $\alpha^{CP}_{max}$  obtained with different methods.
With the classification method comparable performance is achieved when learning
spin weight $wt$, coefficients $C_0, C_1, C_2$ or directly  $\alpha^{CP}_{max}$.
For the regression method  learning directly $\alpha^{CP}_{max}$ is significantly less performant.
Otherwise, is no clear winner between different methods.

\begin{table}
 \vspace{2mm}
  \caption{
    The mean and standard deviations of $\Delta C_i$, the difference between generated and predicted $C_i$, obtained
     from {\it DNN} with classification and regression methods for $N_{class} = 51$.
    \label{Tab:c012sCoeff}}
      \begin{center}
    \begin{tabular}{|l|c|c|}
    \hline
    Coefficients  & Classification & Regression \\
    \hline \hline 
    $\Delta C_0$   & mean = 0.000   & mean = 0.004\\
                   & std  = 0.038  & std = 0.029 \\
    \hline
    $\Delta C_1$   & mean = 0.001  & mean = -0.004\\
                   & std  = 0.051  & std = 0.042 \\
    \hline
    $\Delta C_2$   & mean = -0.003  & mean = -0.04\\
                   & std  = 0.051  & std = 0.042 \\
    \hline
    \end{tabular}
  \end{center}
%\end{table}	
%\begin{table}
 \vspace{2mm}
  \caption{
    The mean and standard deviation of $\Delta \alpha^{CP}_{max}$, the difference between true and predicted  $\alpha^{CP}_{max}$,
    obtained from {\it DNN} with classification and regression methods.
    \label{Tab:DeltalphaCPmax}}
      \begin{center}
    \begin{tabular}{|l|c|c|}
    \hline
    Method  & Classification & Regression \\
    \hline \hline 
    Using $wt$                  & mean = -0.006 $\pm$ 0.001 [rad]  & mean = 0.000 $\pm$ 0.001 [rad]\\
                                & std  = 0.126 [rad]              & std = 0.137 [rad] \\
    \hline
    Using $C_0, C_1, C_2$        & mean = 0.000 $\pm$ 0.001 [rad]  & mean = -0.001 $\pm$ 0.001 [rad]\\
                                & std  = 0.153 [rad]              & std =  0.138 [rad] \\
    \hline
    Direct                      & mean =- 0.003 [rad]              & mean = 0.020 [rad]\\
                                & std  = 0.139 [rad]              & std = 0.458 [rad] \\
    \hline
    \end{tabular}
  \end{center}
\end{table}

\section{Summary}
\label{Sec:Summary}

We have performed a proof-of-concept for the {\it DNN} methods in the measurement of Higgs boson  $H \to \tau \tau$ 
CP mixing angle dependent coupling.
That  extends work of refs.~\cite{Jozefowicz:2016kvz, Lasocha:2018jcb} of classification
between  scalar and pseudoscalar Higgs CP state.
Several solutions  of classification and of regression types were prepared and numerical results
were collected.
For the measurement we have studied approaches where; (i) spin weights, (ii)
coefficients for the functional form of the spin
weight (iii) directly  the mixing angle at which the weight has its maximum, were targeted.
In cases (i) and (ii)
the classification approach seemed comparable to the regression,
but the comparisons relied on the discretised and normalized quantities due to classification limitations.
The regression approach seems more natural for continuous observables and does not have such limitations.
On the other hand, regression approach has performed much worse in the case of direct $\alpha^{CP}_{max}$ prediction.

For the  feature list we have chosen  idealistic case, assuming that complete set of  $\tau$ decay
products 4-momenta is known, including challenging to reconstruct neutrinos.
We have exploited  then the $\tau \to \rho \nu$ decay mode.
The results are encouraging, the understanding of environment for future discussion of measurement ambiguities
was not compromised with respect to what was achieved in previous publications for scalar/pseudoscalar classifications.

The mean value of the preferred mixing angle $\alpha^{CP}_{max}$ can be constrained by the
trained {\it DNN} with per-event resolution better than 0.15~[rad] using a classification approach.
Both classification and regression approaches allow to learn spin weight with uncertainties (average $l_2$ norm) better than 15\%.
Both approaches allow also to learn coefficients $C_0, C_1, C_2$ of the functional spin weight form.
The coefficients are directly related to the polarimetric vectors of decaying $\tau^\pm$ leptons.
This provides interesting possibility for the future studies of experimental ambiguities with samples of the $Z \to \tau\tau$ decays,
much more abundant and available for the LHC measurements. Departure from SM predictions on $Z\tau\tau$ coupling
can reveal itself in the observables build from polarimetric vectors of decaying $\tau^\pm$ leptons too.

We plan, following \cite{Jozefowicz:2016kvz, Lasocha:2018jcb}, to  extend our studies to more realistic feature lists
and other decay modes. Already now, the variety of ML methods
for the determination of most preferred CP state mixing angle, demonstrated  potential and  robustness
for future experimental analyses and measurements with  the LHC data.

\vskip 1 cm
\centerline{\bf \Large Acknowledgments}
\vskip 0.5 cm
We would like to thank J. Kurek and  P. Winkowska for help with technical
implementation and testing of the analysis code used for this paper preparation. 

%\vskip 0.5 cm

\providecommand{\href}[2]{#2}
\appendix

\section{Deep Neural Network}
\label{App:DNN}

The structure of the simulated data and the {\it DNN} architecture follows what was published in our previous
papers~\cite{Jozefowicz:2016kvz, Lasocha:2018jcb}.
It is prepared for {\tt TensorFlow}~\cite{abadi2015tensorflow}, an open-source machine learning library.

We consider $H\to \tau \tau$ channel of both $\tau^{\pm} \to \rho^{\pm} \nu$ decay.
The data point is thus an event of the Higgs boson production and $\tau$ lepton pair decay products. 
The structure of the event is represented as follows:
\begin{equation}
  x_i = (f_{i,1},...,f_{i,D}),w_{a_i},w_{b_i}, ...,w_{m_i}
\end{equation}
The  $f_{i,1},...,f_{i,D}$ represent numerical features  and $w_{a_i},w_{b_i},w_{m_i} $ are weights proportional
to the likelihoods that an event comes from a class $A, B, ..., M$, each representing different $\alpha^{CP}$
mixing angle.
The $\alpha^{CP} = 0, 2\pi$ corresponds to scalar CP state and  $\alpha^{CP} = \pi$ to pseudoscalar CP state.
The weights calculated from the quantum field theory matrix elements are available and stored
in the simulated data files.
This is a convenient situation, which does not happen in many other cases of ML classification.
The $A, B, ... M$ distributions highly overlap in the $(f_{i,1},...,f_{i,D})$
space, the more detailed discussion in case of two hypotheses only, scalar and pseudoscalar,
can be found in~\cite{Jozefowicz:2016kvz}.

Thanks to similar {\it DNN} architecture, we have prepared three implementations for
measuring Higgs boson CP state: binary classification, multiclass classification and regression:
\begin{itemize}
  \item
For binary classification the aim is to discriminate between two hypothesis,
$\mathscr{H}_0$ and $\mathscr{H}_{\alpha^{CP}}$.
\item
For multiclass classification, the aim is to simultaneously learn weights (probabilities) for several
$\mathscr{H}_{\alpha^{CP}}$ hypotheses; learn coefficients of the weight functional
form or directly learn the mixing angle at which spin weight has its maximum, $\alpha^{CP}_{max}$.
A single class can be either  single discretised $\alpha^{CP}$ or a range for the $C_i$ parameters.
The system is learning probabilities for classes to associate with the event. 
\item
For regression case, the aim is similar as for multiclass classification case, but now problem is defined as
a continuous case. The system is learning value to associate with the event. The value can be a vector of
spin weights for a set of $\mathscr{H}_{\alpha^{CP}}$ hypotheses, set of $C_i$ coefficients or $\alpha^{CP}_{max}$. 
\end{itemize}

\begin{figure}
  \begin{center}
     {
       \includegraphics[width=7.5cm,angle=0]{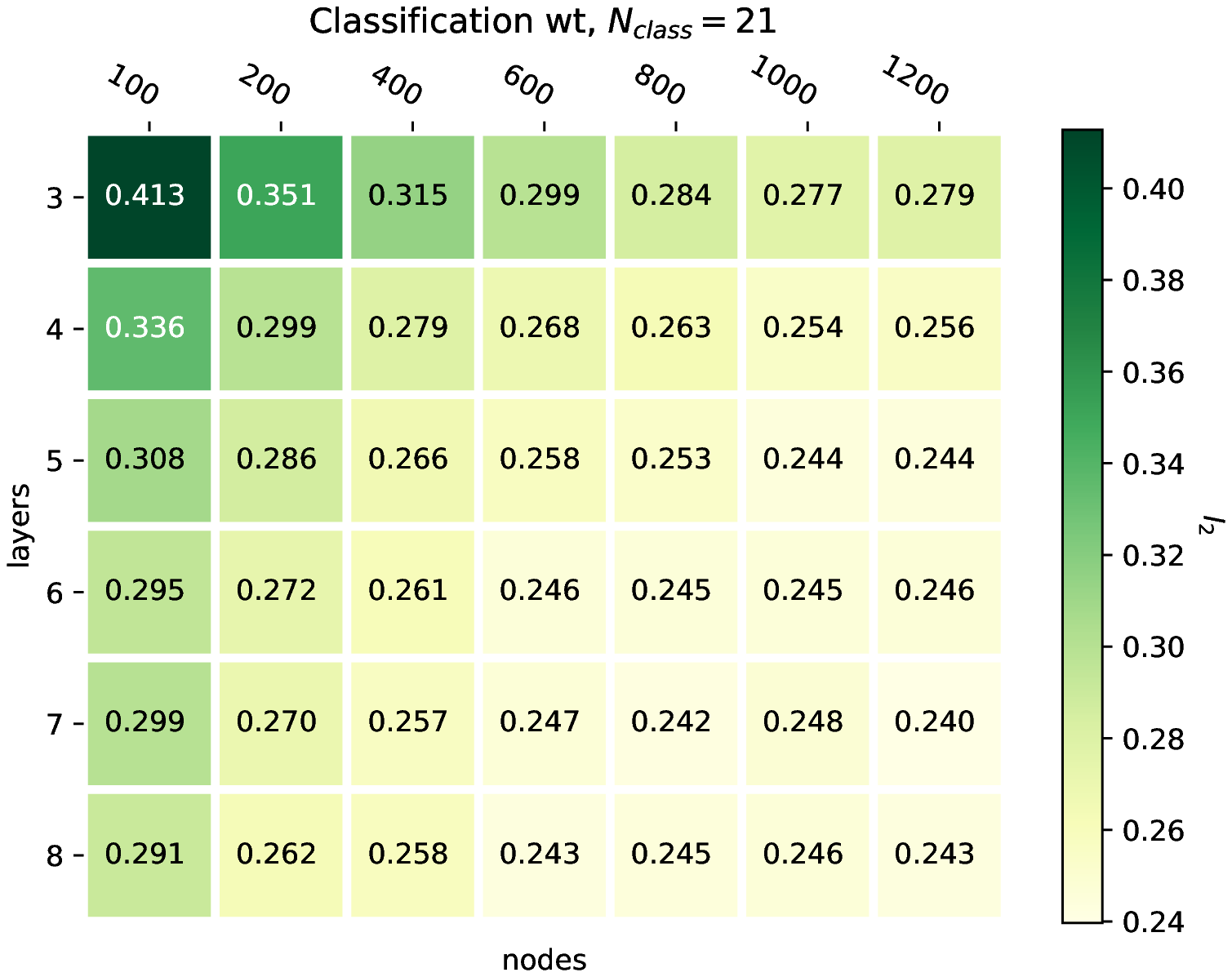}
       \includegraphics[width=7.5cm,angle=0]{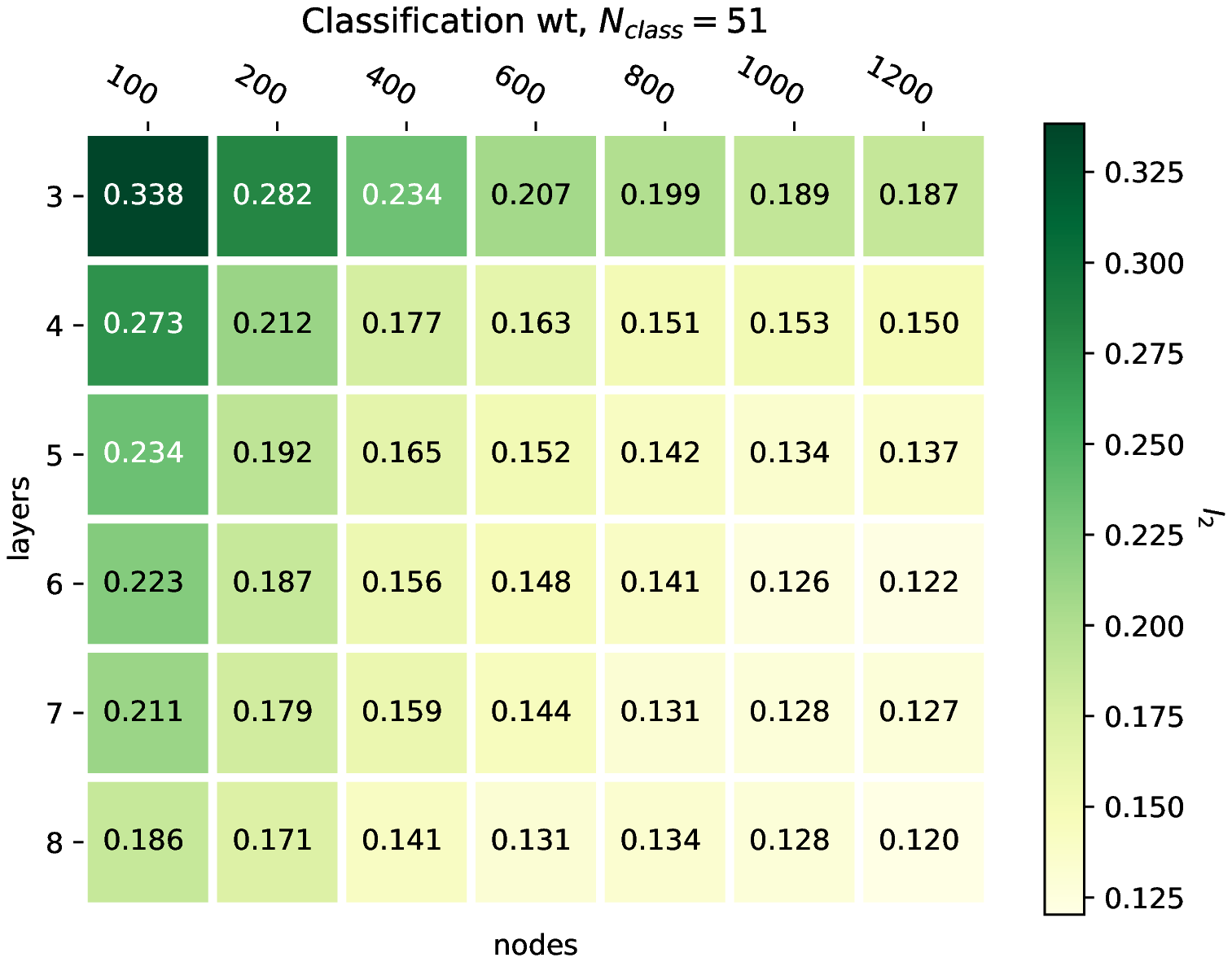}
     }
  \end{center}
  \caption{ Performance of $wt$ fitting ($l_{2}$) for different number of layers and nodes, assuming $N_{class}$= 21 (left-side) and $N_{class}$= 51 (right-side).
   \label{figApp:arch_comp}}
 \end{figure}

The network architecture consists of 6 hidden layers, 1000 nodes each  with
ReLU activation functions and is initialized with
random weights. Such architecture has been found as a good tradeoff between the performance and computation time,
what can be seen in Fig.~\ref{figApp:arch_comp}. Learning procedure is optimized using a variant of stochastic gradient descent algorithm
called Adam~\cite{kingma2014adam} and Batch Normalization~\cite{ioffe2015batch}.

The last layer is specific to the implementation case, different is dimension of the output vector,
activation function and a loss function. In the following, we will describe details.

{\bf Classification:}
The loss function used in stochastic gradient descent is a cross entropy of valid values and
neural network predictions~\cite{dlbook}. It is a common choice in case of binary
or multiclass classification models. The loss function for sample of $N_{evt}$ events and
classification for $N_{class}$ reads as follows:
\begin{equation}
  Loss = \sum_{k=1}^{N_{evt}} \sum_{i=1}^{N_{class}} y_{i,k} log(p_{i,k}),
\end{equation}
where $k$ stands for consecutive event and $i$ for class index. The $y_{i,k}$ represents neural-network
predicted probability for event $k$ being of class $i$ while $p_{i,k}$ represents true
probability used in supervised training.

{\bf Regression:}
In case of predicting $wt$ the last layer of {\it DNN} is $N$ dimensional output
(granularity with which we want to discretise it). For predicting $C_0, C_1, C_2$ the
last layer of {\it DNN} is N=3 dimensional output, i.e. values of  $C_0, C_1, C_2$.
Activation of this layer is a linear function.
Loss functions is defined as Mean Squared Error (MSE) between true and predicted parameters
\begin{equation}
  Loss = \sum_{k=1}^{N_{evt}} \sum_{i=1}^{i=N} (y_{i,k} - p_{i,k})^2,
\end{equation}
where $k$ stands for event index and $i$ for index of function form parameter. The $y_{i,k}$ represents 
predicted value of $C_i-th$ parameter for event $k$ while $p_{i,k}$ represents true value.
For predicting the $\alpha^{CP}_{max}$  the last layer of {\it DNN} is N=1 dimensional output,
i.e. values of   $\alpha^{CP}_{max}$.

The {\tt tf.reduce\_mean} method of {\tt TensorFlow} is used,
with the loss function 
\begin{equation}
  Loss = \sum_{k=1}^{N_{evt}} (1 - cos(y_{k} - p_{k})),
\end{equation}
where $y_{k}, p_{k}$ denotes respectively predicted and true value of $\alpha^{CP}_{max}$.

\begin{figure}
  \begin{center}
     {
       \includegraphics[width=7.5cm,angle=0]{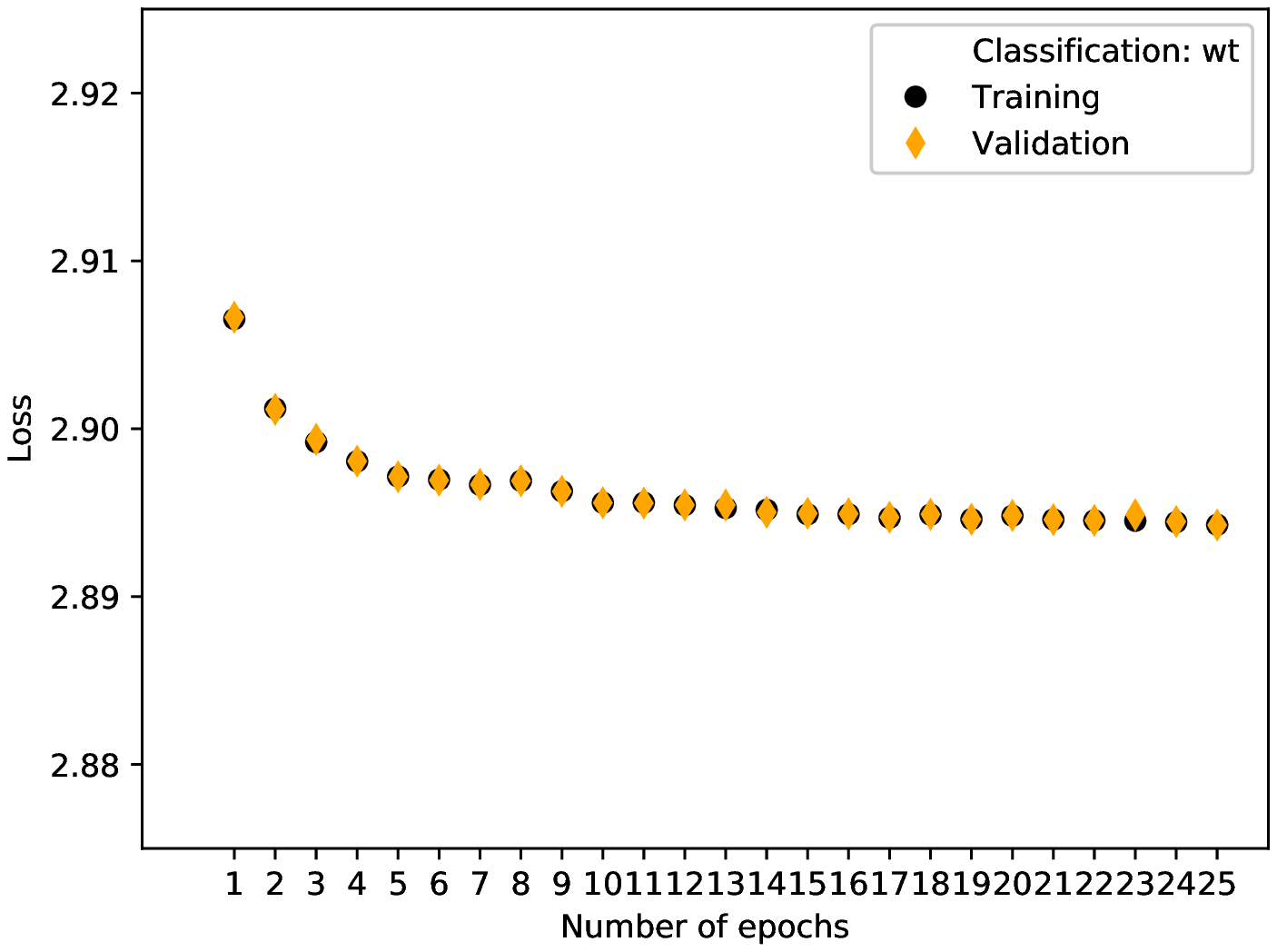}
       \includegraphics[width=7.5cm,angle=0]{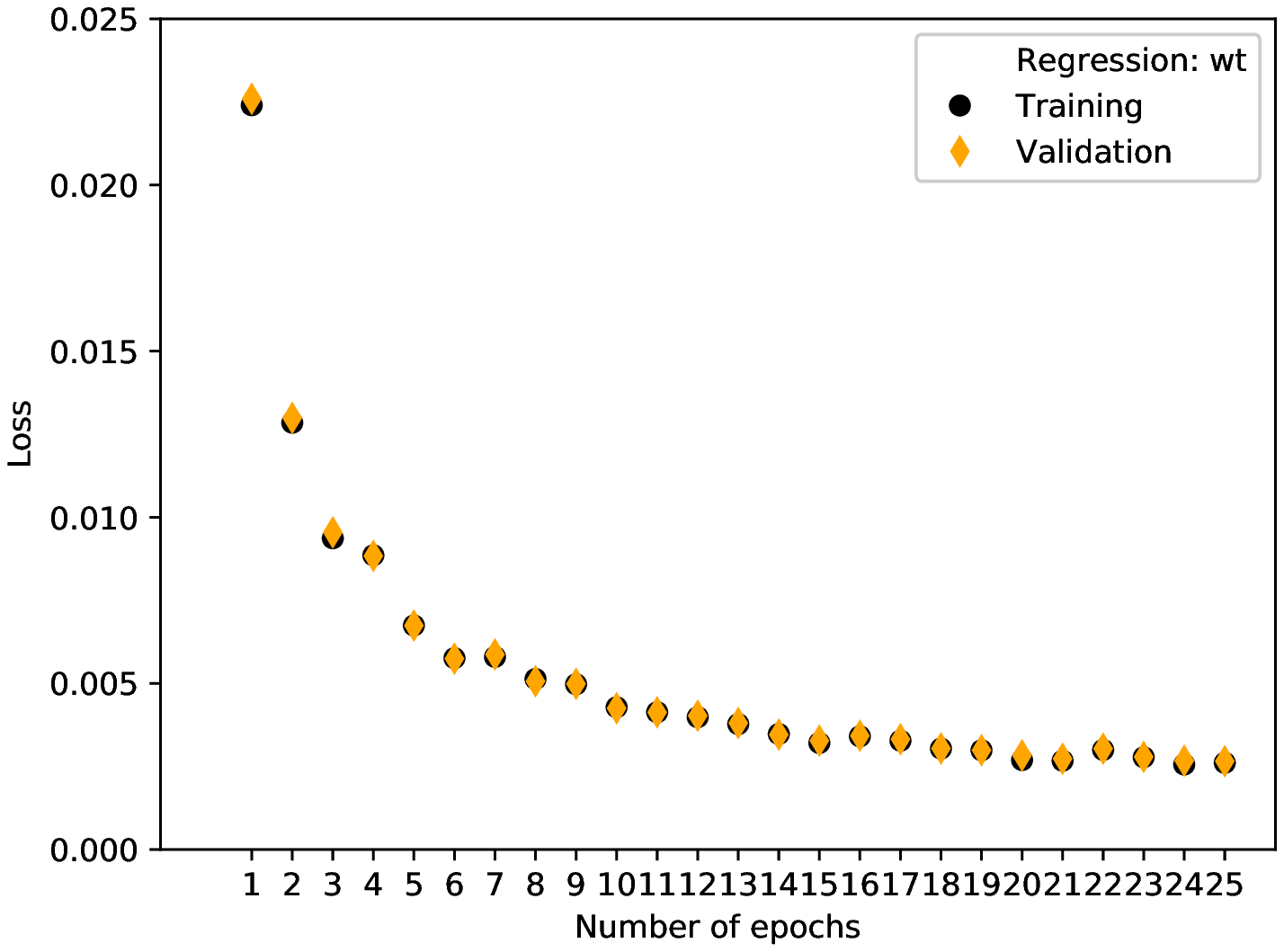}
       \includegraphics[width=7.5cm,angle=0]{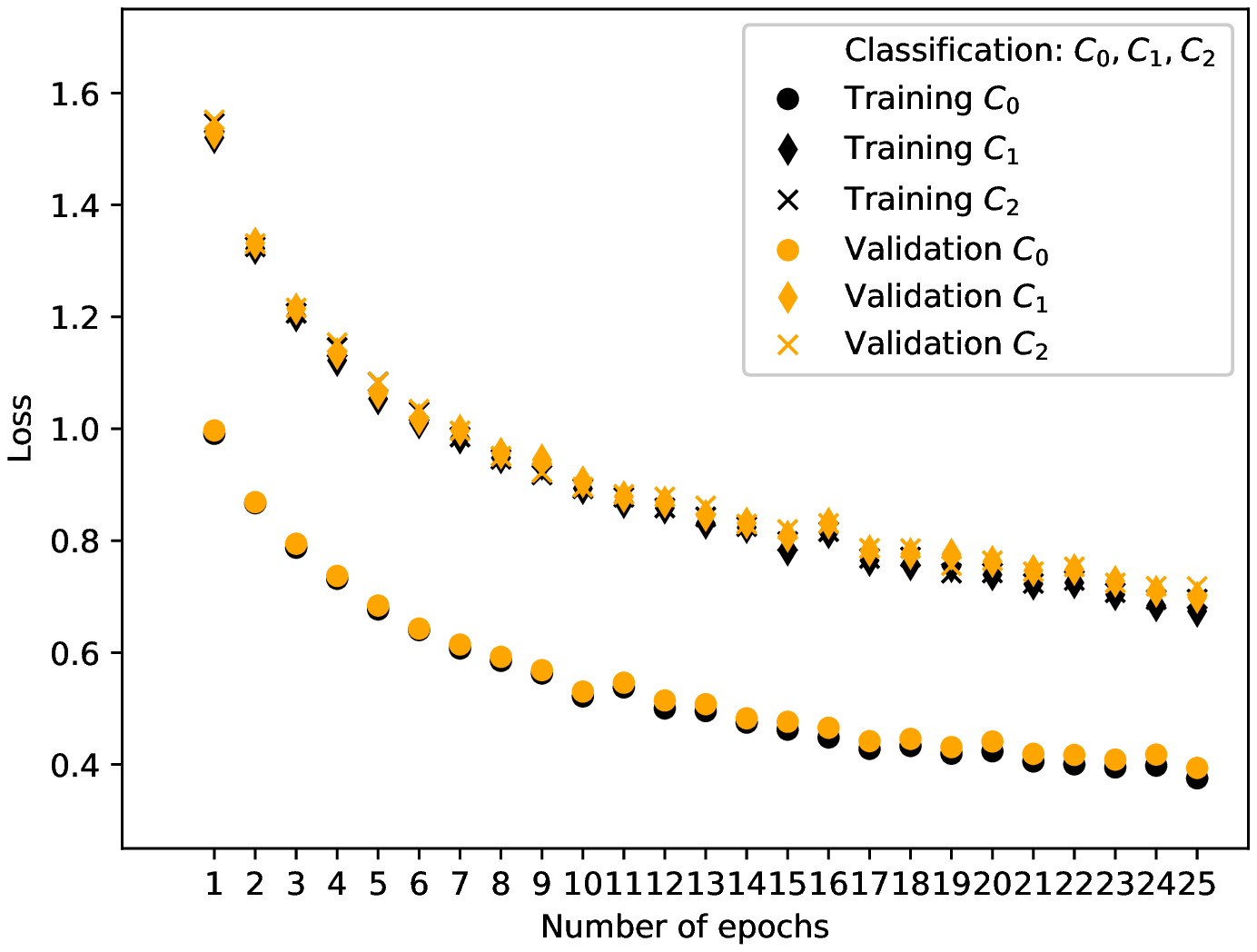}
       \includegraphics[width=7.5cm,angle=0]{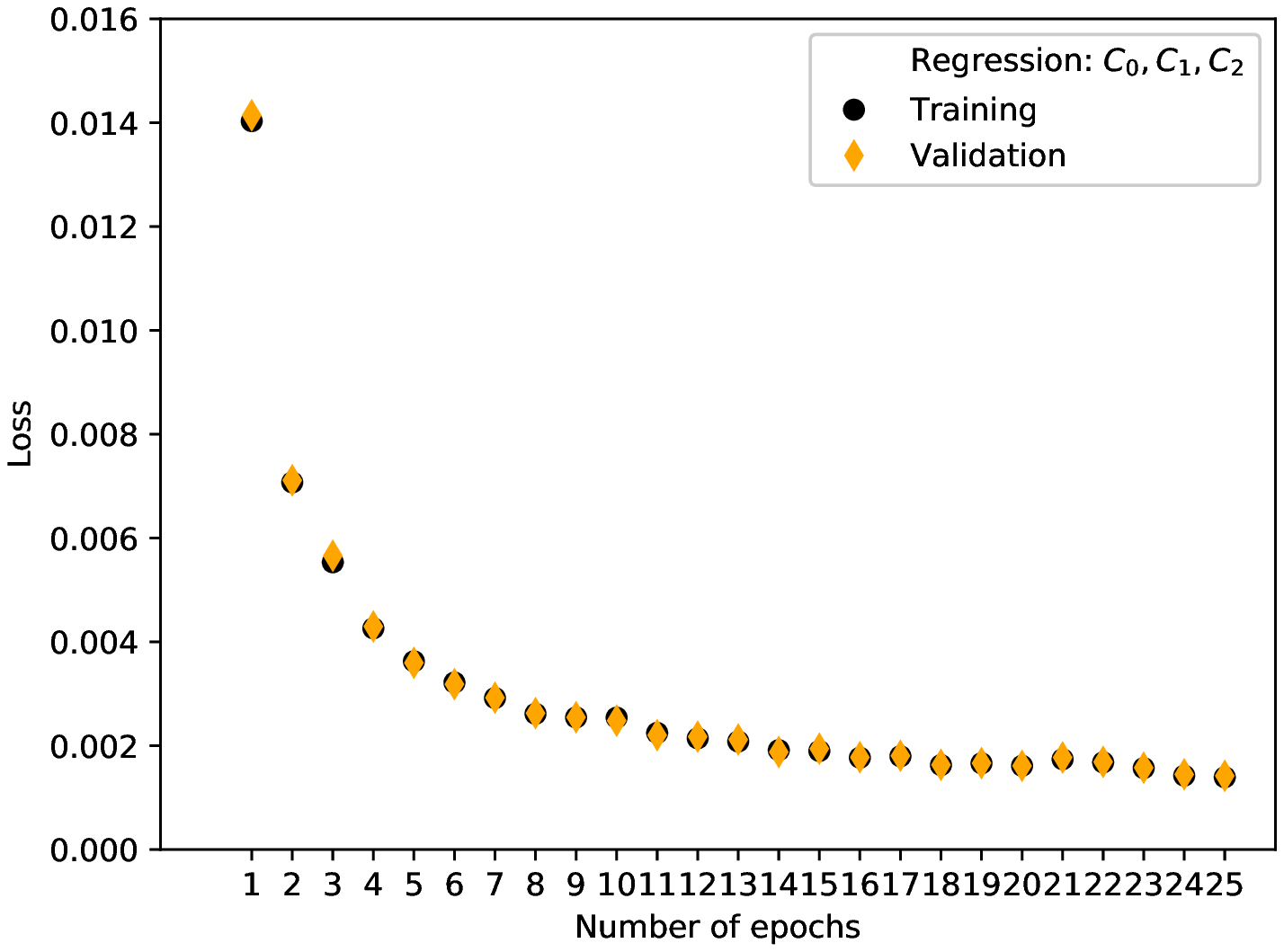}
       \includegraphics[width=7.5cm,angle=0]{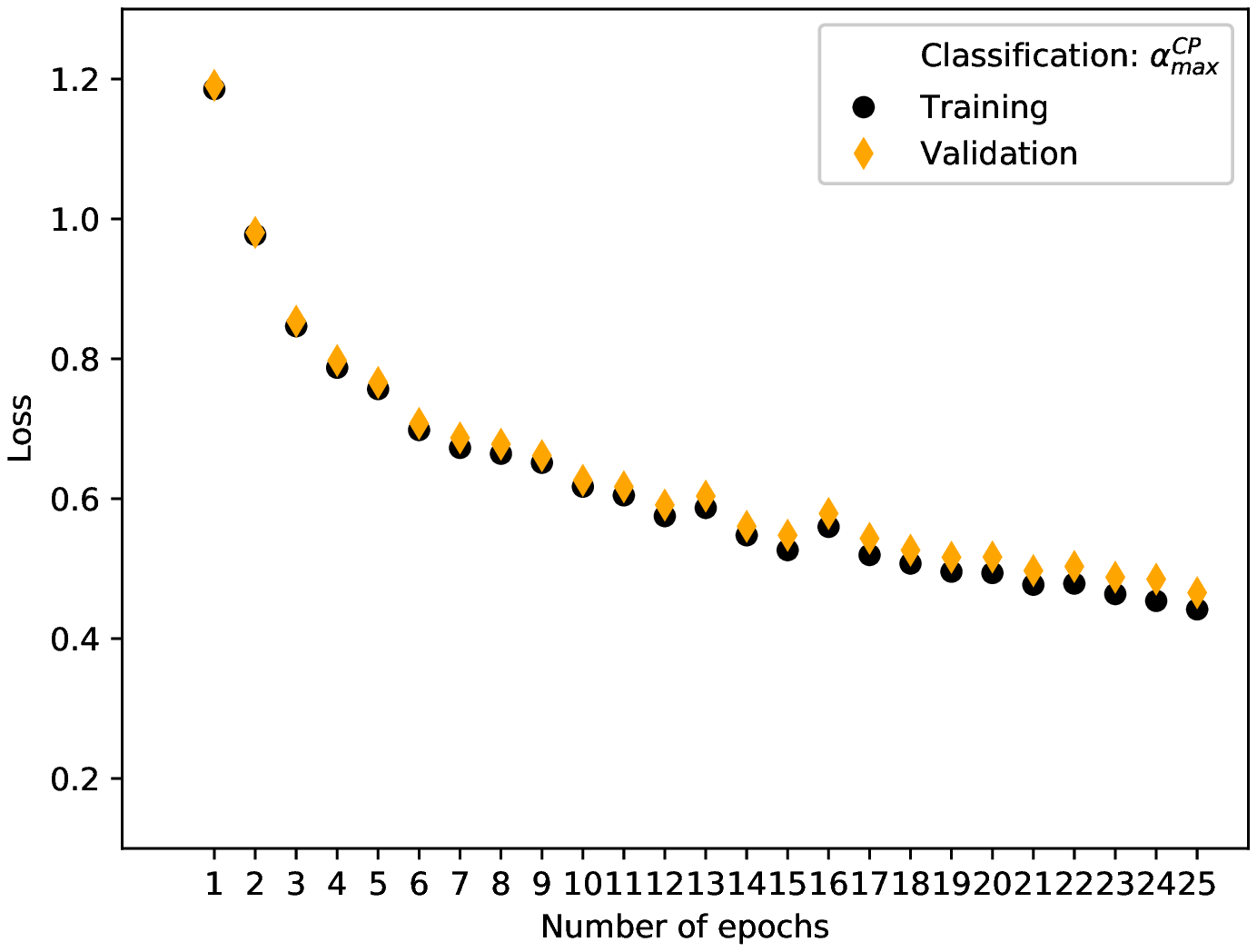}       
       \includegraphics[width=7.5cm,angle=0]{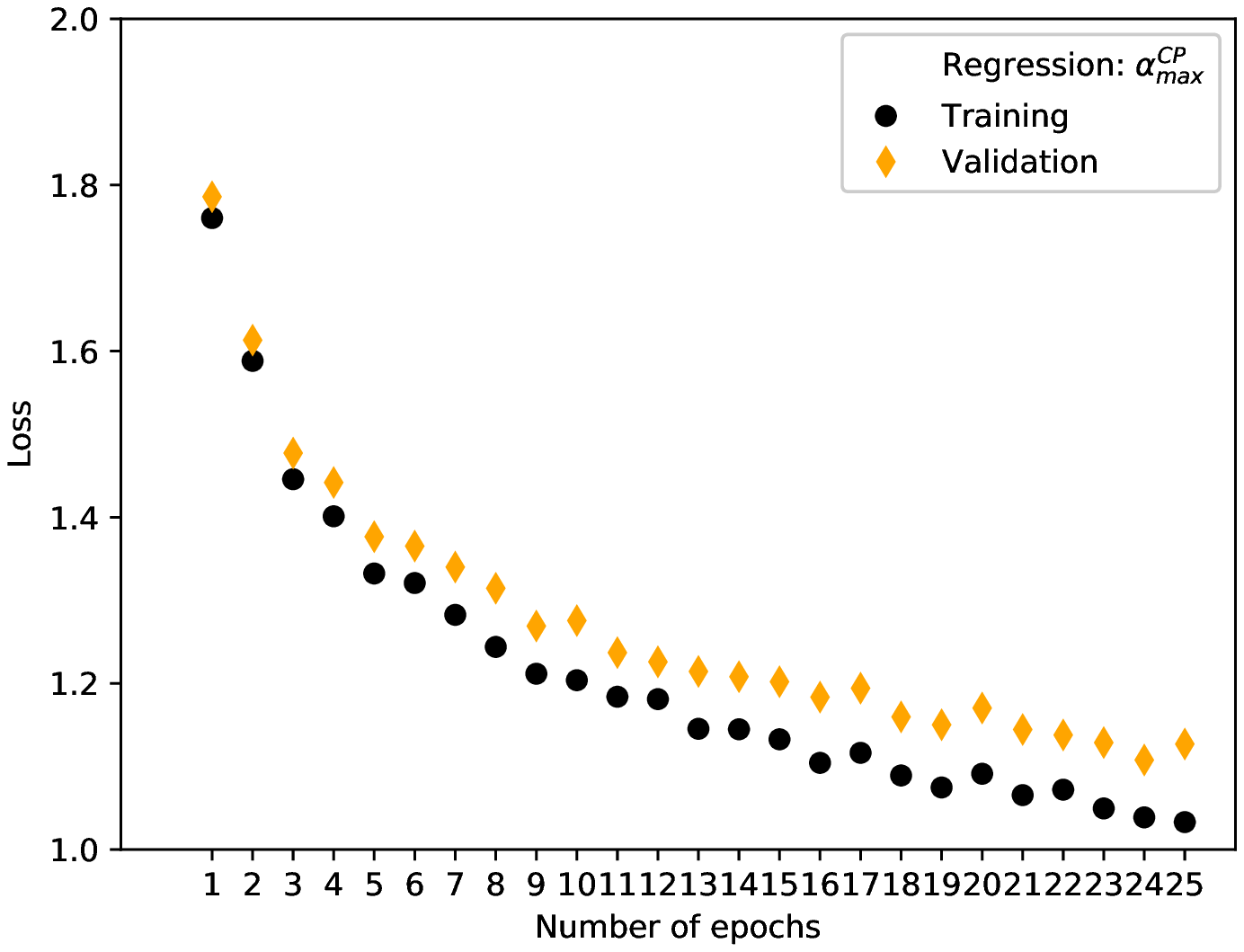}
     }
  \end{center}
  \caption{ The {\it DNN} loss for classification (left-side) and regression (right-side), as function of number
     of epochs used for training. It is shown for learning spin weight (top plots), $C_i$ coefficients (middle plots) and
     most likely  mixing angle $\alpha^{CP}_{max}$ (bottom plots).
     For the classification, $N_{class}$= 21 was used.
   \label{figApp:DNN_loss}}
 \end{figure}

In Fig.~\ref{figApp:DNN_loss}, for all problems considered, distributions of the
loss functions on the training and validation samples, as a function of number of epochs
used for training are shown. Left plots are for the classification and right plots for the corresponding
regression. The values of the loss are case specific and should not be directly
compared, their shape is monitoring the training process. For all cases the loss is decreasing
with number of epochs, both on training and validation samples. It is overlapping 
for all cases except [{\tt Regression:$\alpha^{CP}_{max}$}] (bottom right plot),
for that single case one small loss in performance is observed for validation sample
compared to training sample.
Training with 25 epochs seems sufficient for both classification and regression for all presented scenarios.

\end{document}